\documentclass[acmsmall,authorversion,nonacm]{acmart}

\usepackage{xcolor}
\usepackage{etoolbox}
\newcommand{\etal}{\textit{et al}. }
\newcommand{\ie}{{\em i.e., }}
\newcommand{\eg}{{\em e.g., }}
\usepackage{xcolor}
\usepackage{adjustbox}
\usepackage{float}
\usepackage{xspace}
\usepackage{subfloat}
\usepackage{subfig}
\usepackage{enumitem}

\newcommand\setItemnumber[1]{\setcounter{enumi}{\numexpr#1-1\relax}}
\AtBeginDocument{%
  \providecommand\BibTeX{{%
    \normalfont B\kern-0.5em{\scshape i\kern-0.25em b}\kern-0.8em\TeX}}}

\begin{document}

\title[Modeling Classroom Occupancy using Data of WiFi Infrastructure in a University Campus]{Modeling Classroom Occupancy using Data of WiFi Infrastructure in a University Campus}

\author{Iresha Pasquel Mohottige}
\email{i.pasquelmohottige@unsw.edu.au}
\author{Hassan Habibi Gharakheili}
\email{h.habibi@unsw.edu.au}
\author{Vijay~Sivaraman}
\email{vijay@unsw.edu.au}
\affiliation{%
  \institution{The University of New South Wales, Sydney, Australia.}
  \city{Sydney}
  \state{Australia}
  \postcode{2052}
}

\author{Tim Moors}
\affiliation{%
  \institution{Oiklo Pty Ltd.}
  \country{Australia}}
\email{tim@oiklo.com}

\renewcommand{\shortauthors}{Iresha Pasquel Mohottige, Hassan Habibi Gharakheili, Vijay Sivaraman, and Tim Moors}

\begin{abstract}
	Universities worldwide are experiencing a surge in enrollments, therefore campus estate managers are seeking continuous data on attendance patterns to optimize the usage of classroom space. As a result, there is an increasing trend to measure classrooms attendance by employing various sensing technologies, among which pervasive WiFi infrastructure is seen as a low cost method. In a dense campus environment, the number of connected WiFi users does not well estimate room occupancy since connection counts are polluted by adjoining rooms, outdoor walkways, and network load balancing. 
	In this paper, we develop machine learning based models to infer classroom occupancy from WiFi sensing infrastructure. Our contributions are three-fold: (1) We analyze metadata from a dense and dynamic wireless network comprising of thousands of access points (APs) to draw insights into coverage of APs, behavior of WiFi connected users, and challenges of estimating room occupancy; (2) We propose a method to automatically map APs to classrooms using unsupervised clustering algorithms; and (3) We model classroom occupancy using a combination of classification and regression methods of varying algorithms. We achieve 84.6\% accuracy in mapping APs to classrooms while the accuracy of our estimation for room occupancy is comparable to beam counter sensors with a symmetric Mean Absolute Percentage Error (sMAPE) of 13.10\%.
	
\end{abstract}




\maketitle

\section{Introduction}\label{S1}

	In a large university, enrollments are steadily increasing but resources such as lecture rooms do not grow at the pace of enrollment. As  classrooms are allocated to courses in advance based on enrollment, it is becoming increasingly challenging for estate managers of university campus to allocate the growing enrollment to limited available classroom spaces. However, class attendance often deviates from the class enrollment and widely vary depending on the factors like time-of-day, lecture engagement and availability of virtual learning environments. Therefore, campus management is giving increasing attention to infrastructure that can monitor and maximize use of campus resources. This has led to concepts such as smart campus that incorporates the available infrastructure in the decision making process in order to optimally use the limited resources at minimal costs. 
	
	It is expected~\cite{sensormarket} a significant growth (\ie almost 14\%) in use of occupancy sensors for next four years. However, special-purpose hardware sensors have a high upfront cost and require efforts in deployment and maintenance whereby limiting their adoption only to commercial spaces as opposed to university campuses having large number of buildings. To this end, there is an emerging need for affordable, reliable, low-cost occupancy sensors in the context of large university campuses. As the wireless infrastructure pervades modern campuses and usage of mobile devices is growing, metadata from network of WiFi APs can be used to estimate classroom occupancy  in many university campuses. 
	However, using WiFi APs to estimate occupancy can be challenging and so requires a careful analysis. WiFi signals are not limited to indoor space but pass through walls, and thus devices carried by users in nearby rooms or outside walkways (bystanders) may connect to APs inside rooms -- this can corrupt the occupancy estimations that use WiFi session data. Furthermore, errors occur due to the WiFi users connecting with multiple devices, room occupants connecting to APs outside the room and room occupants who do not show any presence in WiFi.
	
	The focus of our study is to use WiFi sessions data for estimating the occupancy of rooms where formal teachings take place, and enrolled students (class list) are known. Note that in addition to WiFi session data, our method requires two more sources of information namely timetabling and class-list as input. Therefore, estimating the occupancy of social spaces or meeting/seminar rooms is beyond the scope of this work since the additional data sources are not available for these rooms.
	The contributions of our paper are three-fold:(1) We analyze metadata from a dense and dynamic wireless network of our university campus comprising of thousands of APs to draw insights into coverage of APs, behavior of WiFi connected users, and challenges of estimating room occupancy; (2) We propose a method to automatically map APs to classrooms using unsupervised clustering algorithms; and (3) We model classroom occupancy using a combination of classification and regression methods of varying algorithms. Our solution builds upon our previous work \cite{wifiLCN2018} by extending our data analysis to highlight coverage of WiFi APs and dynamics of WiFi clients, and also developing a method to automatically map APs to classrooms. 
	New contributions have helped us improve the performance of our occupancy estimation method presented in the prior conference paper \cite{wifiLCN2018}. We achieve 84.6\% accuracy in mapping APs to classrooms along with a symmetric Mean Absolute Percentage Error (sMAPE) of 13.10\% in estimating room occupancy. Our estimation results are comparable to prior methods which employed dedicated and specialized sensors for room occupancy.
	
	The rest of this paper is organized as follows: \S\ref{sec:prior} describes prior work in estimating occupancy using various sensing technologies, and  \S\ref{sec:WiFiSensing} describes the analysis of data from WiFi sensing. In \S\ref{sec:mapAP}, we present our learning-based approach for mapping APs to classrooms, while in \S\ref{sec:WiFiOccupancy} we develop a model to estimate classroom occupancy. The paper concludes in \S\ref{sec:concl}.

\section{Related Work}\label{sec:prior}
	
	This section briefly presents related prior work and indicates how our work differs from existing approaches. The number of occupants in a room is useful information for a variety of applications such as optimal resource allocation, efficient energy consumption, crowd handling, adaptive network load balancing and security surveillance in residential, commercial and campus buildings. The studies in \cite{Ardakanian, Ghai2012, Candanedo2017, McKenna2015, Zou2017, Cali2015, Mao2016} include related previous work on such applications which define the importance of our work.

	\subsection{Use of Specialized Sensors}
		There are studies that estimate room occupancy using specialized occupancy detection hardware. In~\cite{Dong2014a} researchers used machine learning techniques such as Support Vector Machine (SVM), Neural Networks (NN) and Hidden Markov Models (HMM) to process the data collected from a network of sensors consisting CO$_{\textup{2}}$ monitors and ambient sensors. HMM gave the most realistic results in predicting the number of occupants in offices with $73\%$ accuracy, however it was only tested in small rooms with less than 10 occupants. In their approach to determine occupancy using single passive infrared sensor combined with machine learning techniques Raykov \etal\cite{Raykov2016} proposed a low-cost occupancy estimation solution that produced a mean absolute error (MAE) of $1$, but was tested only in rooms with 14 or less occupants. Golestan \etal\cite{Golestan2018} developed time series neural networks to estimate the number of room occupants with a RMSE of $0.8$ for rooms with maximum 67 occupants. They used a set of occupancy indicative sensors including BLE (Bluetooth Low Energy) beacons. Sgouropoulos \etal in~\cite{Sgouropoulos} achieved a MAE of $1.2$ by employing camera image processing techniques. Paci \etal\cite{Paci2015} utilized camera sensors and thermal comfort sensors combined with Support Vector Regression (SVR) to count number of people inside large lecture rooms. Their approach produced a MAE of $7$ people for rooms with 0 - 150 occupants, but worked well only when there is less movement. However, in image processing-based methods complex processing algorithms require heavy computational resources and if explicit consent is not obtained, privacy remains an issue. Yoshida \etal\cite{Yoshida2015} installed a number of devices (\eg Raspberry Pi) in a room to collect RSSI from WiFi networks. They estimated room occupancy by analyzing changes in signal propagation between APs and installed devices. They employed linear regression (LR) and SVR algorithms and achieved a MAE of $0.47$ in estimating occupancy in rooms with maximum 8 people. All of these approaches are based on special-purpose hardware sensors which require new sensor installations, therefore, have the disadvantage of associated costs in deployment and maintenance.
		
		Another drawback of the existing proposals for occupancy estimation is their requirement of user cooperation. Authors in \cite{Yang2016} employed a specific mobile app to collect Received Signal Strength Indication (RSSI) data from beacons transmitted from Apple's iBeacons. The work in \cite{Conte2015} proposes to estimate room occupancy by modifying the iBeacon protocol. Both approaches require users to install a mobile app on their device to collect and transfer data from the device to a remote processing server. We believe that it can be quite challenging to encourage a reasonable number of users to install a new app which can drain their mobile battery faster due to underlying Bluetooth communication. On the other hand, our use of WiFi session data does not need any changes in either user devices or existing infrastructure.
		
		Furthermore, existing occupancy estimation approaches are tested only in rooms with a limited setting (\ie in one classroom for small number of occupants etc.) hence not clearly indicate how they would scale. Therefore, it is necessary to develop a reliable method which does not require cumbersome dedicated sensor deployment and maintenance.
		Consequently, a number of research efforts emerge in using existing infrastructure such as wireless APs to estimate room occupancy.

	\subsection{Use of Existing Infrastructure}
		Most light-weight approaches for occupancy estimation use existing infrastructure as occupancy sensors. In~\cite{Akkaya2015}, Akkaya \etal highlighted the growing trend to employ implicit sensing infrastructure (\eg electricity/lighting systems, or enterprise computer network) to estimate occupancy due to the associated high costs in deployment and maintenance of special-purpose hardware sensors. They also emphasized the challenges in estimating room occupancy with WiFi AP infrastructure, especially in areas such as lecture theater in a university. 
		Melfi \etal\cite{Melfi2011} employed occupancy sensing methods such as monitoring of MAC and IP addresses in routers and WiFi APs. Although accuracy was within a $10\%$ confidence interval around the ground truth occupancy for whole buildings, it was unacceptably erroneous at floor or room granularity due to the overlap of AP coverage and inconsistent wireless connectivity of devices. Balaji \etal\cite{Balaji2013} attempted to improve the accuracy issues identified in~\cite{Melfi2011} by using occupant identity. They used WiFi MAC address and AP location from WiFi session data and achieved $86\%$ accuracy in determining occupancy in office spaces in a commercial building. 
		Using a combination of number of WiFi devices, electrical energy demand and water consumption, Das \etal\cite{Das2017} achieved an overall MAPE of only $13.22\%$. Ouf \etal\cite{Ouf2017} captured $70\%$ of the variability in room occupancy explained by WiFi device counts in a study that evaluated effectiveness of using WiFi AP data to estimate occupancy as opposed to CO$_{\textup{2}}$ sensors.
		
		Slightly similar to our work, Eldaw \etal \cite{Eldaw2016} attempted to estimate class attendance retrospectively by considering the WiFi traces of selected classrooms for the entire semester as input. Authors associate user ids to a class based on the number of their “revisit” over a semester – in other words, bystanders are filtered out if they appear in WiFi logs of a room less than 50\% of the semester. In our method, instead, we develop a model that estimates the attendance of a classroom using the WiFi log over the class time-slot only, without a need for data of the entire semester. Work by Redondi \etal \cite{Redondi2016} primarily aimed to determine whether a classroom is occupied or not (instead of estimating the count of occupants) by considering WiFi connections from devices inside a classroom. Authors applied a threshold value on RSSI of WiFi connections to filter bystanders. However, they do not provide any insights into the impact of the used threshold values on their estimation (\eg comparing results with/without threshold-based filtering).
			
		WiFi localization is another well-discussed area of occupancy research. Work in \cite{Vasisht2016} develops a method to localize WiFi clients using a single AP and achieves an accuracy up to centimeters – the proposed method required changes inside the WiFi AP. Authors do not attempt estimating room occupancy since they require classroom coordinates to map localized clients to the classrooms. Instead, we estimate classroom occupancy by classifying WiFi users as inside and outside, and thus we do not require any changes in existing infrastructure. Another work by \cite{Jiang2012} localize people using WiFi fingerprints, however required people to install a mobile app to collect channel information.
		
		It is important to note that relying upon purely APs located in a room to estimate occupancy introduce errors in a university campus with high density of APs where occupants in a room may connect to APs both in and around the room. To the best of our knowledge, our work is the first to develop a practical method for mapping APs to rooms using real data and use metadata in WiFi session logs combined with machine learning techniques to estimate occupancy in classrooms with a large number of occupants in a university campus.

\section{WiFi Sensing of Occupants}\label{sec:WiFiSensing}
	In this section, we begin with our method for sensing occupants and describe our dataset. We then clarify challenges of inferring the count of people in a room by touching upon the wide coverage of WiFi APs in a large university campus and drawing basic insights into user connections footprint.

	\subsection {Data Collection}
		We collected daily dumps of WiFi session logs from the IT department of our university for 70 WiFi APs located in 7 lecture theaters on the campus, for the period of 2017-July-31 to 2017-October-27 (\ie sem2-2017) and 2018-February-26 to 2018-June-1 (\ie sem1-2018) -- in our university there are about 5000 APs operational across the entire campus. We chose to select two buildings (teaching-focused) in which majority of lecture theaters are located, and obtained WiFi traces from 70 APs covering selected rooms of various sizes.

		We show in Table \ref{t_sampleWifiLog} a sample of WiFi session logs. Each row of our dataset contains several fields including a unique \textit{User ID} (\ie a unique identifier and password is required for WiFi authentication with enterprise-level security) and the \textit{MAC address} of user device, time when the device  is \textit{associated/}\textit{disassociated} to/from the corresponding AP, \textit{Session duration}, \textit{AP name}, several counters (\ie \textit{Tx/Rcvd Bytes})
		and performance metrics such as the signal strength as shown in Table \ref{t_sampleWifiLog}. Due to the sensitive nature of such information that we used in our work, the research was approved by university Human Research Ethics Advisory Panel under the approval number HC17140 to use the anonymized personal information.

		\begin{table*}[t]
			\renewcommand{\arraystretch}{1.3}
			\caption{Sample of WiFi session logs.}
			\vspace{-2mm}
			\label{t_sampleWifiLog}
			\centering
			\adjustbox{max width = 1\textwidth}
			{
				\begin{tabular}{|c|c|c|c|c|c|c|c|c|c|c|c|}
					\hline
					\textbf{User ID} & \textbf{MAC address} & \textbf{Association time} & \textbf{Disassociation time} & \textbf{Session duration} & \textbf{AP name} & \textbf{Bytes Tx} & \textbf{Bytes Rcvd}  & \textbf{SNR} & \textbf{RSSI} & \textbf{Status}\\
					\hline
					145e7e26 & 00:08:22:60:fb:fe & 31/07/2017 10:40& 31/07/2017 11:15 &35 min & mattap1 & 2717397 &1717397 &31&-63&Disass\\
					\hline
					145e7e26 & 00:1e:64:d5:43:e6 & 31/07/2017 10:55& 31/07/2017 11:20 &25 min & mattap14 & 473749 &2456743 &27&-68&Disass\\
					\hline
					b6c72a33 & 00:34:5c:fb:8d:2b & 31/07/2017 11:15& 31/07/2017 11:20 &5 min & mattap13 & 1465373 &6293826 &35&-61&Disass\\
					\hline
					490801c0 & 00:3b:21:5d:fb:80 & 31/07/2017 20:40& - &20 min & clb17 & 156318 &3462431 &49&-45&Ass\\
					\hline
				\end{tabular}
			}
		\end{table*}

		As shown in Table \ref{t_sampleWifiLog}, records of the first two rows correspond to a user with student identifier \textit{145e7e26} (we obfuscated student IDs). This user connected to the WiFi network from two different devices on day 2017-July-31 -- the second column lists the MAC address of the connected device. For each session, WiFi logs consist of various information related to time including \textit{associated time}, \textit{disassociated time} and also \textit{session duration} -- note that the session duration of each record can be computed from disassociation time minus association time. Next we see the \textit{AP name}. Our university IT department uses a unique name for each WiFi AP and this name often reflects the location (\eg building name, floor) of APs which makes it easier to be used as an identifier compared to AP's MAC address. Furthermore, WiFi session logs include bytes and packets exchanged between WiFi user and AP during each session. The \textit{RSSI} column shows the average strength of signal received during the session while average signal to noise ratio is recorded in \textit{SNR} column. In \textit{Status} column, \(Disass\) indicates a disassociation if the session has been disconnected at the time of generating the report while \(Ass\) indicates an ongoing session at the time of report generation. Since WiFi session logs (reports) are generated at a fixed time everyday (\ie at 9pm) majority of the sessions are recorded as \(Disass\). An example with \(Ass\) status is shown in the fifth row of the Table \ref{t_sampleWifiLog}. For such records, Disassociation Time is not applicable and session duration is computed as the duration of time between the association time and the report generation time. The column \textit{Retries} indicates the number of times data frames have been resent to the receiver until the AP received an ACK (acknowledgment) during the session. The large values seen in the WiFi logs for retries are not surprising since interference and multi-path fading are common in a dense WiFi network.
		
		In addition to WiFi session data, we obtained data of timetabling information containing course timeslots allocated to rooms. Note that we do not have access to course-related information (\eg course name, faculty) which had been filtered due to privacy reasons.

		\begin{figure*}[t!]
			\begin{center}
				\mbox{	
					\subfloat[Student in MatC (10am-11am).]{
						{\includegraphics[width=0.42\textwidth,height=0.3\textwidth]{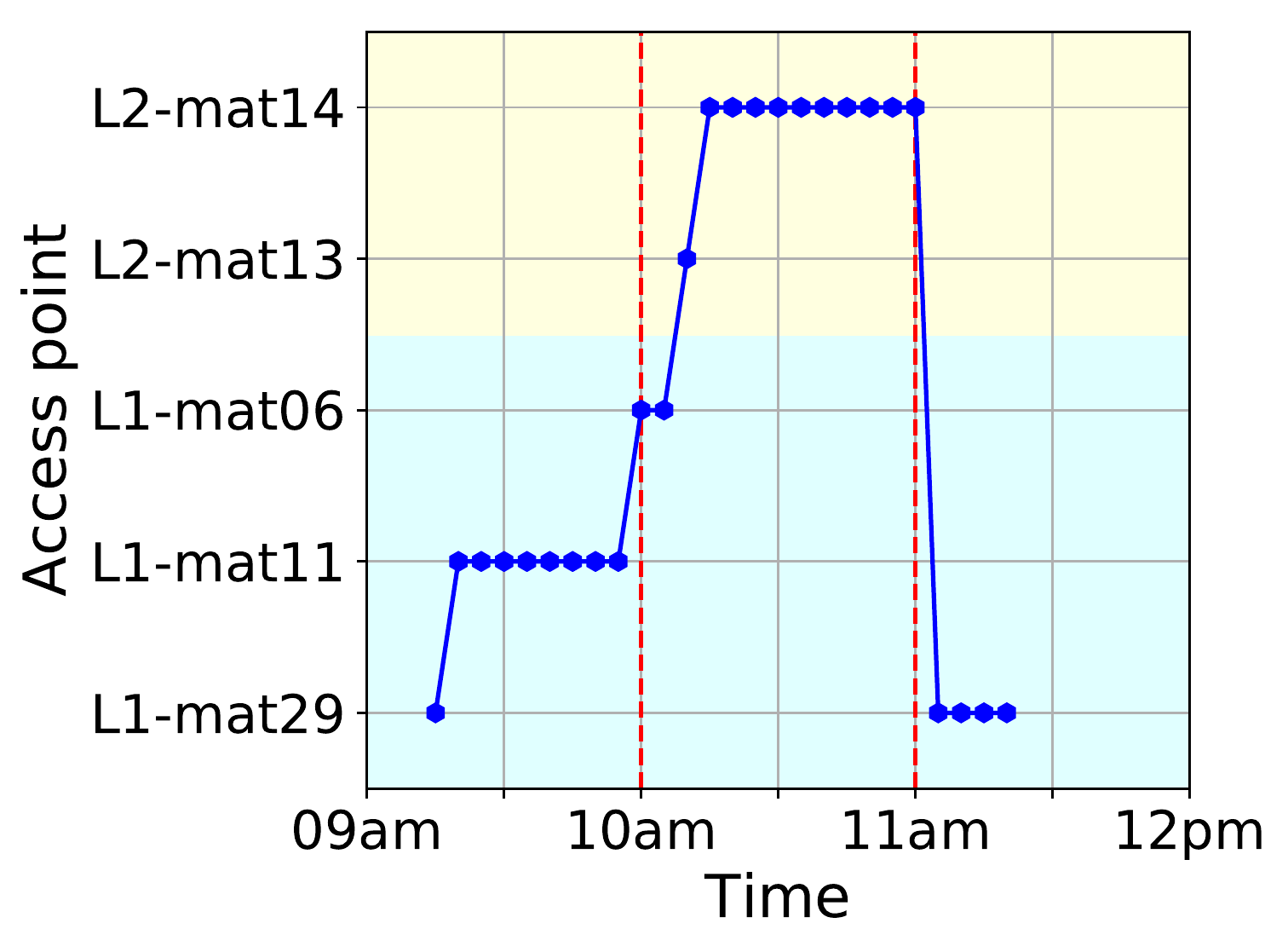}}\quad
						\label{f_studentWifiTrace}
					}
					\subfloat[MatB (L1).]{
						{\includegraphics[width=0.22\textwidth,height=0.30\textwidth]{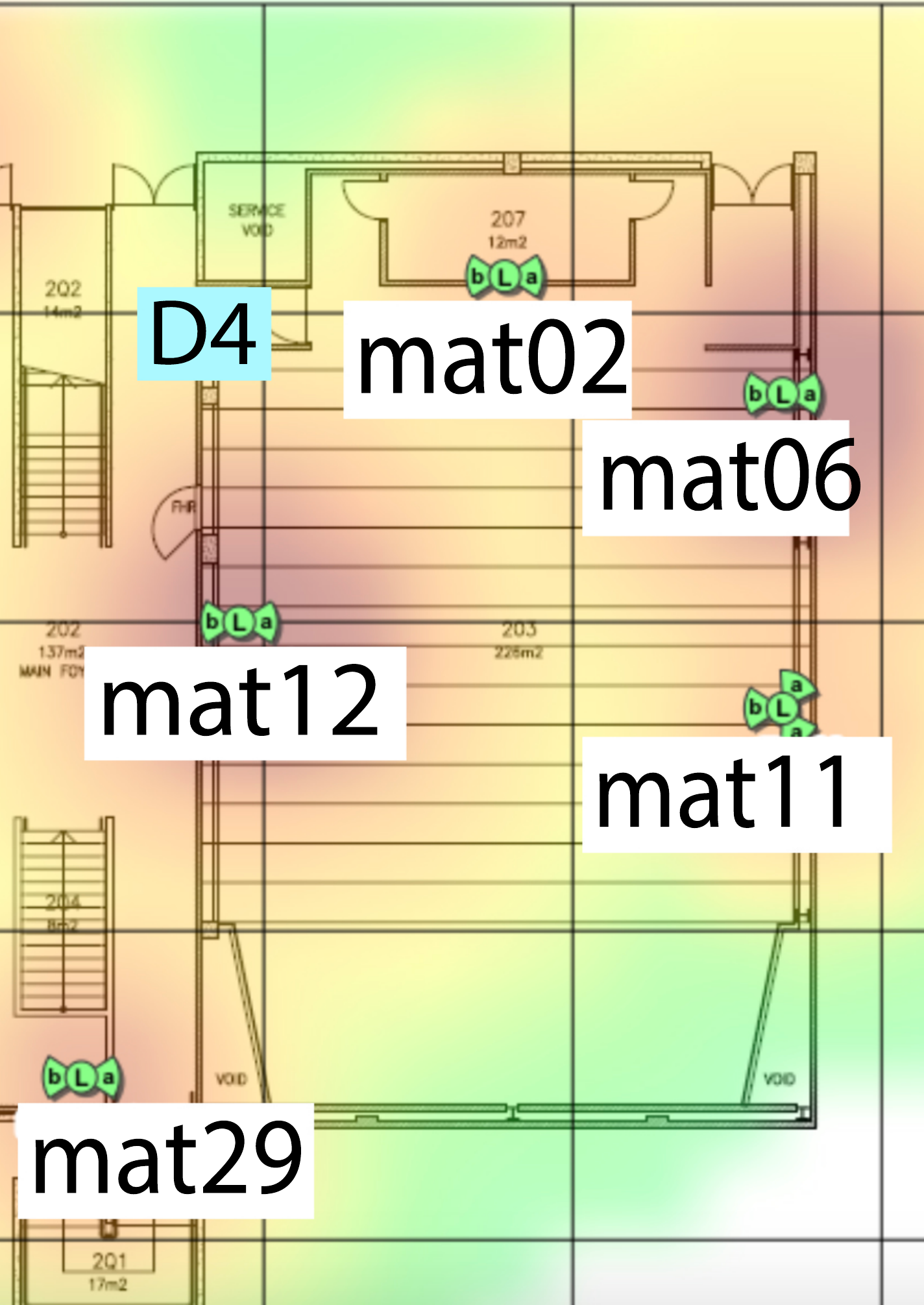}}\quad
						\label{f_MAT_LEVEL1}
					}
					\subfloat[MatC and MatD (L2).]{
						{\includegraphics[width=0.25\textwidth,height=0.30\textwidth]{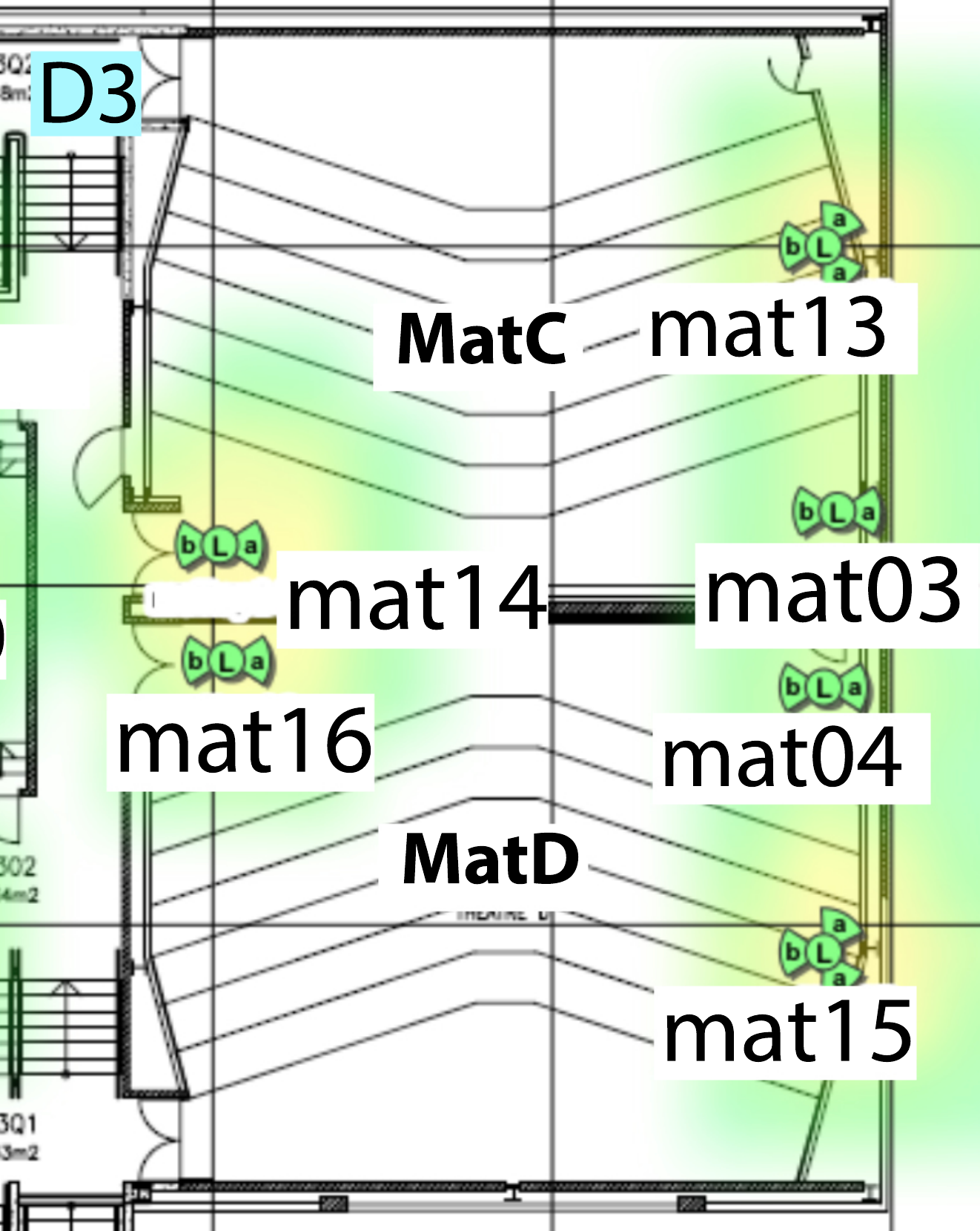}}\quad
						\label{f_MAT_LEVEL2}
					}
				}
				\caption{Time-trace of connected APs for an enrolled student and building AP layout.}
				\label{fig:studentTimeTrace}
			\end{center}
		\end{figure*}
			
	\subsection{Challenges of Inferring Classroom Occupancy using WiFi Traces}\label{subsec:challenges}
	
		In this subsection, we look at a few examples of WiFi users and the variety of their connections due to overlapping coverage of APs found in the dataset from our university campus to show that estimating classroom occupancy requires more knowledge than merely counting  unique user identifiers connected to APs in a room.
	
		\subsubsection{Identifying WiFi Users:\\}\label{sec:identifyUsers}
			By analyzing WiFi session data during a class, we see different types of WiFi users. In Fig. \ref{f_studentWifiTrace}, we plot a time trace of AP connections for a student (with student identifier \textit{a4636cd1}) between 9am and 12pm. This student enrolled in a tutorial class of Course-101 held from 10am to 11am on Fridays in Semester 2, 2017 in classroom MatC. MatC is located at the second floor of a two-story building, \ie Mathews building. APs of level 1 (L1) and level2 (L2) are shaded in light-blue and light-yellow (in Fig. \ref{f_studentWifiTrace}) respectively. Each solid-blue dot indicates the AP to which this student is connected at every 5 minutes. 
			We can see that the student was consistently connected to the WiFi network throughout the period from 9:15am-11:20am -- there were no trace time samples in our dataset (\ie{9am-9:10am and 11:25am-12pm}) during this period of focus, since our dataset only covers 70 out of 5000 APs across our university campus. The student was first seen connected to AP \textit{mat29} located in walkway of L1 in Mathews building, as shown in Fig. \ref{f_MAT_LEVEL1}. The student then got connected to \textit{mat11} in the room MatB in L1 and maintained the connection for about 40 minutes. The student probably attended another class (for which we do not have the ground-truth data) held in MatB from 9am to 10am. At 10am and 10:05am the student connected to \textit{mat06}, still in MatB. Next, at 10:10am the student was seen connected to \textit{mat13} in MatC (shown in Fig. \ref{f_MAT_LEVEL1}) as expected. Then at 10:15am the student connected to AP \textit{mat14} and stay connected to it till 11am. Lastly, the student was captured by \textit{mat29} located at L1 walkway, leaving the building after the class.
			
\begin{figure*}[t!]
				\begin{center}
					\mbox{
						\subfloat[{With two devices.}]{
							{\includegraphics[width=0.30\textwidth]{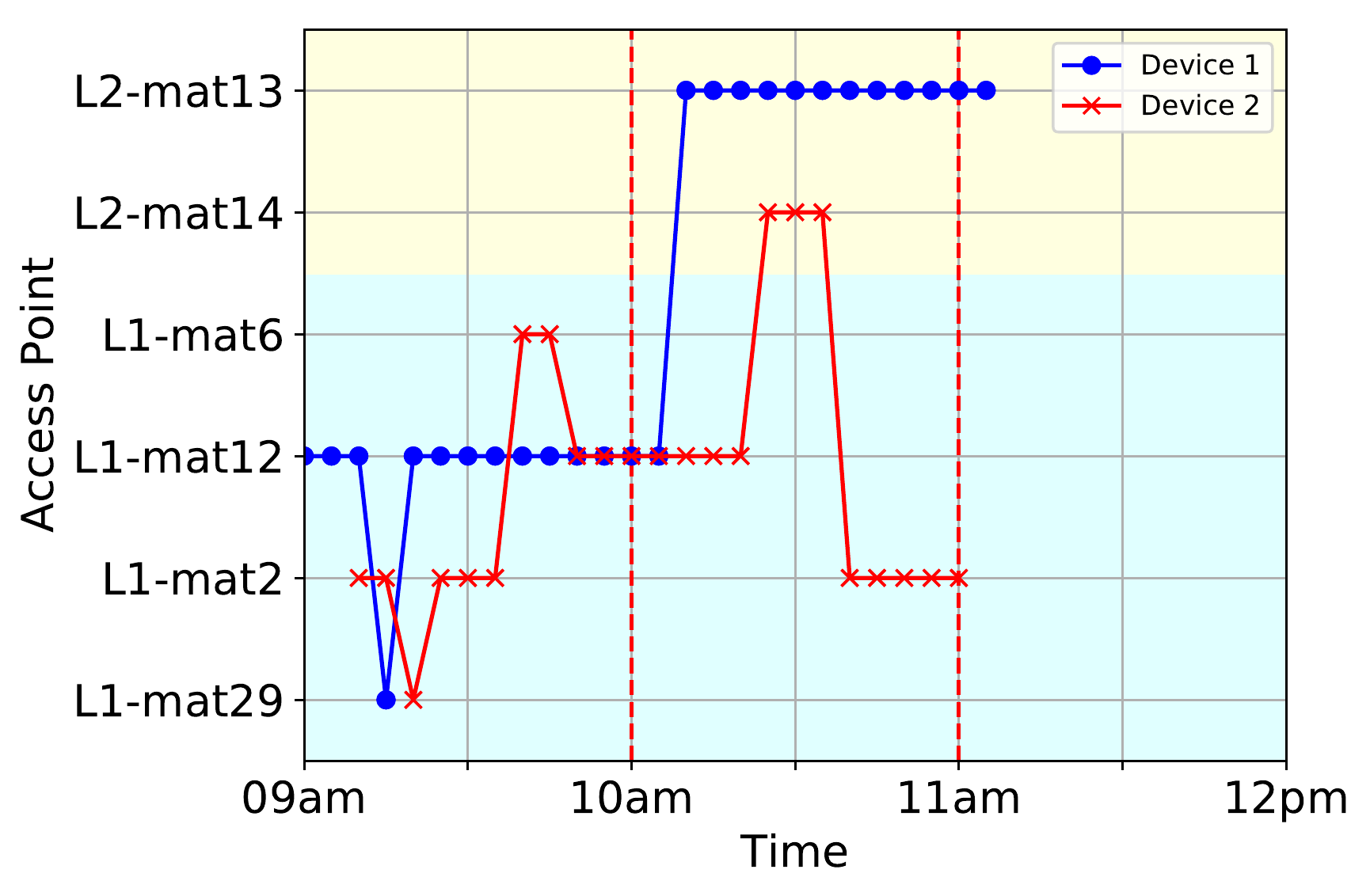}}\quad
							\label{f_multipleDevices}
						}
					}
					\hspace{-5mm}
					\mbox{
						
						\subfloat[{Non-enrolled.}]{
							{\includegraphics[width=0.30\textwidth]{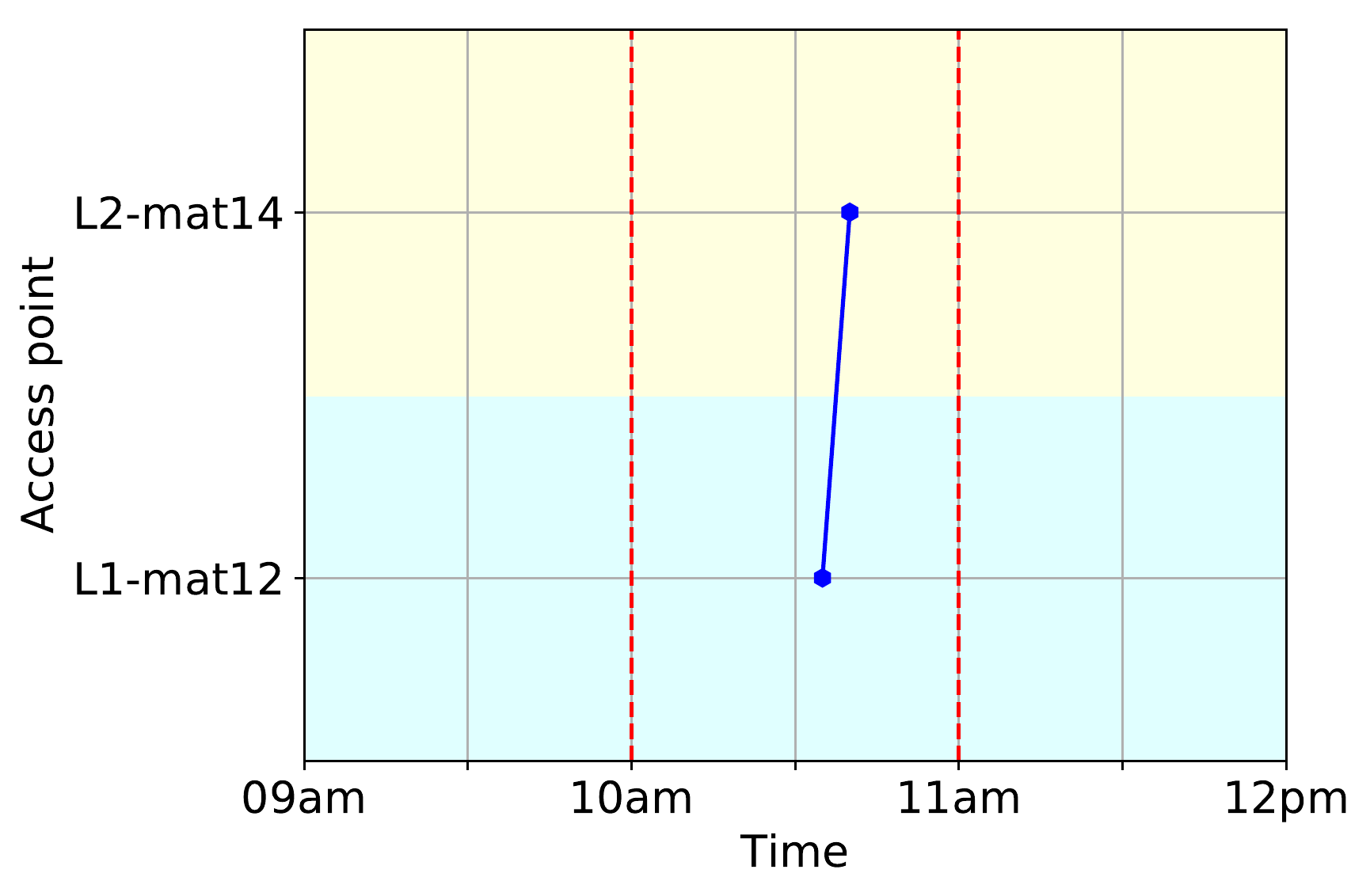}}\quad
							\label{f_studentnotenrolled}
						}
					}
				\hspace{-5mm}
					\mbox{
						
						\subfloat[{Interrupted.}]{
							{\includegraphics[width=0.30\textwidth]{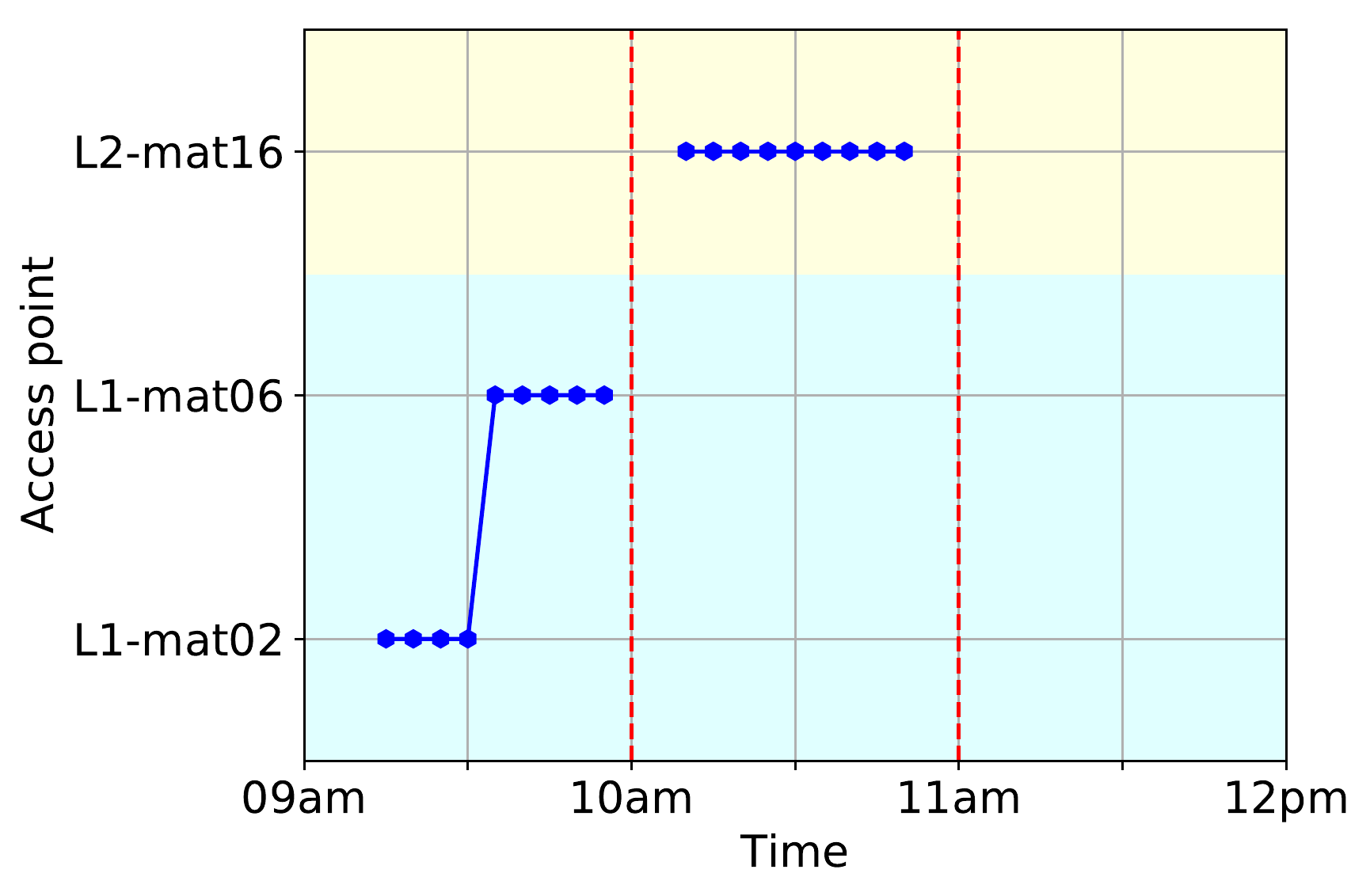}}\quad
							\label{ff_connected_outsideAP}
						}
					}
					\hspace{-5mm}
					%
					\caption{{Time-trace of users connection to WiFi APs in Mathews building.}}
					
					\label{f_deviceAssociations}
				\end{center}
			\end{figure*}

			We show in Fig.~\ref{f_deviceAssociations} various patterns of WiFi connected users during a tutorial class scheduled for 10am-11am in room MatC (top yellow ribbon in these plots corresponds to APs in this room): Fig.~\ref{f_multipleDevices} illustrates a WiFi user connected via two devices, \ie Device1 remains permanently connected to the inside AP \textit{mat13}, and Device2 enters the room with its already established connection to an outside AP \textit{mat12} located at L2, joins (after about 20 minutes) the inside AP \textit{mat14} in this room, and later switches to an outside AP \textit{mat2} located at L2; Fig.~\ref{f_studentnotenrolled} shows a passerby WiFi user temporarily connected to an AP in this classroom; and lastly, Fig.~\ref{ff_connected_outsideAP} shows a WiFi user who is an enrolled student of the tutorial class held in room MatC, but connected to AP $mat16$ located in adjacent room, MatD.  
				This example highlights the variety of users connections that need to be accounted for estimating room occupancy -- multiple connections in Fig.~\ref{f_multipleDevices} are to be aggregated as a single user; the user in Fig.~\ref{f_studentnotenrolled} should be filtered out; and the user in Fig. \ref{ff_connected_outsideAP} should be accounted in estimating the occupancy of the subject class.
	
		\subsubsection{Coverage of WiFi APs:\\}\label{sec:coverageofAPs}
		
\begin{figure*}[t!]
				\mbox{
					\subfloat[Connections to campus-wide WiFi APs.]{
						{\includegraphics[width=0.45\columnwidth, height=.4\columnwidth]{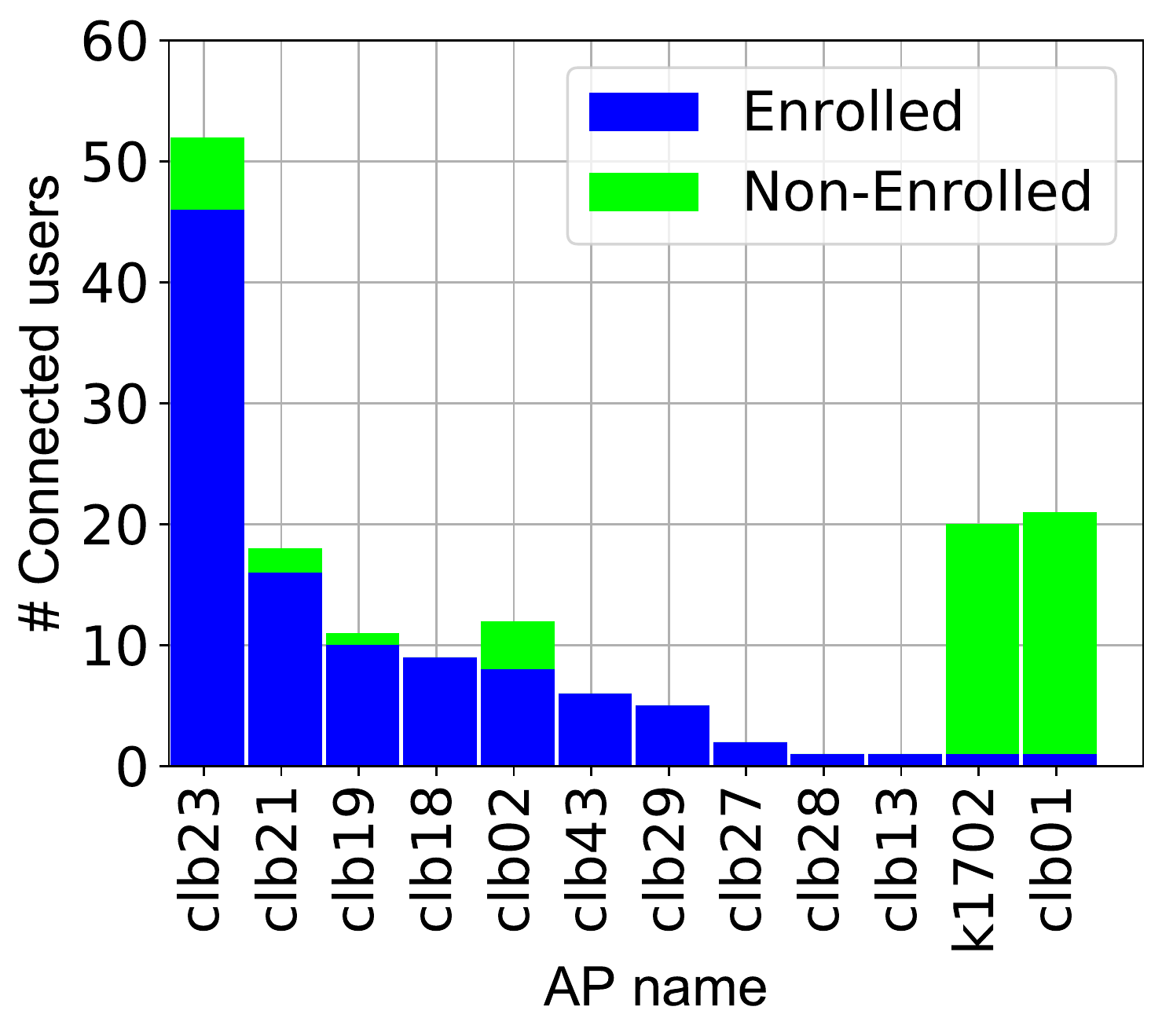}}\quad
						\label{fig:f_APconnections}
					}
				}
				\hspace{1cm}
				\mbox{
					\subfloat[$CLB8$.]{
						{\includegraphics[width=0.2\columnwidth, height=.38\columnwidth]{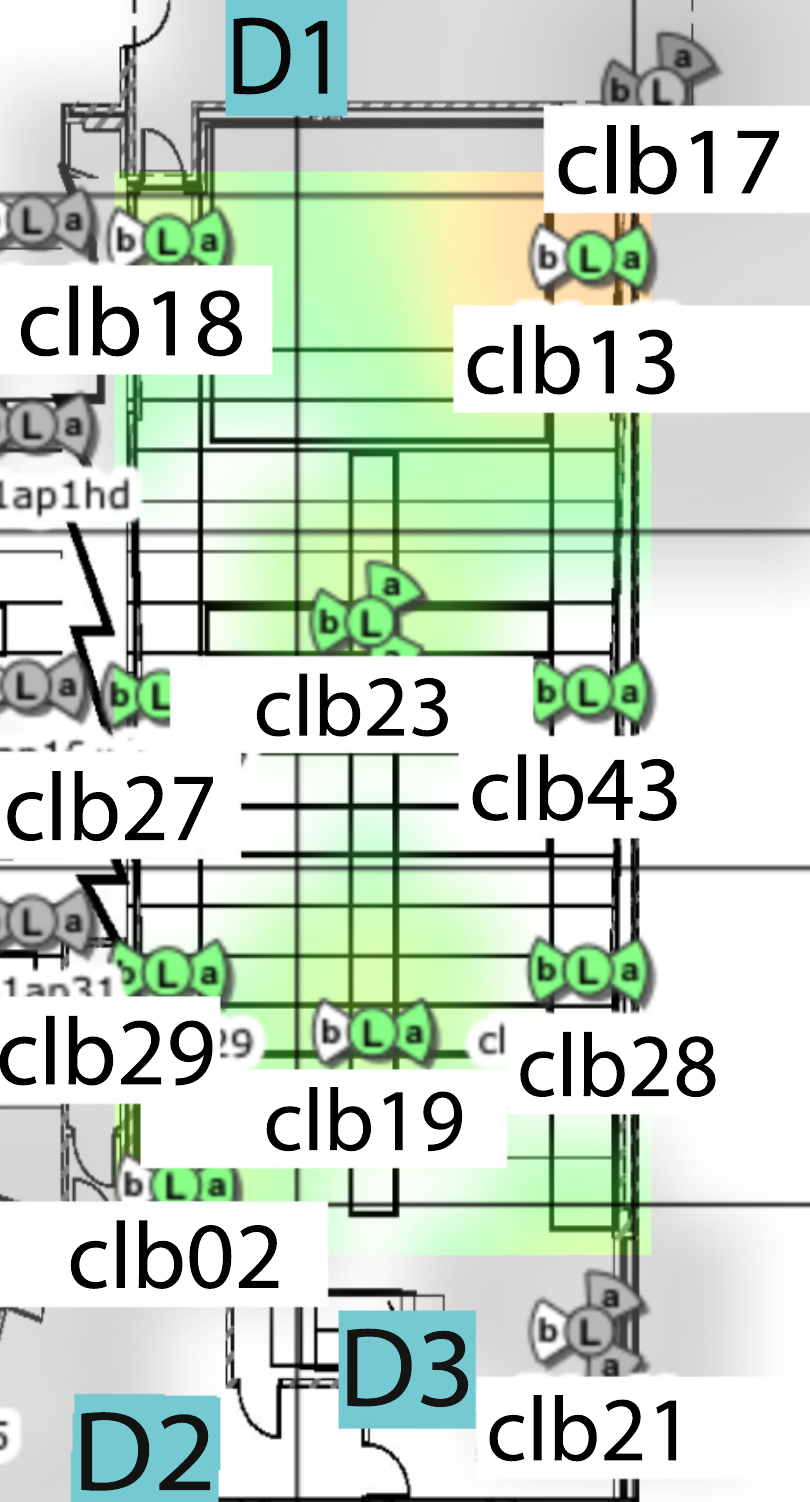}}\quad
						\label{fig:CLB8}
					}
				}
				\caption{{Enrolled/non-enrolled user connections to APs in $CLB8$ (lecture theater with 9 APs and three doorways, \ie D1, D2 and D3) during a selected class and room AP layout.}}
				\label{fig:APConnections}
				
			\end{figure*}

			We performed several spot measurements in real classes to correlate attendees' layout (their seating pattern) in classroom and the corresponding WiFi session logs. As an example, we show in Fig.~\ref{fig:f_APconnections}, our observation for a class in the theater $CLB8$. We show the layout of APs for $CLB8$ with 9 APs and three doorways in Fig.~\ref{fig:CLB8}. This selected class had 212 students enrolled and was scheduled on Tuesdays from 1pm to 2pm during semester 2, 2017 -- the observation was made at 1.30pm. We see the number of WiFi users connected at each AP (all APs to which at least one enrolled student is connected) at that time. In the Fig.~\ref{fig:f_APconnections} the enrolled and non-enrolled students are shown by blue and green bars respectively. For this measurement, we observed that many students were clustered in the middle of the class as indicated by the highest number of room occupants connected to AP \textit{$clb23$} located in the middle of the room, as shown in Fig.~\ref{fig:CLB8}. 
			Another observation was that a group of students sat near doorways and thus got connected to their nearest APs, \ie \textit{$clb19$} close to D3, \textit{$clb18$} close to D1, and \textit{clb2} close to D2 as shown in Fig. \ref{fig:CLB8} -- each of these APs serve about 10 WiFi users.    
	
			Interestingly, $clb21$ shows the second highest number of occupant connections, though it is located outside the room (but close to doorway D3 at back). This is probably because students who enter the room from entrance D3 has a high chance of sitting at the back and kept their connection to the same AP -- they connected to \textit{clb21} while entering the room. 
			This observation shows that just considering those APs located inside a classroom may result in missing out a significant number of occupants connected to an external close-by AP (\ie \textit{$clb21$} in our example). Therefore, it is important for each classroom to identify APs (inside and outside but close-by) that serve attendees. In other words, we need to map WiFi APs to campus  classrooms. This becomes useful to count enrolled students connected to APs associated with the corresponding classroom. 
			We also note that there are enrolled students who may not always attend the class and connect to other APs far from the room during the class, (\ie connections to \textit{k1702}, shown by the second blue bar from the right in Fig.~\ref{fig:f_APconnections}), thus should not be mapped to the room of interest. Similarly, AP \textit{clb01} and \textit{clb28} shows a few connections from enrolled students -- these APs are located inside a classroom  next to CLB8.
				Therefore, it is important to account for all the WiFi connections made to APs both inside and outside the classroom, especially those that cover a significant number of room occupants \eg $clb21$ in Fig.~\ref{fig:APConnections}.

	\begin{figure}
	\centering
	\begin{minipage}{.4\textwidth}
		\centering
		\includegraphics[width =0.8\textwidth, height=3.5cm]{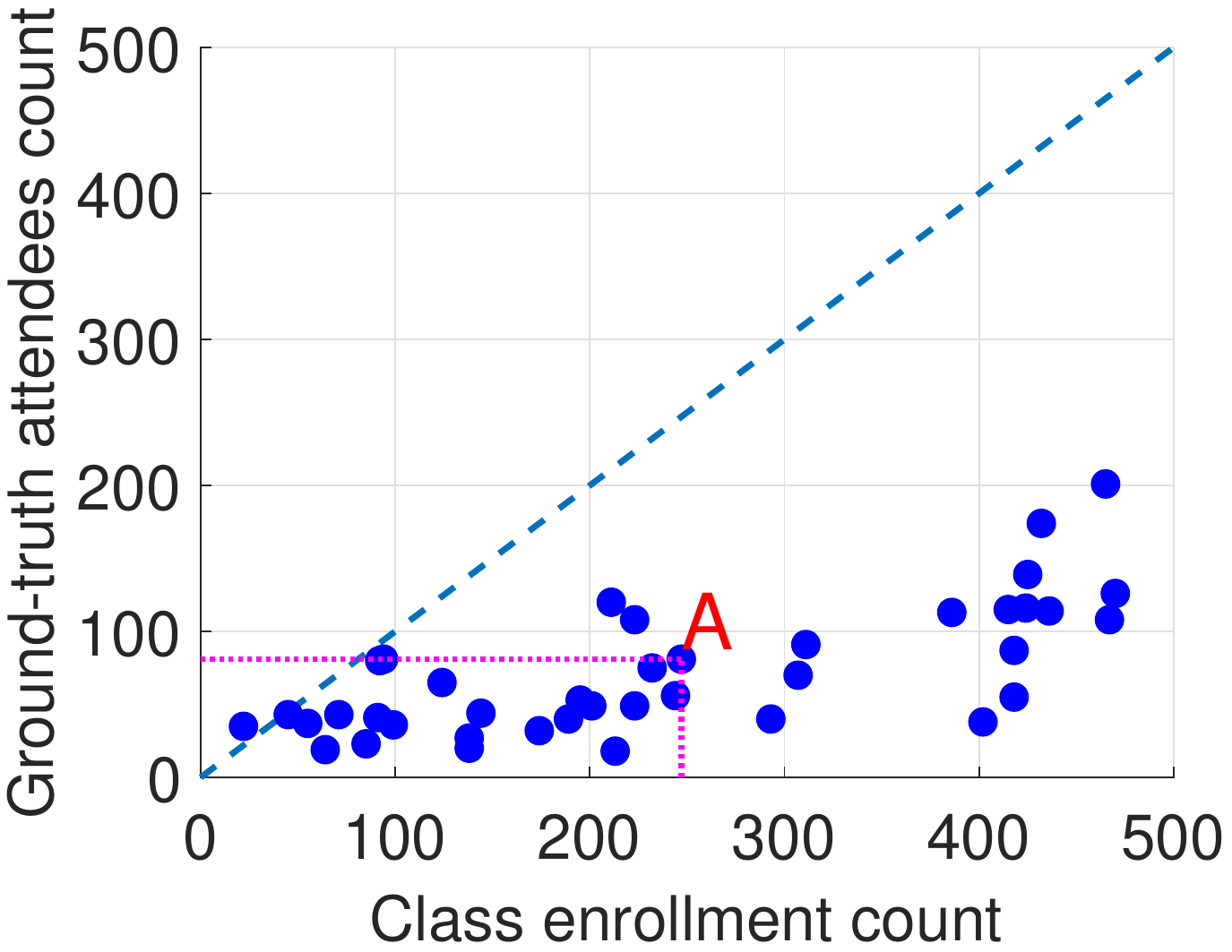}
		\caption{Ground truth data showing class attendance is often lower than its enrollment.}
		\label{f_attendeesVSenrollments}
	\end{minipage}%
	\hspace{7mm}
	\begin{minipage}{0.4\textwidth}
		\centering
		\includegraphics[width =1.1\textwidth, height=3.5cm]{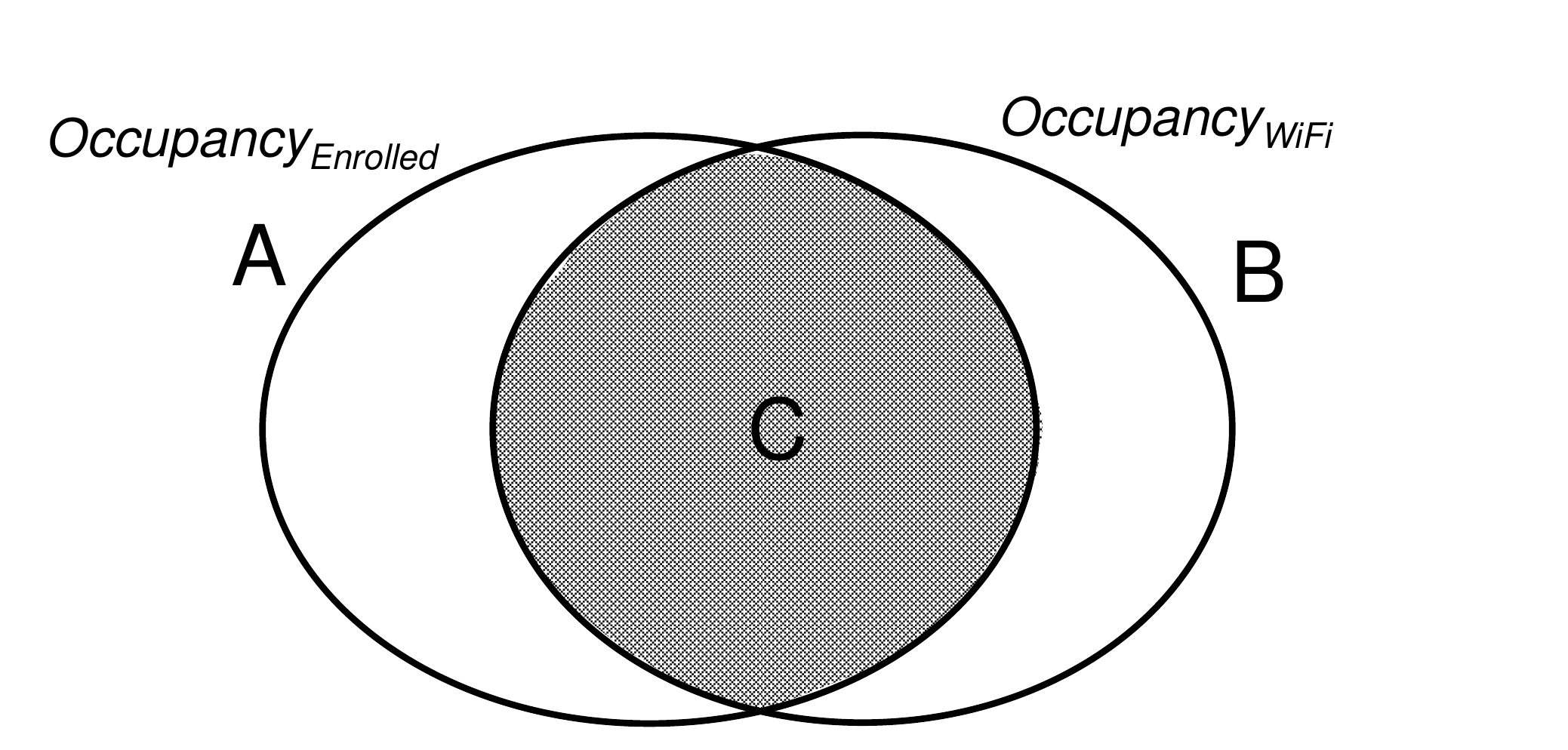}
		\caption{Classroom occupancy is inferred from the intersecting of enrolled WiFi users and all WiFi users connected to class APs.}
		\label{f_occupancyVenDiagram}
	\end{minipage}		
\end{figure}		
	
		\subsection{Why Filtering Bystanders and Mapping APs to Rooms?}
	
		We collected ground-truth attendance data of 40 classes held on campus -- our samples cover a variety of courses and classroom locations, from different days of week as well as different times of day. We plot in Fig.~\ref{f_attendeesVSenrollments} the count of attendees versus enrollments across classes -- each blue circle represents a class. It is seen that the attendance count is well below the enrollments for most of the classes (\ie circles fall under the line $y=x$), especially for larger classes with enrollment counts of more than 200. For example, the class highlighted by red letter ``A'' in Fig.~\ref{f_attendeesVSenrollments} has an enrollment of 247 while the attendance was only 81 students.
		This clearly highlights the need for measuring class attendance patterns automatically and continuously, enabling university estate managers to optimize the usage of classroom spaces.
		
		In addition to ground-truth data, we obtained WiFi session logs and class lists (enrolled students) for the above 40 classes to analyze class attendance and count of WiFi users (connected to APs inside these individual classrooms). For each class, the WiFi session data of all APs in a classroom (where the class is run) was considered. 
		We denote: (a) class enrollment count by ``\(Occupancy_{Enrolled}\)'' that is obtained from class list; (b) measured attendee count by ``\(Occupancy_{WiFi}\)'' which is the total number of WiFi users during a class; (c) measured enrolled count by ``\(Occupancy_{EnrolledWiFi}\)'' that is the number of enrolled students connected to WiFi during that class. As illustrated in Fig.~\ref{f_occupancyVenDiagram}, \(Occupancy_{EnrolledWiFi}\) (set C) is the intersection of the other two sets namely \(Occupancy_{Enrolled}\) (set A) and \(Occupancy_{WiFi}\) (set B).
		
\begin{figure}
			\centering
			\begin{minipage}{0.27\textwidth}
					\centering
					\includegraphics[width =\textwidth]{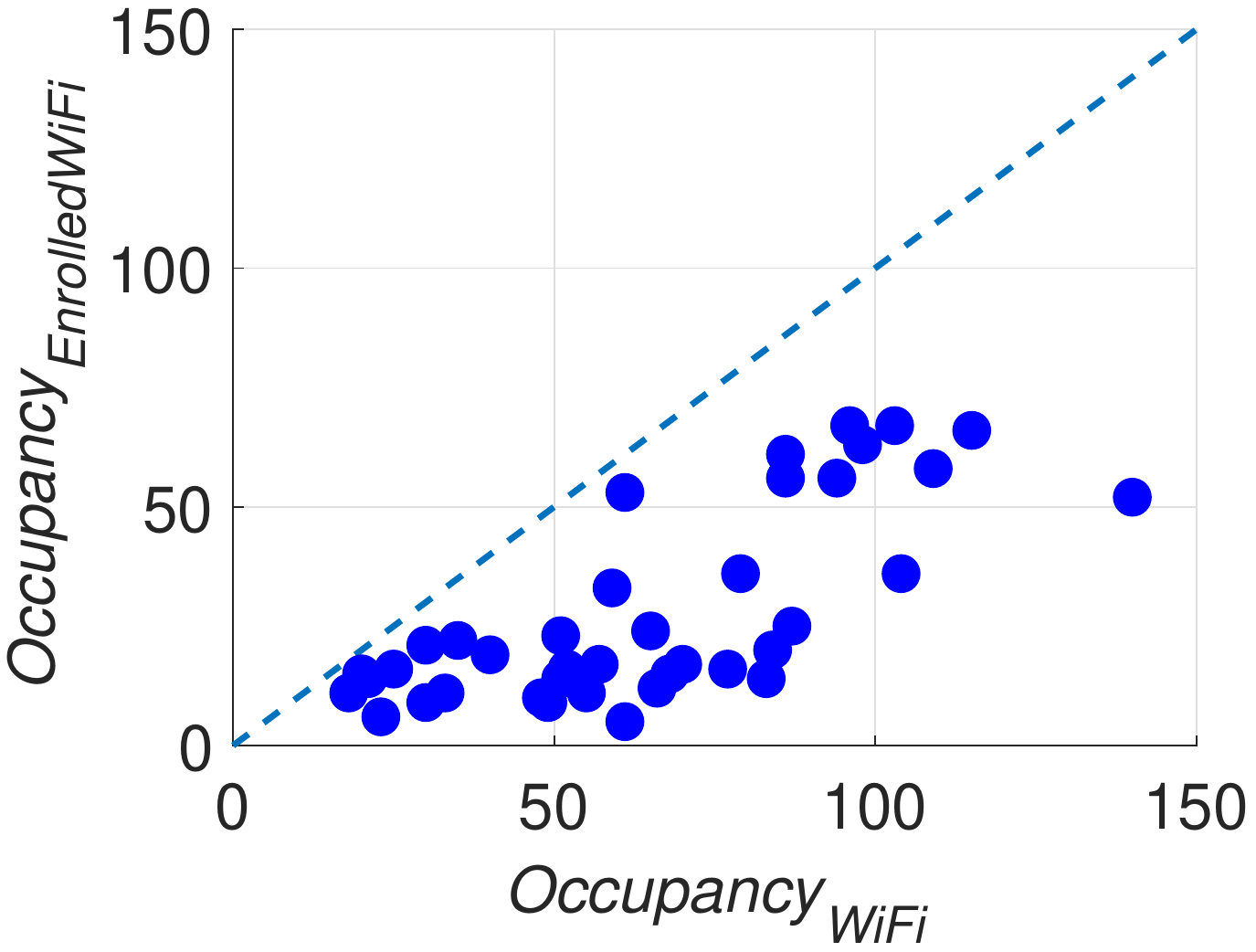}
					\caption{WiFi users connected to APs corresponding to a given room include both occupants and bystanders.}
					\label{f_allAttendeesVSenrolledAttendees}
			\end{minipage}
			\hspace{2mm}
			\begin{minipage}{0.27\textwidth}
				\centering
				\includegraphics[width =\textwidth]{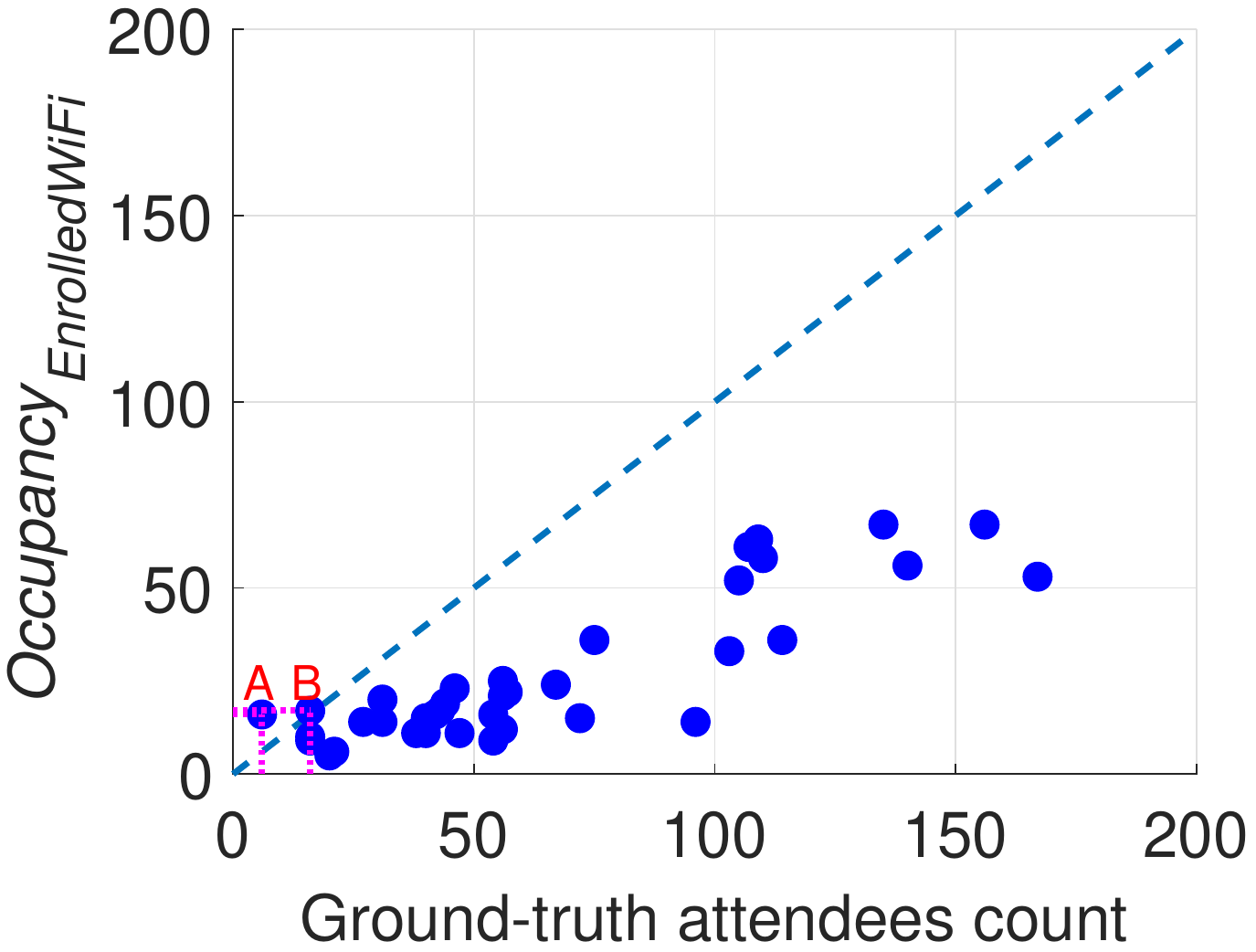}
				\caption{Ground-truth occupancy does not exactly match \(Occupancy_{EnrolledWiFi}\).}
				\label{f_enrolledAttendeesVSgroundTruth}
			\end{minipage}
			\hspace{2mm}
			\begin{minipage}{0.27\textwidth}
				\centering
				\includegraphics[width =\textwidth]{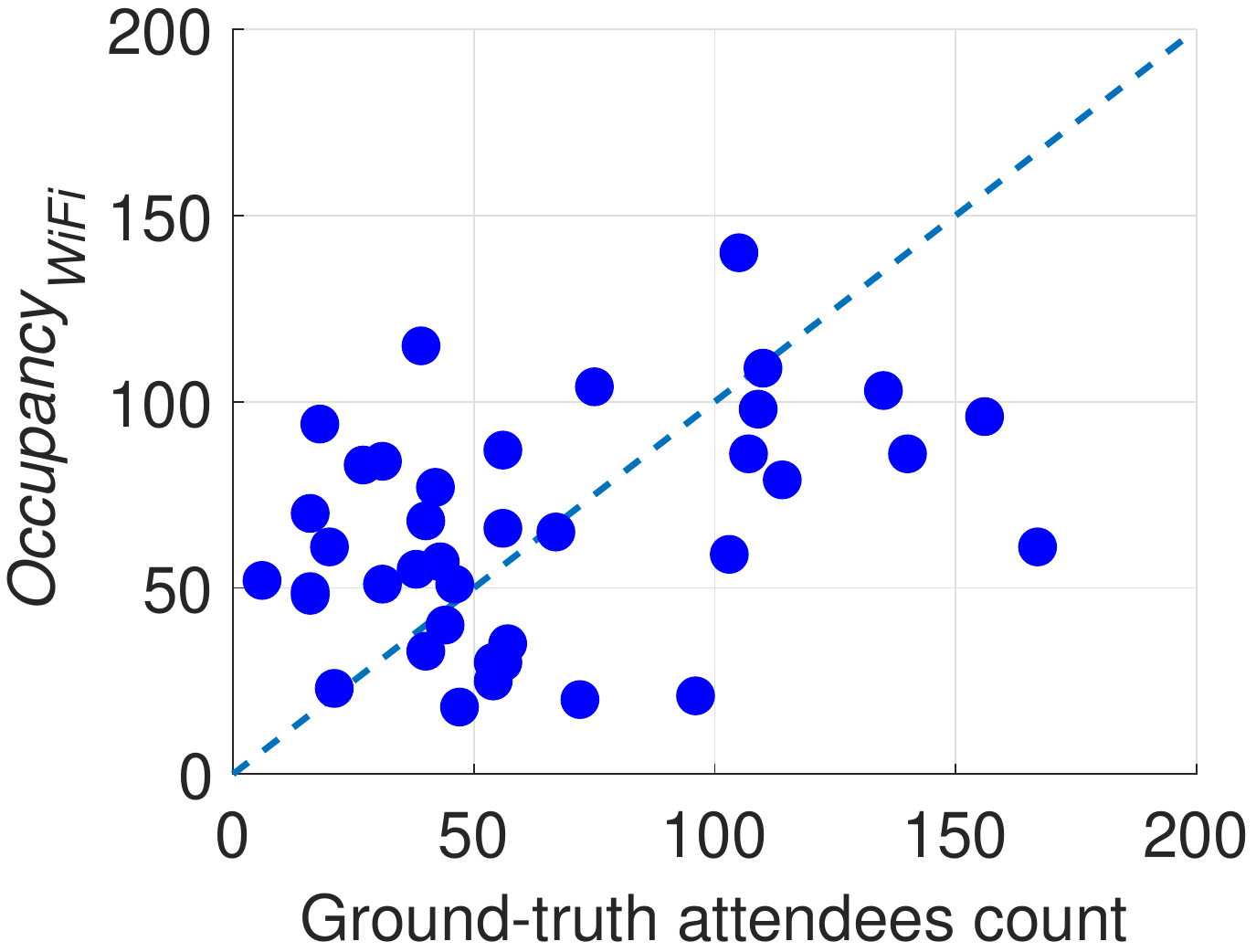}
				\caption{Ground-truth occupancy correlates better with \(Occupancy_{EnrolledWiFi}\) than \(Occupancy_{WiFi}\).}
				\label{f_allAttendeesVSgroundTruth}
			\end{minipage}%
			
		\end{figure}

		We found that \(Occupancy_{WiFi}\) was always higher than the \(Occupancy_{EnrolledWiFi}\), as shown by the scatter plot in Fig.~\ref{f_allAttendeesVSenrolledAttendees}. This indicates that the \(Occupancy_{WiFi}\) covers a variety of WiFi users including enrolled students in a class, students in adjacent rooms, and also passersby/bystanders  as discussed in \S\ref{sec:identifyUsers}. 
		Furthermore, we plot \(Occupancy_{EnrolledWiFi}\) versus ground-truth attendees in Fig.~\ref{f_enrolledAttendeesVSgroundTruth} to show that  \(Occupancy_{EnrolledWiFi}\) was lower than the observed actual occupancy in many classes. Such cases occur when some of the room occupants connect to APs outside of the classroom, or they do not connect to university WiFi network (\eg{instead they may turn off devices during lectures, or use Internet via their personal mobile 3G/4G}). In \S\ref{sec:coverageofAPs} we showed that  certain APs located outside of the classroom may cover a significant number of occupants inside \eg $clb21$ in Fig.~\ref{fig:APConnections}. Also, we observe two outliers (highlighted by A, B in Fig.~\ref{f_enrolledAttendeesVSgroundTruth}), where the ground-truth occupancy is less than the \(Occupancy_{EnrolledWiFi}\). This is possible when enrolled students connect to APs in the classroom from outside but within close proximity.

		Comparing Fig.~\ref{f_enrolledAttendeesVSgroundTruth} with Fig.~\ref{f_allAttendeesVSgroundTruth} shows visually that the ground-truth attendees count displays a better correlation with \(Occupancy_{EnrolledWiFi}\) than with \(Occupancy_{WiFi}\). 
		We verified this by computing the Pearson's correlation coefficient for these two pairs that are found to be $0.77$ and $0.35$ for  \(Occupancy_{EnrolledWiFi}\)  and \(Occupancy_{WiFi}\), respectively.
		Also, we computed sMAPE when occupancy is estimated by measuring the count of all WiFi users versus the count of enrolled WiFi users. As shown in Table~\ref{t_initialanalysis}, a slightly lower error is obtained when enrolled students are considered for class occupancy. Therefore, filtering out non-enrolled user from from the WiFi session logs would enhance the estimation.     
		
		\begin{table}[t!]
			\renewcommand{\arraystretch}{1.3}
			\caption{Correlation between ground-truth attendees count and measured count considering all WiFi users and enrolled WiFi users.}
			\label{t_initialanalysis}
			\centering
			\adjustbox{max width = \columnwidth}{
				\begin{tabular}{|l|c|c|}
					\hline
					\textbf{Measured occupancy} & \textbf{Pearson's correlation coefficient} & \textbf{sMAPE}\\
					\hline
					All WiFi users & $0.35$ & $26.3$\%\\
					\hline
					Enrolled WiFi users  & $0.77$ & $24.1$\%\\
					\hline
			\end{tabular}}
			\vspace{-3mm}
		\end{table}

		Based on our findings  so far, in \S\ref{sec:mapAP} we will develop a method to automatically map campus APs to their corresponding classrooms. Next, in \S\ref{sec:WiFiOccupancy} we will use WiFi session data of the APs mapped to individual classrooms to estimate room occupancy using machine learning techniques.
	
	
\section{Mapping WiFi APs to classrooms}\label{sec:mapAP}
	
	{In a large university campus with nearly 100 acres of real estate, and over 50,000 students and staff, the IT department of the university operates a dense and dynamic network comprising thousands of wireless APs to provide an improved WiFi experience for users. We use WiFi AP logs to estimate classroom occupancy, therefore knowing what APs cover the room occupants is important. We saw in \S\ref{sec:coverageofAPs} even APs outside a room can largely cover occupants of the room because WiFi signals go through walls. Although wireless site surveys provide records of AP locations in an area, it is cumbersome to manually combine such data with a system that counts room occupants, especially in a dense university campus where there is a large number of buildings and APs, considering the time to be spent and errors that may occur. On the other hand, surveys do not provide up-to-date information on how room occupants are covered by the APs located in and around the rooms. In this section, we present our method to automatically map APs to classrooms of a university campus.} We develop a practical application based on realistic data collected from 70 APs on the campus to map these APs to their corresponding classrooms -- note that some APs do not associate with any rooms since they are located in corridors or walkways.

	\subsection{Feature Selection for WiFi APs}\label{pd}
		It is possible to compute how many users are connected to a particular AP at a given time using the WiFi session logs that provide the unique user identifiers, time of associations, time of disassociation, and the connected AP for each session. Similarly, the number of enrolled students connected to a particular AP can also be computed using the enrollment list (\ie class lists) of the class held in the room at the time of interest. During a particular class, at every fixed time interval (\eg every 10 minutes) we compute the following two features:

		\begin{enumerate}
			\item \(fracClass\): Fraction of connections made by students enrolled in the class to a particular AP, \eg 25\% of the students enrolled in the class might connect to $AP_{k}$ giving $AP_{k}$ a \(fracClass\) of 25\%.

			\begin{equation}
			\label{eq_fracClass}%
			{fracClass_{AP_{k}}=\frac{No.\:of\:enrolled\:connections\:to\:AP_{k}}
				{\sum_{i}\:enrolled\:connections\: to\:AP_{i}}}
			\end{equation}
			
			\item \(classFrac\): Fraction of connections to an AP that were made by students enrolled in the class, \eg 60\% of the connections to $AP_{k}$ might be made by students enrolled in the course.
			
			\begin{equation}
			\label{eq_classFraq}%
			{classFracs_{AP_{k}} =\frac{No.\:of\:enrolled\:connections\:to\:AP_{k}}
				{No.\:of\:connections\: to\:AP_{k}}}
			\end{equation}
			
		\end{enumerate}
		
		The parameter \(fracClass\) is a measure of how each AP covers the connections of enrolled students. For an AP located faraway from the room, the number of connected enrolled students is typically smaller than that for an AP located in or around the room, hence a lower \(fracClass\) is expected. The other parameter we define is \(classFrac\) which indicates how vulnerable each AP is to the connections from WiFi users located outside the room of interest.
		
\begin{figure*}[t!]
				\begin{center}
					\mbox{
						\subfloat[{every 2 minutes.}]{
							{\includegraphics[width=.32\textwidth,]{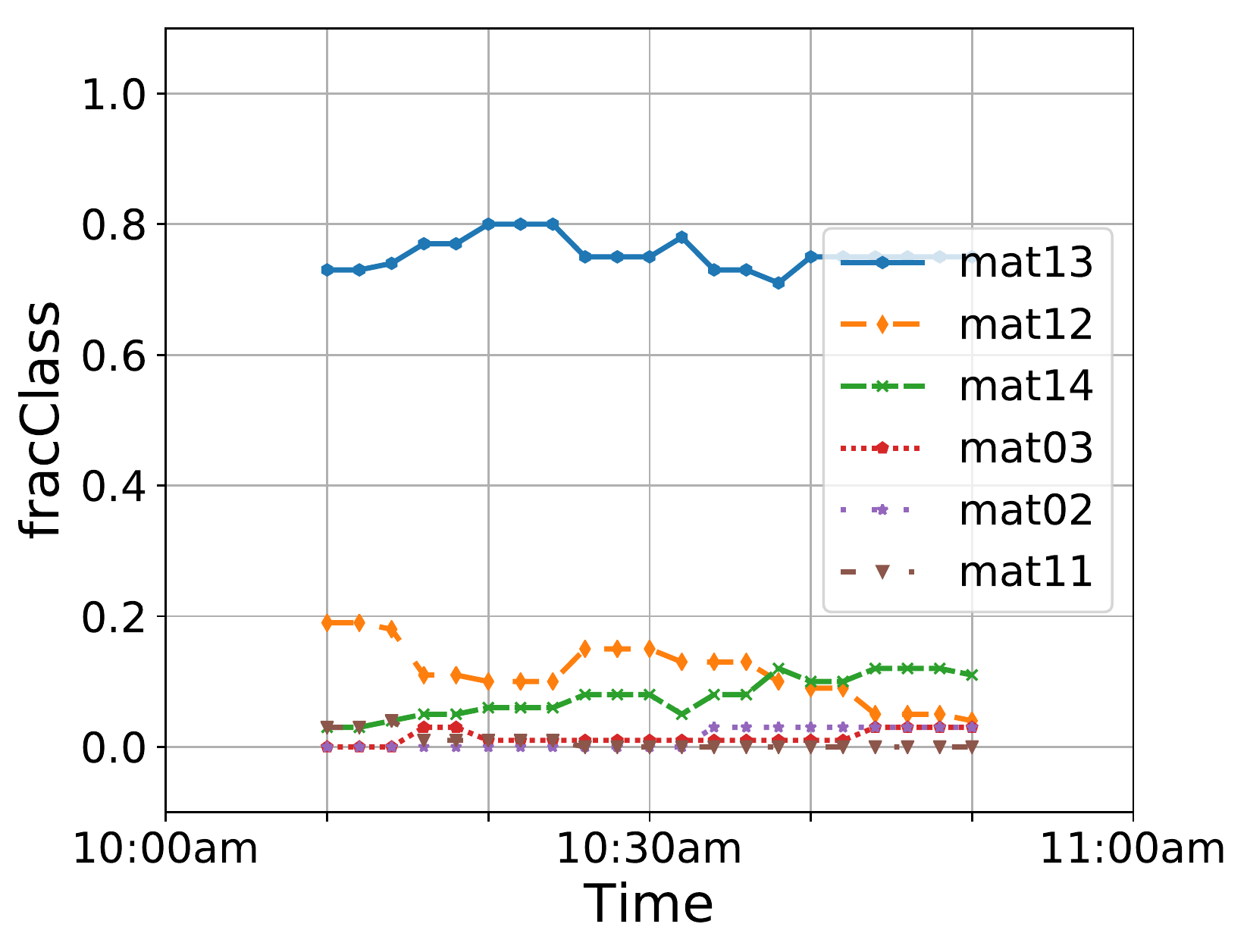}}\quad
							\label{fracClass_2}
						}
					}
					\hspace{-.8cm}
					\mbox{
						
						\subfloat[{every 5 minutes.}]{
							{\includegraphics[width=0.32\textwidth]{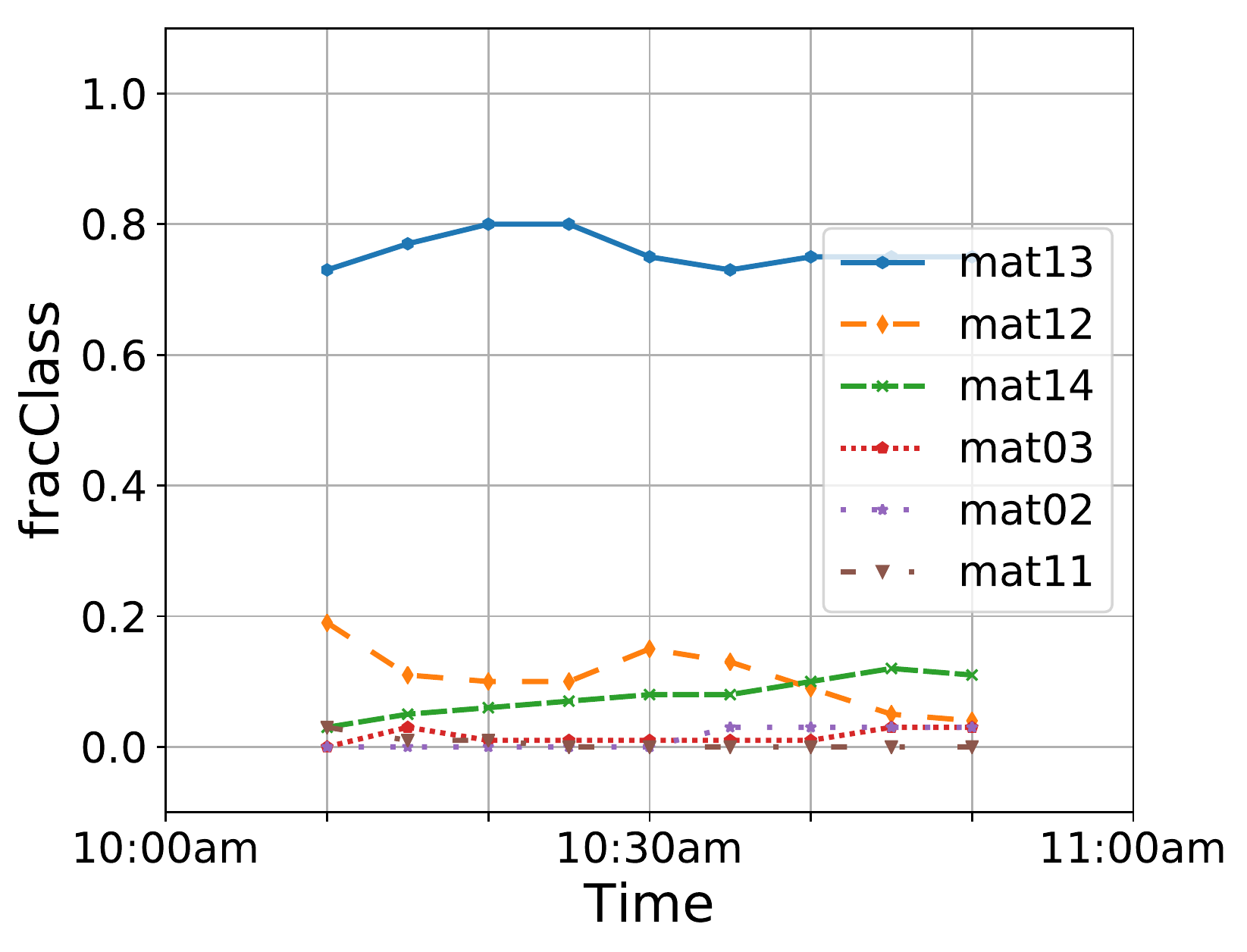}}\quad
							\label{fracClass_5}
						}
						\hspace{-.8cm}
					}
					\mbox{
						
						\subfloat[{every 10 minutes.}]{
							{\includegraphics[width=0.32\textwidth]{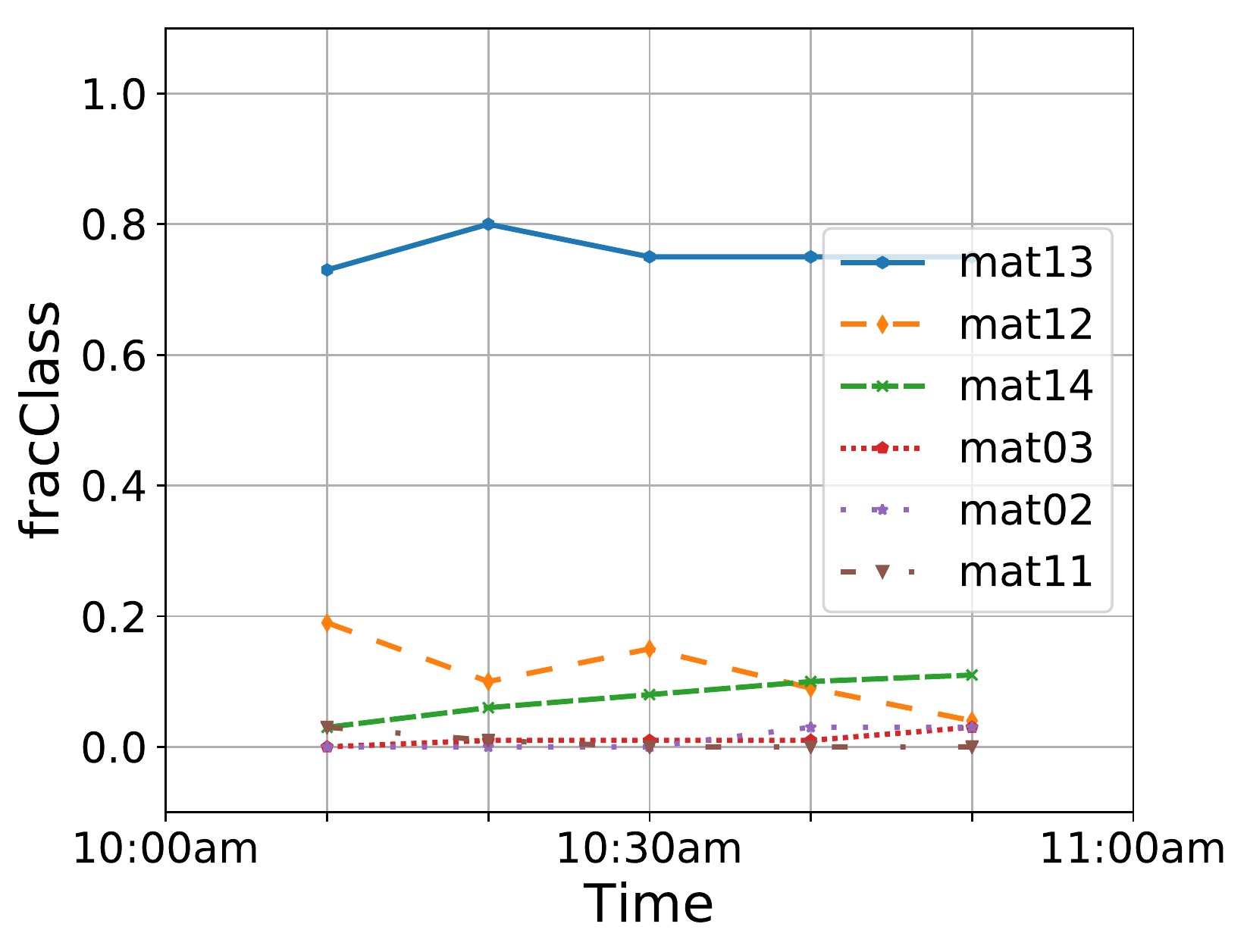}}\quad
							\label{fracClass_10}
						}
					}
					\caption{\(fracClass\) computed during 10:10am-10:50am for class Course-100 scheduled on 10am-11am in theater MatC.}
					\label{feature1_fracClass}
				\end{center}
			\end{figure*}

\begin{figure*}[t!]
				\begin{center}
					\mbox{
						\subfloat[{every 2 minutes.}]{
							{\includegraphics[width=.32\textwidth,]{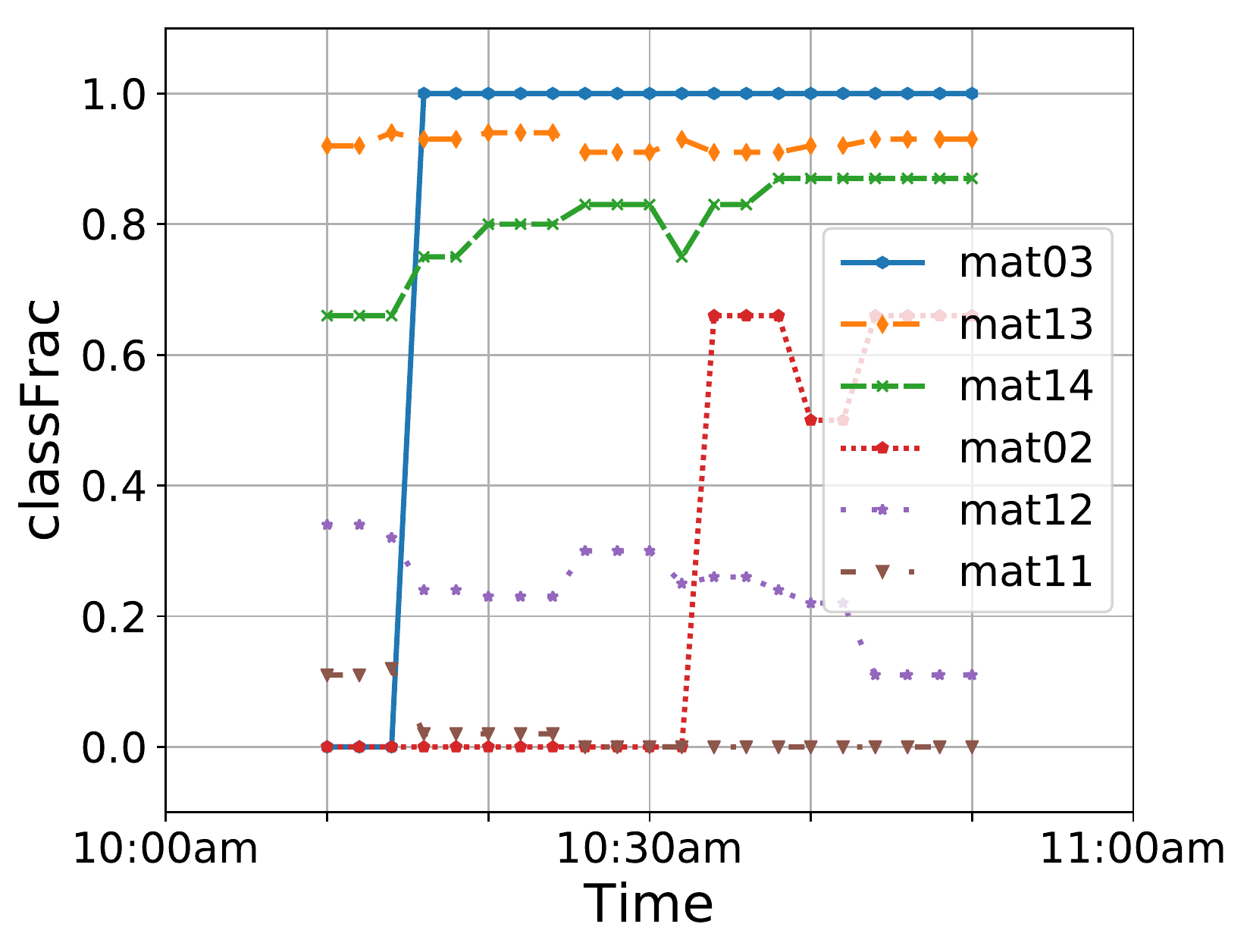}}\quad
							\label{classFrac_2}
						}
					}
					\hspace{-.8cm}
					\mbox{
						
						\subfloat[{every 5 minutes.}]{
							{\includegraphics[width=0.32\textwidth]{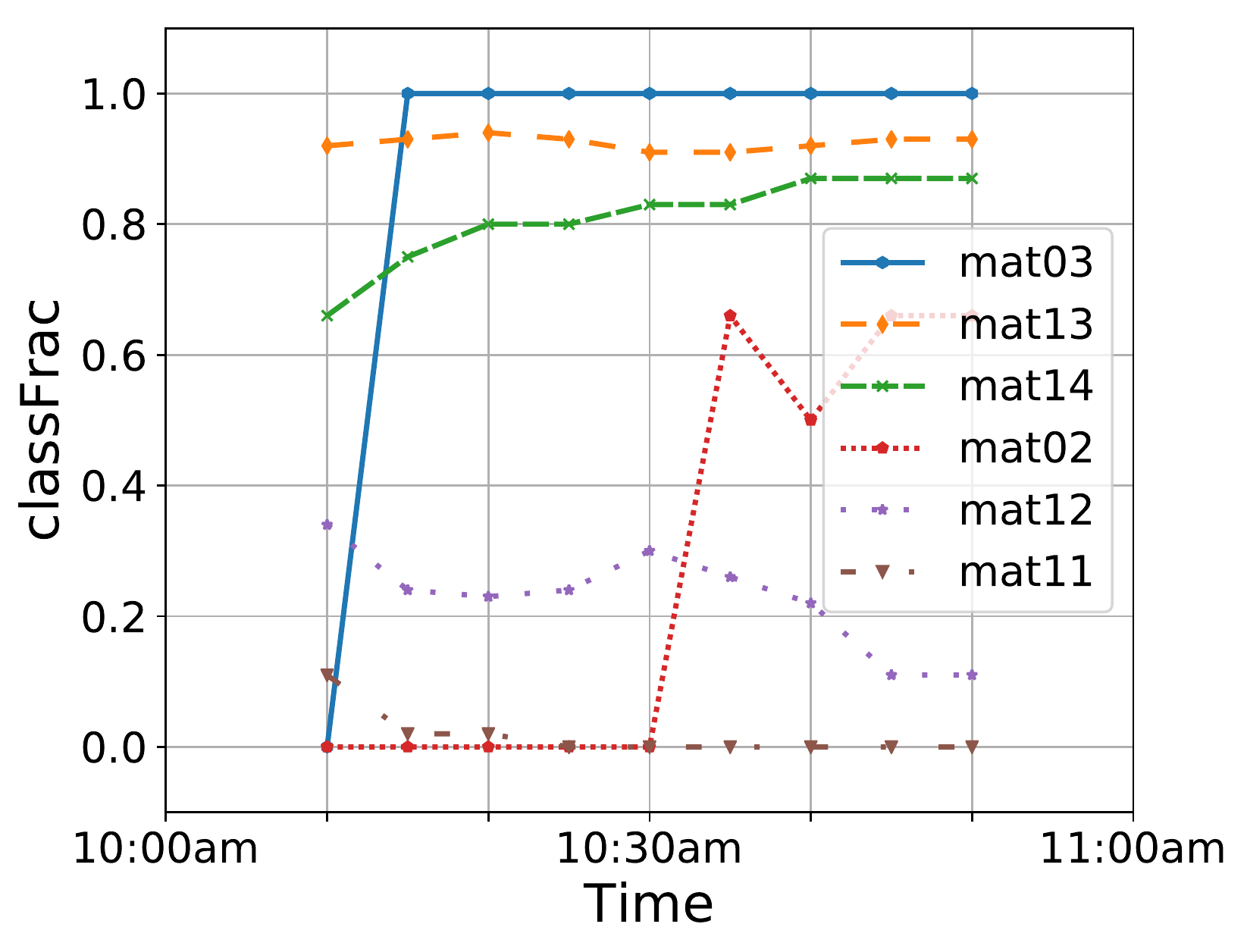}}\quad
							\label{classFrac_5}
						}
					}
					\hspace{-.8cm}
					\mbox{
						
						\subfloat[{every 10 minutes.}]{
							{\includegraphics[width=0.32\textwidth]{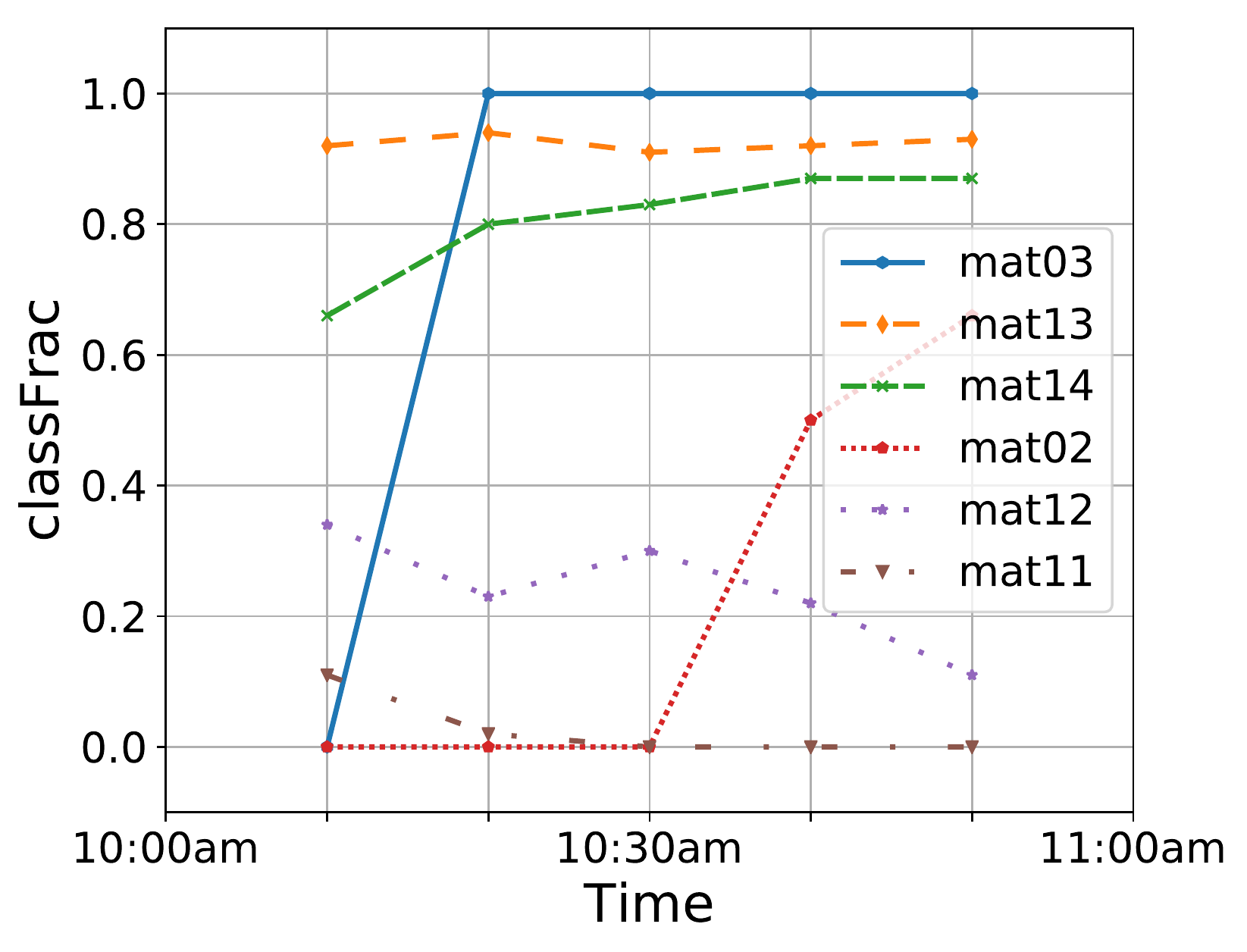}}\quad
							\label{classFrac_10}
						}
					}
					\caption{\(classFrac\) computed during 10:10am-10:50am for class Course-100 scheduled on 10am-11am in theater MatC.}
					\label{feature2_classFrac}
				\end{center}
			\end{figure*}

			To better understand these two key features, we compute them for a sample class (\ie a lecture of Course-100) in theater MatC, across APs to which enrolled students of the class connect and at varying time resolutions (\ie 2-min, 5-min, and 10-min), shown in Fig.~\ref {feature1_fracClass} and Fig.~\ref {feature2_classFrac}. Due to flux of students entering/exiting the class during the first and last few minutes of lectures,  features are computed for the interval between 10 minutes 	after the scheduled lecture time and 10 minutes prior to
			end of the scheduled lecture time. Unsurprisingly, profiles for both \(fracClass\) and \(classFrac\) get smoother by reducing the resolution of sampling (Fig.~\ref {feature1_fracClass} and Fig.~\ref {feature2_classFrac}), but the profile trend is  largely maintained from 2-minute resolution on the left to 10-min resolution on the right. We will look closely at the impact of sampling rate on accuracy and time complexity of APs mapping in \S\ref{S4_B}.
			
			Looking at Fig.~\ref {feature1_fracClass}, AP \textit{mat13} (located inside the room) contributes to most of connections (\ie more than $80\%$) made by enrolled students followed by \textit{mat12} and \textit{mat14}. Note that \textit{mat12} is located at L1 while our subject class is held at L2. This is probably due to a one-hour tutorial class of the same course which is held at L1 (just prior to this class) and thus users devices maintain their connections made in the previous class, though users moved to a new room which is located just above the previous room. In the middle of the class (\ie around 10:30am), we see that \textit{mat03} (located in MatC) starts getting connections from enrolled students while connections count of \textit{mattap12} (located at L1) starts falling. This is probably because new connections from users closer to \textit{mat12} cause connections from the class of MatC to switch to their nearby AP \textit{mat03}.

			Now moving to Fig.~\ref {feature2_classFrac}, connections made by enrolled students to each of those APs located inside the room MatC (\ie \textit{mat03}, \textit{mat13}, and \textit{mat14}), account for more than $60\%$ of the total connections while this metric (\ie \(classFrac\)) is $20\%$ for \textit{mat12} which is located at L1. We note that the profile of \(classFrac\) for APs \textit{mat12} and \textit{mat11} falls during the class time since the count of enrolled students connected to those APs drops as explained above -- \ie a rise in connections from nearby users probably forces connections from users inside the classroom to migrate.
			Surprisingly, \(classFrac\) for AP \textit{mat02} located at L1 starts rising to a value of about $60\%$ after 10:45am, since the number of non-enrolled students connected to it drops (\ie possibly due to end of another class), and thus the contribution of enrolled students of Course-100 becomes significant.

			This example shows that the two features (\ie \(fracClass\) and \(classFrac\)) are collectively needed to associate an AP to its corresponding room.
			In what follows, we feed these temporal features to a model that learns how to distinguish APs (to which class occupants get connected) located in and around a given classroom, from other APs spread across the campus.
			
	\subsection{Unsupervised Clustering of APs}  \label{S4_B}
	
\begin{figure}
			\centering
			\begin{minipage}{\textwidth}
				\centering
				\includegraphics[width=0.6\textwidth, height=4cm]{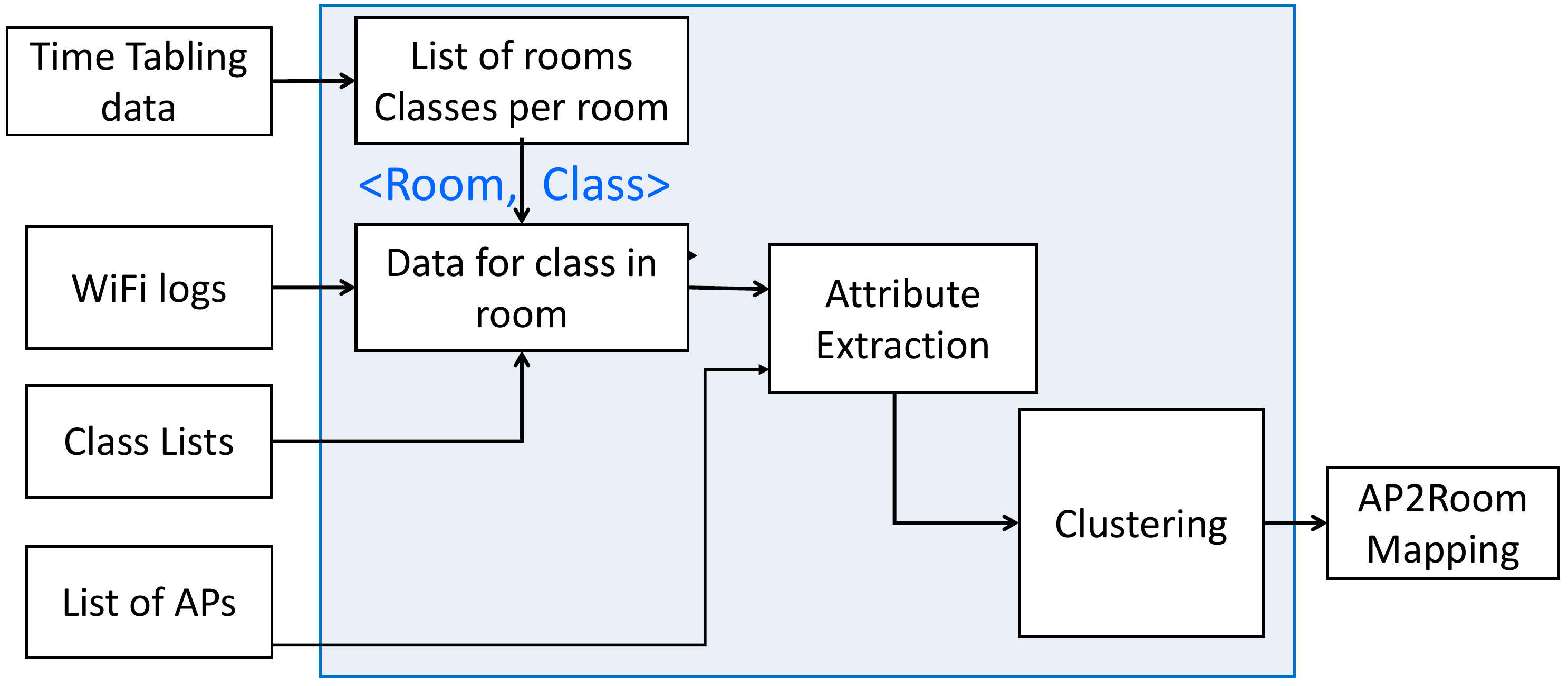}
				\caption{System architecture for mapping APs to classrooms.}
				\label{f_mappingAPsArchitecture}
				
			\end{minipage}
		\end{figure}

		The WiFi session data was collected from IT department of our campus during 2017-July-31 to 2017-October-27 (\ie{sem2-2017}) and 2018-February-26 to 2018-June-1 (\ie{sem1-2018}) while we obtained class lists data for 12 courses held in 5 classrooms.
		The minimal required data to map the APs related to a particular classroom is the  WiFi session data during a single class held in the room of the interest and the list of students enrolled in that class. Additionally, the timetabling information is used to map the classes to rooms where we intend to discover the relevant APs. 
		Our method is scalable across the whole campus at the availability of the input data shown in the method overviews in Fig. \ref{f_mappingAPsArchitecture}.
		
		Our objective is to determine APs in and around a room that cover a significantly large number of the room occupants (mapping APs), and hence two clusters are needed, \ie (a) APs located in and around the room (APs mapped), and (b) APs located far from the room (APs not-mapped). Note that this mapping could be one-to-many especially when an AP in a corridor is close to multiple rooms. We computed the parameters \(fracClass\) and \(classFrac\) at 10-minute resolution for 12 classes across each weeks of the semester (Note that we re-sampled the different length temporal features of classes with varying duration during the clustering). The derived features are then fed as input to clustering algorithms. In the next subsection we evaluate the performance of three widely used clustering algorithms, K-means, EM Clustering using GMMs (EM-GMM) and Hierarchical Clustering (HC). 
	
	\subsection{Clustering Results}
		We evaluate the performance of three clustering algorithms namely, K-means, HC, and EM-GMM. Table~\ref{t_table2} shows results of correct prediction (\ie  true positive and true negative). As opposed to the other two algorithms, EM-GMM is a soft clustering method that computes a probability to associate an instance with each and every cluster. In this work, we cluster our instances by choosing the highest probability derived from EM-GMM. We used the campus-wide layout of WiFi network provided by our university IT department to obtain the ground-truth location of APs, whether they are associated with a room (inside or nearby), or not (faraway outside).	
		
		K-means achieved 85.7\% accuracy in mapping APs associated with rooms and 99.7\% accuracy for APs disassociated with rooms. It is only slightly better than HC and EM-GMM to make a general conclusion on what algorithm performs best for our method. To better visualize clustering features, we first apply Principal Component Analysis (PCA) to our feature set reducing dimensions, and then plot clustering results on two principal components of AP features (for a sample class held in room MatA) in Fig. \ref{f_2Dfeatures}. It is clearly seen that these two PCA components contain enough information to distinguish two clusters of APs, inside and outside, for this example. Also, we observe that all outside APs are correctly classified (blue circles) by K-means while three of inside APs are misclassified as outside. {In terms of response time, K-means takes $53.6$ ms to generate results of mapping APs to classrooms -- this time is $0.84$ ms for HC, and $9.1$ ms for EM-GMM.}
		
		\begin{table}[!t]
			\renewcommand{\arraystretch}{1.3}
			\caption{Performance comparison of clustering algorithms (correct prediction).}
			\label{t_table2}
			
			\centering
			\adjustbox{max width = 0.7\columnwidth}{
				\begin{tabular}{|c|c|c|c|c|}
					\hline
					& Associated room APs  & Not-associated room APs  &  \multicolumn{2}{|c|}{Response time}\\             \cline{4-5}
					& & & average run-time & std. deviation  \\
					\hline
					K-means & $85.7$\% & $99.7$\% & $53.6$ ms&$2.2$ ms\\
					\hline
					HC & $83.1$\% & $99.7$\% & $0.84$ ms&$0.11$ ms\\
					\hline
					EM-GMM &  $81.1$\%& $99.6$\% &  $9.1$ ms& $0.9$ ms\\
					\hline
			\end{tabular}}
			
		\end{table}

		\begin{figure*}[t!]
			\begin{center}
				\mbox{		
					\subfloat[{Avearge time to extract features per AP.}]{
						{\includegraphics[height=4.4cm]{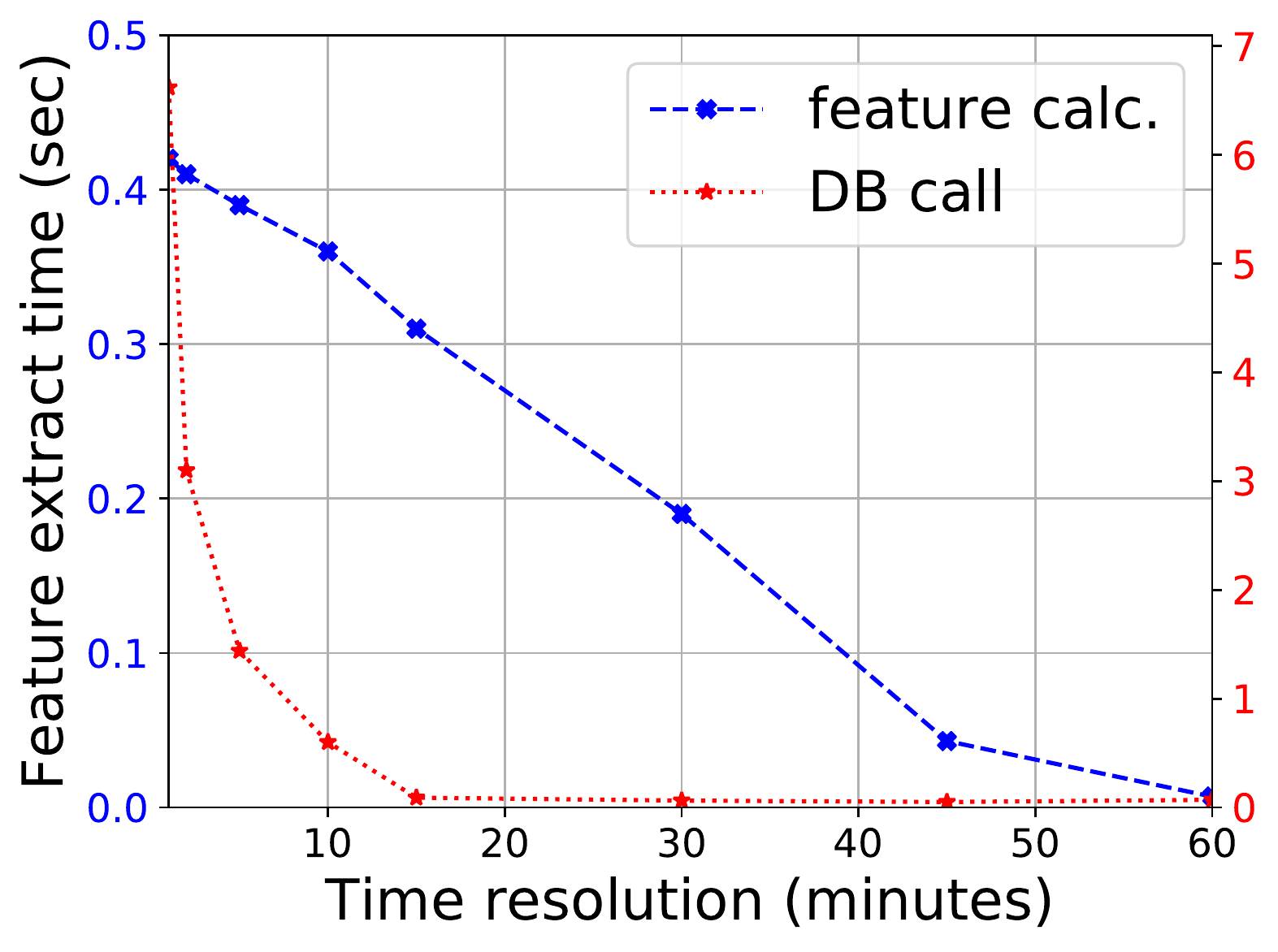}}\quad
						\label{f_kmeansTimeDelay}
					}
				}
				\mbox{
					\subfloat[{Accuracy (using K-means)}]{
						{\includegraphics[height=4.4cm]{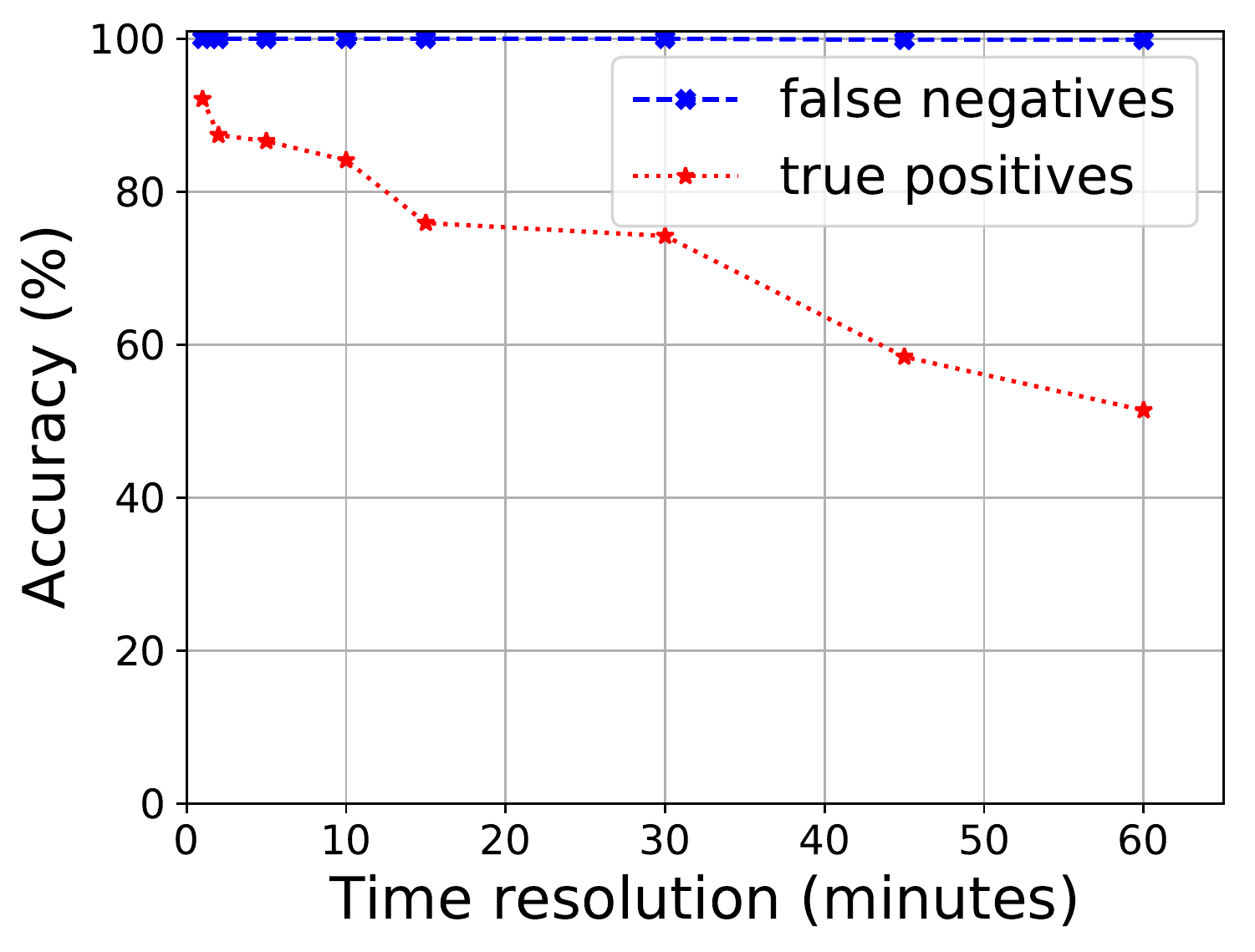}}\quad
						\label{f_kmeansAccuracy}
					}.
				}
				\vspace{-3mm}				
				\caption{{Time for (a) feature extraction (on average) and (b) accuracy of AP clustering at varying temporal resolutions}}
					\label{f_kmeans}
			\end{center}
			\vspace{-6mm}
		\end{figure*}

		We now compute the time complexity of feature extraction and the accuracy of K-means clustering. The temporal features generated at 1, 2, 5, 10, 15, 30, 45 and 60 minute time resolutions.
		Our aim is to estimate the room occupancy in near real-time. With that, the AP mapping algorithm which uses the two features (\ie fracClass and classFrac) becomes more accurate when it is run in real-time since it dynamically captures the WiFi coverage over current room occupants. Fig.~\ref{f_kmeansTimeDelay} shows two components (data retrieval and feature calculation, shown by dashed blue and dotted red lines) of the total time taken to generate features for an AP.  Note that the number of data rows retrieved from the database at coarser resolutions (\eg 60-minute) is hundreds of times less compared to finer resolutions (\eg 1-minute). Therefore, it is seen that the feature extraction time displays a non-linear trend mainly because of the database retrieval time component, and hence the total time of feature extraction rapidly falls with time resolution.  
		
		The accuracy of correctly clustering the APs in and near the room (true positive) is higher when features extracted at higher temporal resolutions  (as shown in Fig.~\ref{f_kmeansAccuracy}). Furthermore, the machine classifies the APs far from the room (true negative) with nearly 100\% at all time resolutions.
		We have 5000 APs on our campus, and it is not practically feasible to generate features for all 5000 APs on campus at high temporal resolution (every 1-minutely) despite of the higher accuracy. 
		Therefore, we select the 10-minute time resolution as it is cheaper at time cost and does not compensate the accuracy which is $92.1\%$ at 1-minute resolution and $84.1\%$ at 10-minute resolution. Note that this value is tuned for a network of 70 APs in our study. The trade-off between accuracy and time-complexity varies by the size of WiFi network.
		Note that features extraction and automatic AP mapping engines run on a machine with 6 CPU cores, 16 GB of memory, and storage of 521GB. 	
	

	\begin{figure}
		\centering
		\begin{minipage}{0.4\textwidth}
			\centering
			\includegraphics[width=\textwidth, height=4cm]{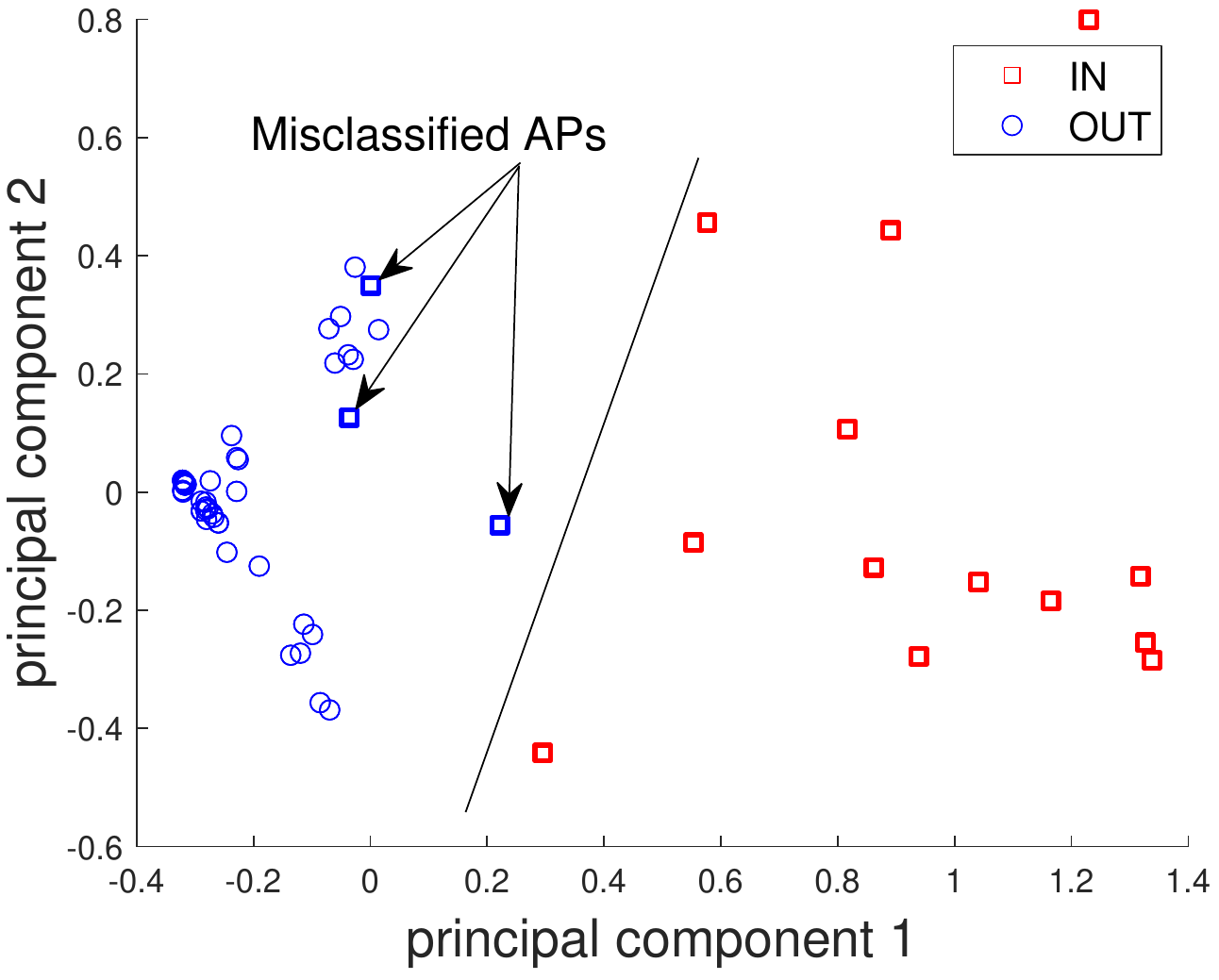}
			\caption{A sample of K-means clustering results on two principal components of AP features, for a class}
			\label{f_2Dfeatures}
		\end{minipage}
		\hspace{7mm}
		\begin{minipage}{.4\textwidth}
			\centering
			
			\includegraphics[width=\textwidth, height=4cm]{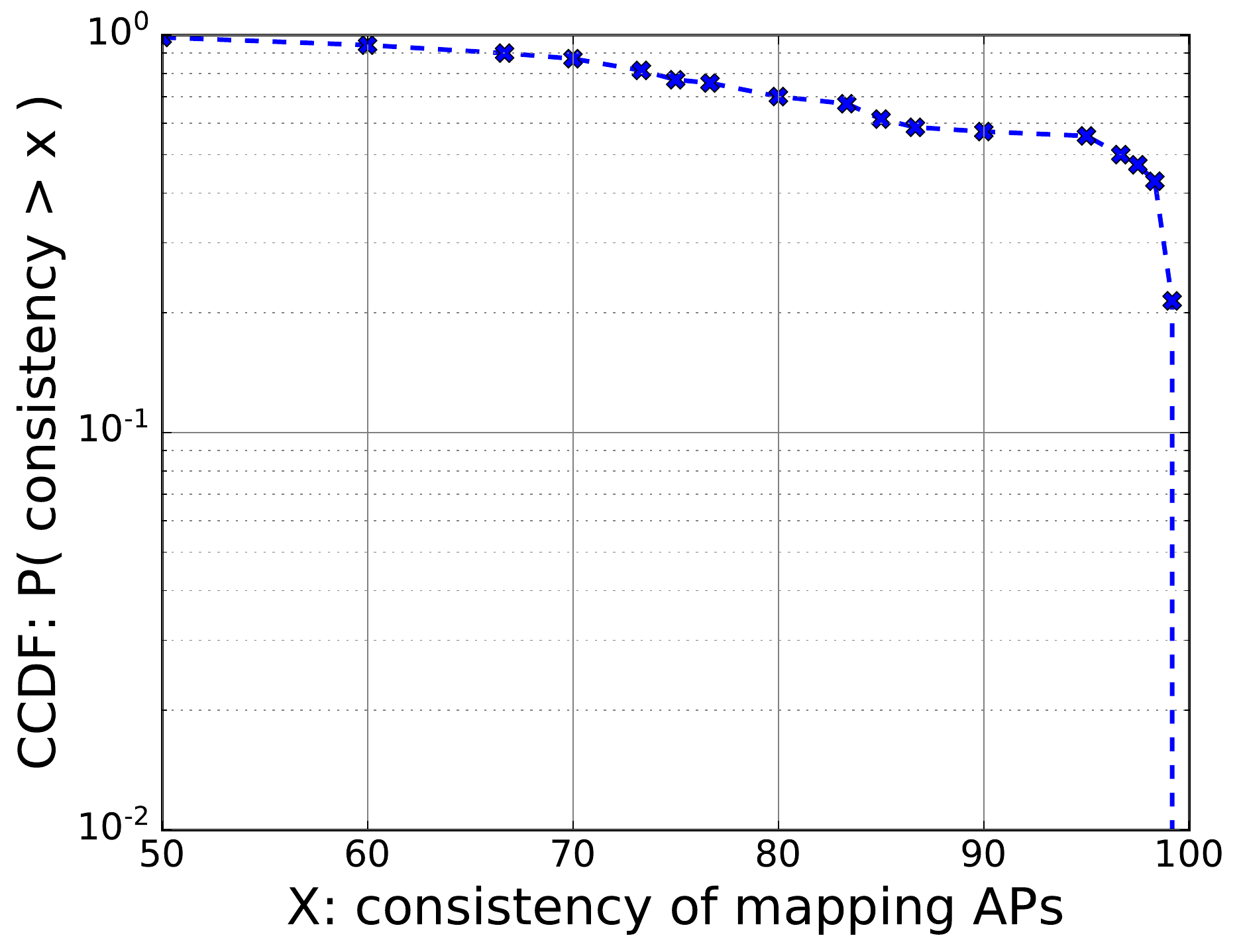}
			\caption{CCDF: mapping consistency.}
			\label{CCDF_Consistency}
		\end{minipage}%
		\vspace{-5mm}
	\end{figure}
	
		\subsubsection{Consistency of AP Mapping:\\}
			We now look at how mapping of APs to classrooms varies across weeks. Note that it is possible to have APs mapped (incorrectly) to their adjacent rooms. Also, in certain circumstances, we may find an AP mapped to a room faraway from its actual location. This  case can only occur if a considerable number of students enrolled in a class do not attend their classroom and connect to an AP located in other side of the campus (far from the classroom) -- also this AP serves no other class (with students) at that particular time.  
			
			We quantify ``consistency'' metric for each AP, computed as fraction of time the AP is correctly mapped to its expected room across all classes  over 10 weeks. Fig.~\ref{CCDF_Consistency} shows the Complementary Cumulative Distribution Function (CCDF) of  consistency for mapping APs.  We see that the chance of having consistency of more than $80$\% is $0.7$. We observe that mapping of APs may fluctuate across weeks, but the chance is fairly low. Note that this is mainly because our mapping algorithm takes WiFi occupancy and enrolled WiFi occupancy as inputs which both are dynamic and fluctuate across weeks.
			Our consistency results illustrate the need for dynamic use of AP mapping (\ie for each class).

		\subsubsection{Impact of Room Size and Class duration on Mapping APs:\\}
		
			We now evaluate the variation of AP mapping accuracy across classrooms and classes of varying duration. 
			
\begin{figure*}[t!]
				\begin{center}
					\hspace{-0.5cm}
					\mbox{
						\subfloat[MatA(v. large).]{
							{\includegraphics[width=0.22\textwidth]{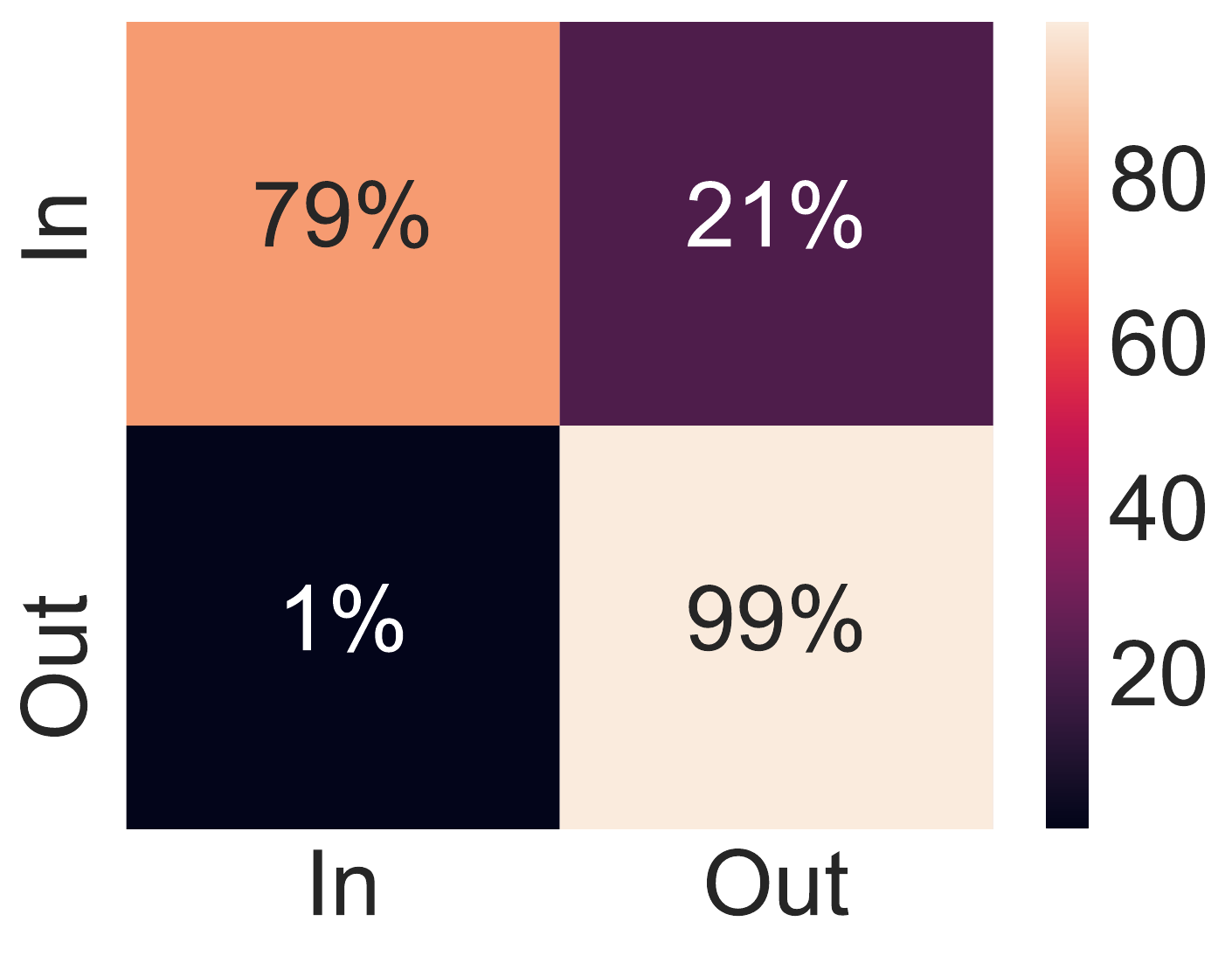}}\quad
							\label{f_MatAConfMat}
						}
					}
					\hspace{-1.2cm}
					\mbox{
						
						\subfloat[CLB8(large).]{
							{\includegraphics[width=0.22\textwidth]{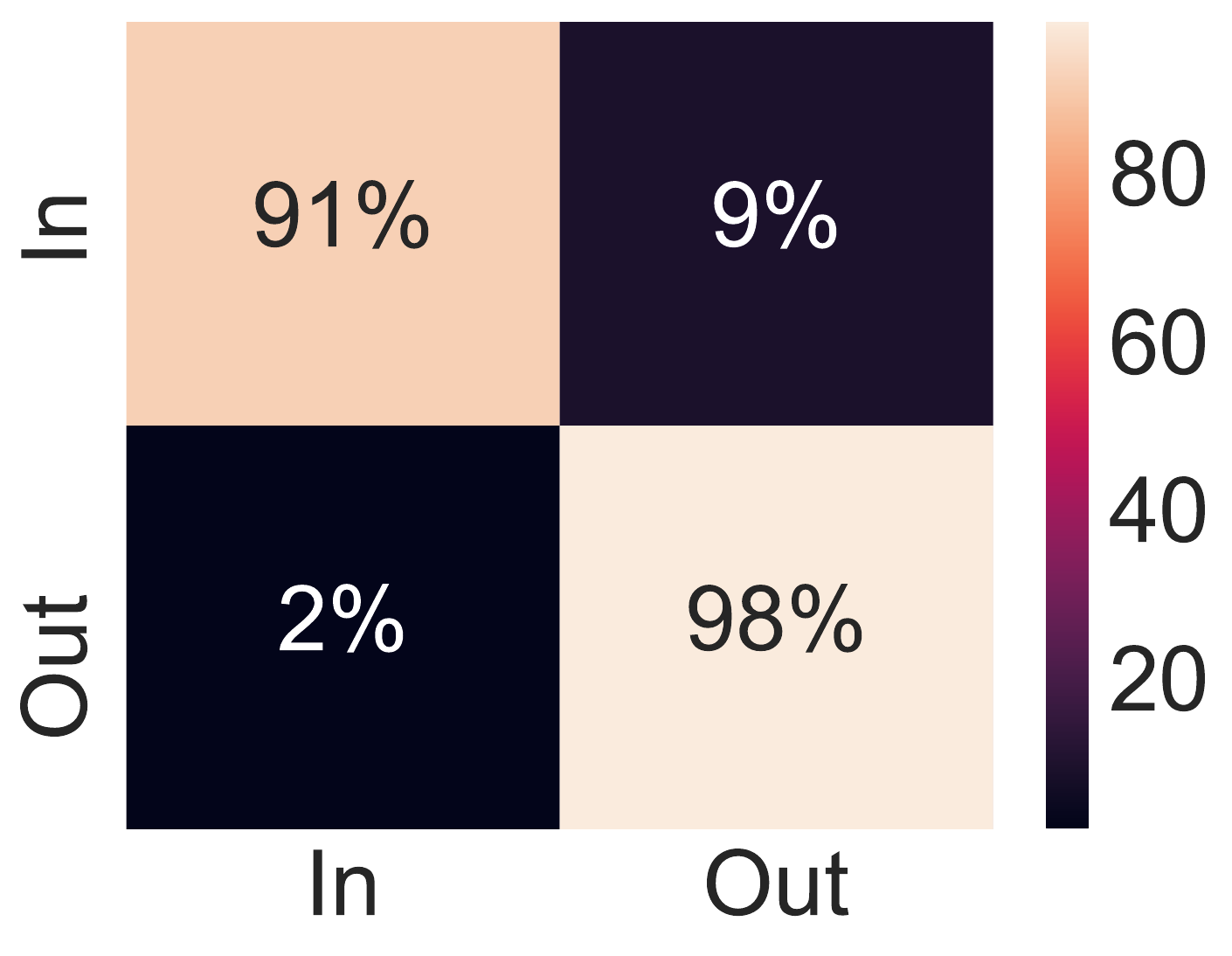}}\quad
							\label{f_CLB8ConfMat}
						}
					}
					\hspace{-1.2cm}
					\mbox{
						
						\subfloat[MatB(large).]{
							{\includegraphics[width=0.22\textwidth]{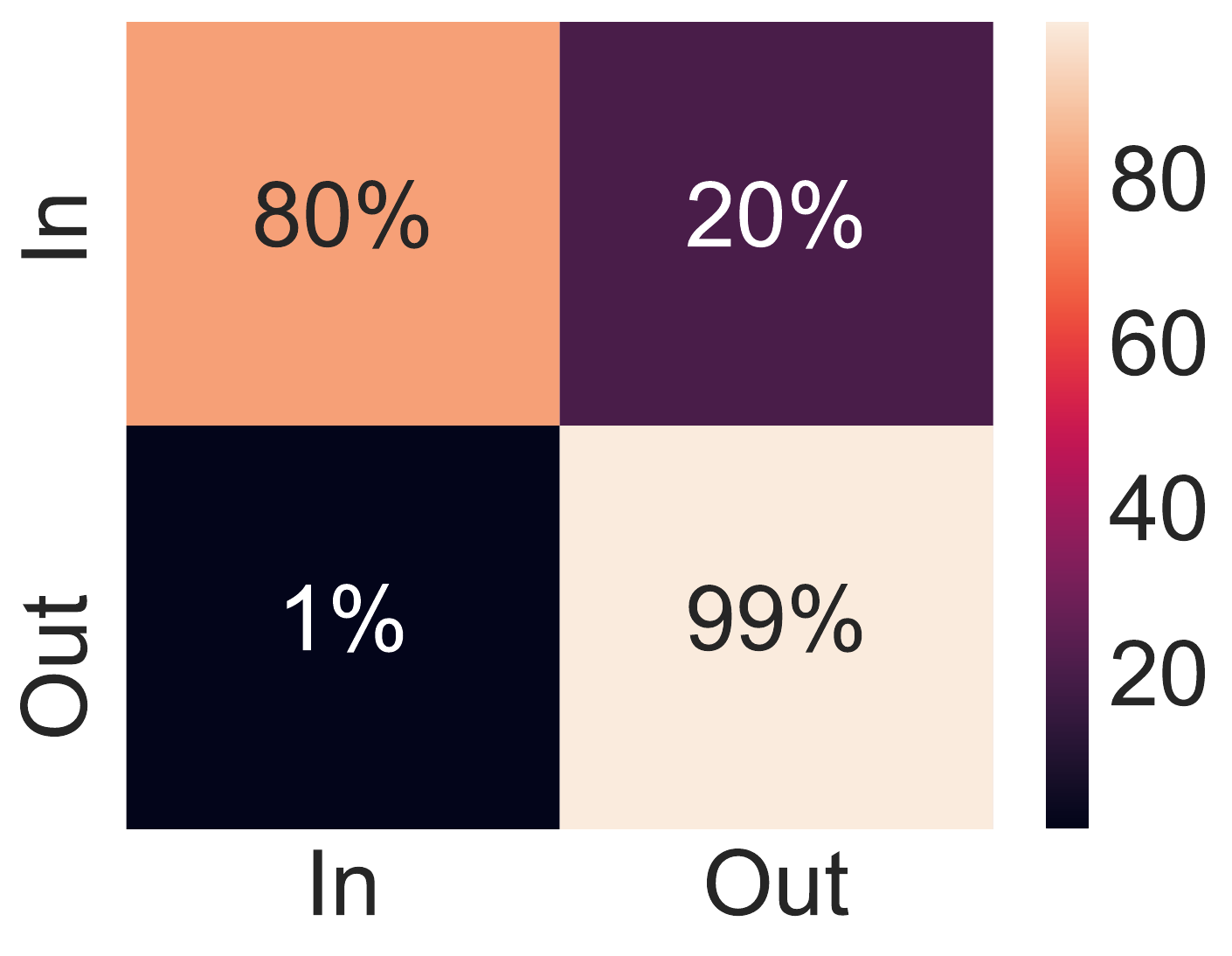}}\quad
							\label{f_MatBConfMat}
						}
					}
					\hspace{-1.2cm}
					\mbox{
						
						\subfloat[MatC(med.).]{
							{\includegraphics[width=0.22\textwidth]{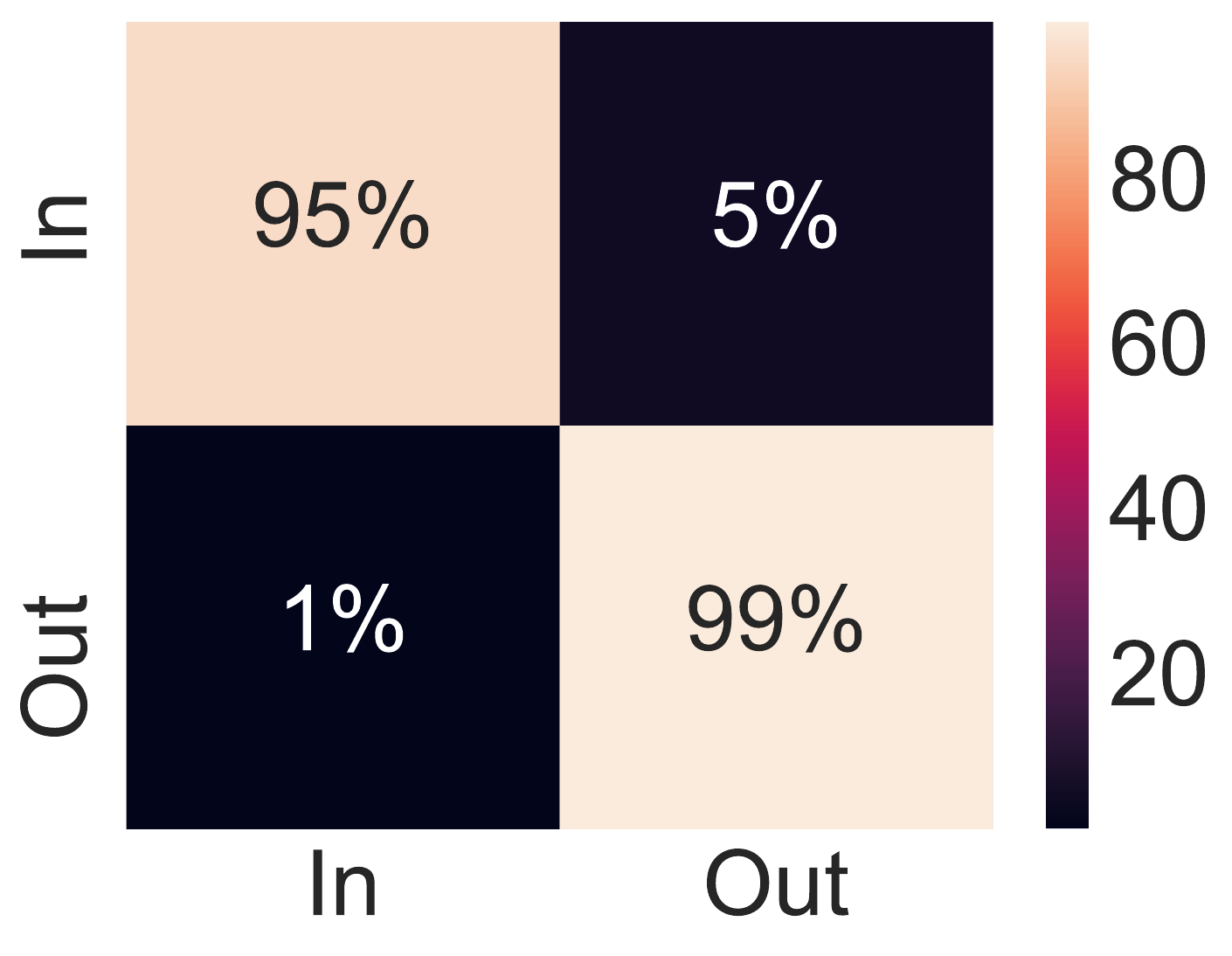}}\quad
							\label{f_MatCConfMat}
						}
					}
					\hspace{-1.2cm}
					\mbox{
						
						\subfloat[Mat228(small).]{
							{\includegraphics[width=0.22\textwidth]{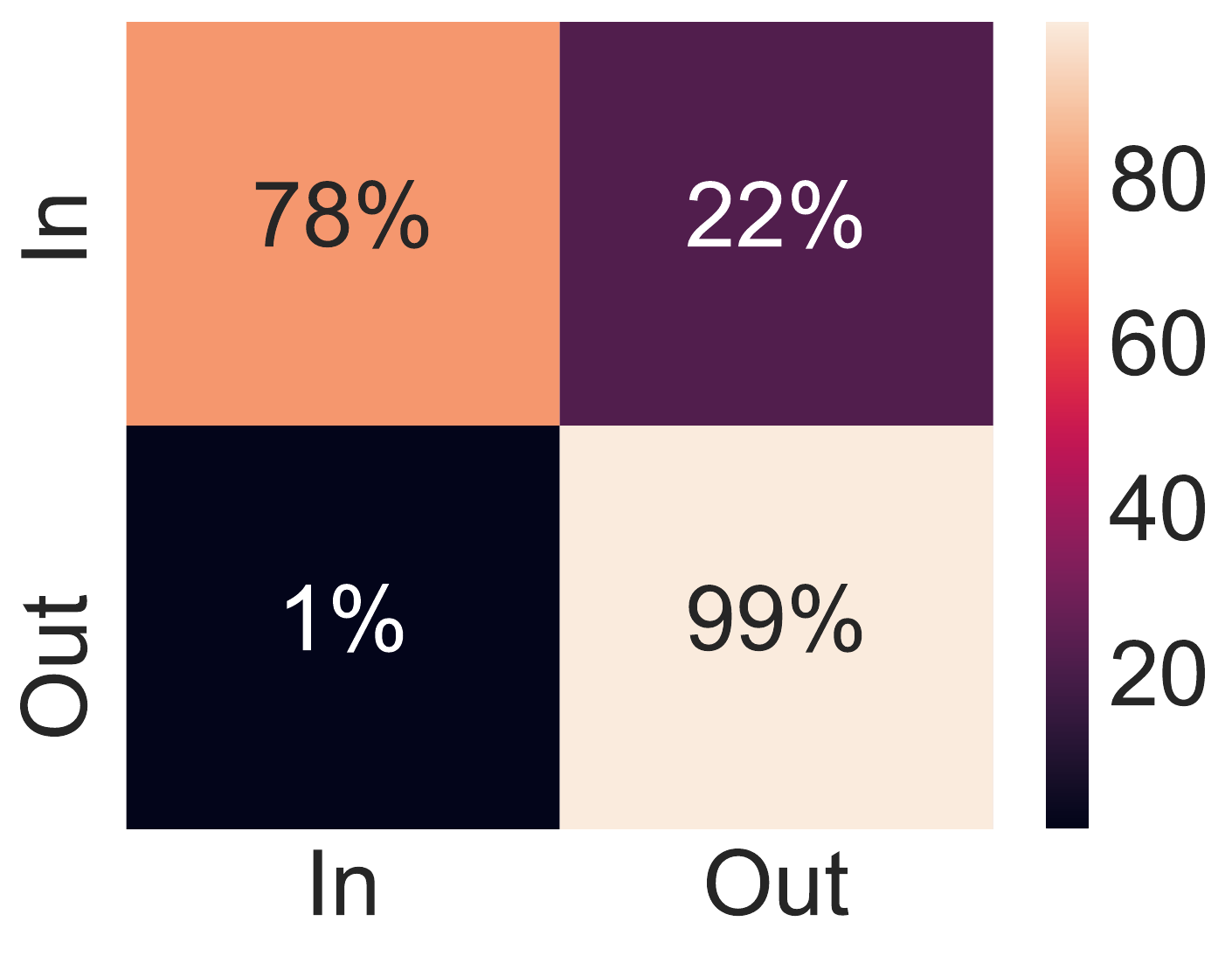}}\quad
							\label{f_Mat228ConfMat}
						}
						
					}
					\vspace{-3mm}				
					\caption{Confusion matrix of AP mapping for five classrooms of varying size.}
					\label{f_confusionMatrices}
				\end{center}
				\vspace{-4mm}
			\end{figure*}

\begin{figure*}[t!]
				\begin{center}
					\mbox{
						\subfloat[{MatA.}]{
							{\includegraphics[height=4.3cm]{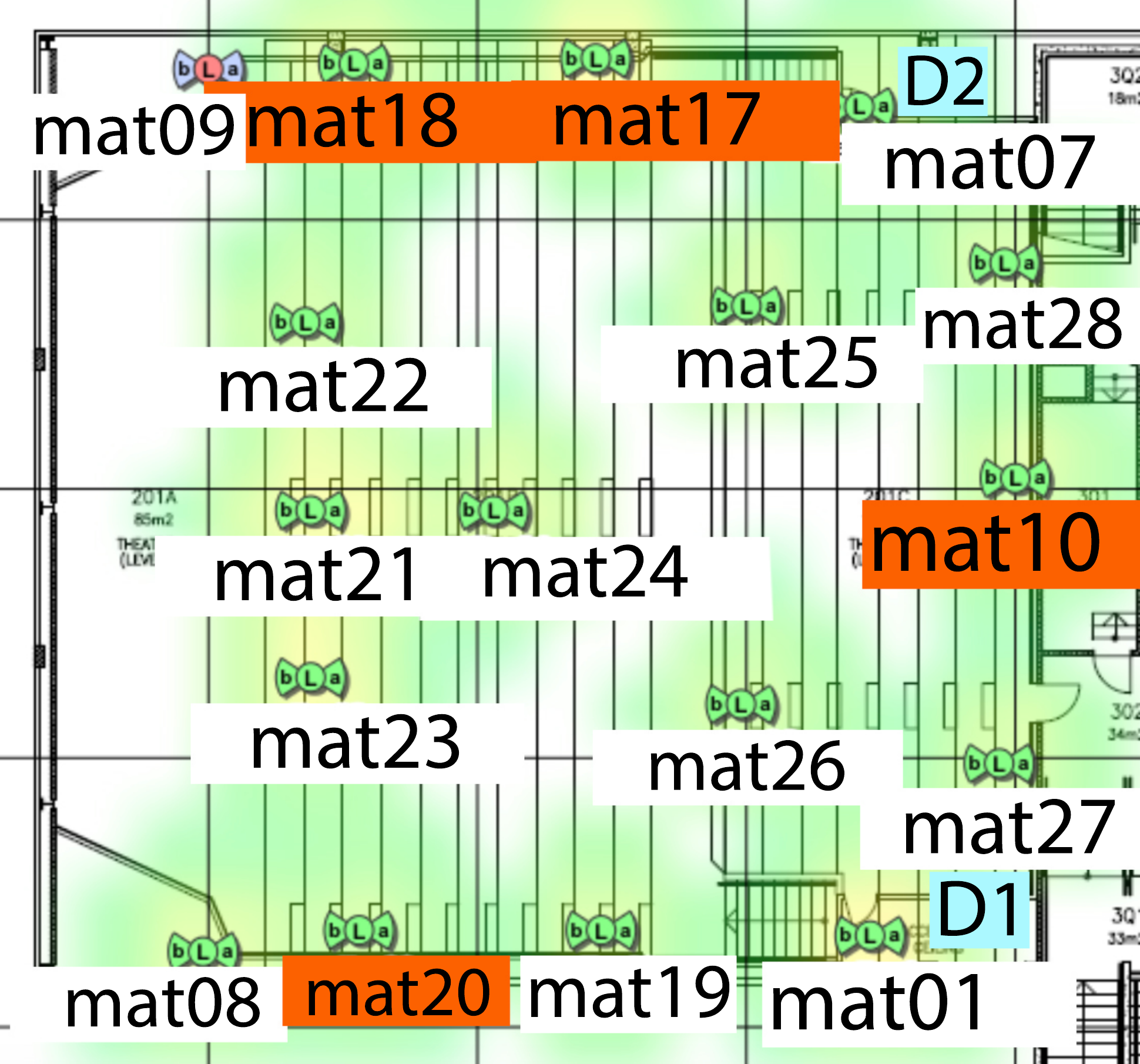}}\quad
							\label{MATA}
						}
					}
					\hspace{.7cm}
					\mbox{
						
						\subfloat[{CLB8.}]{
							{\includegraphics[height=4.3cm]{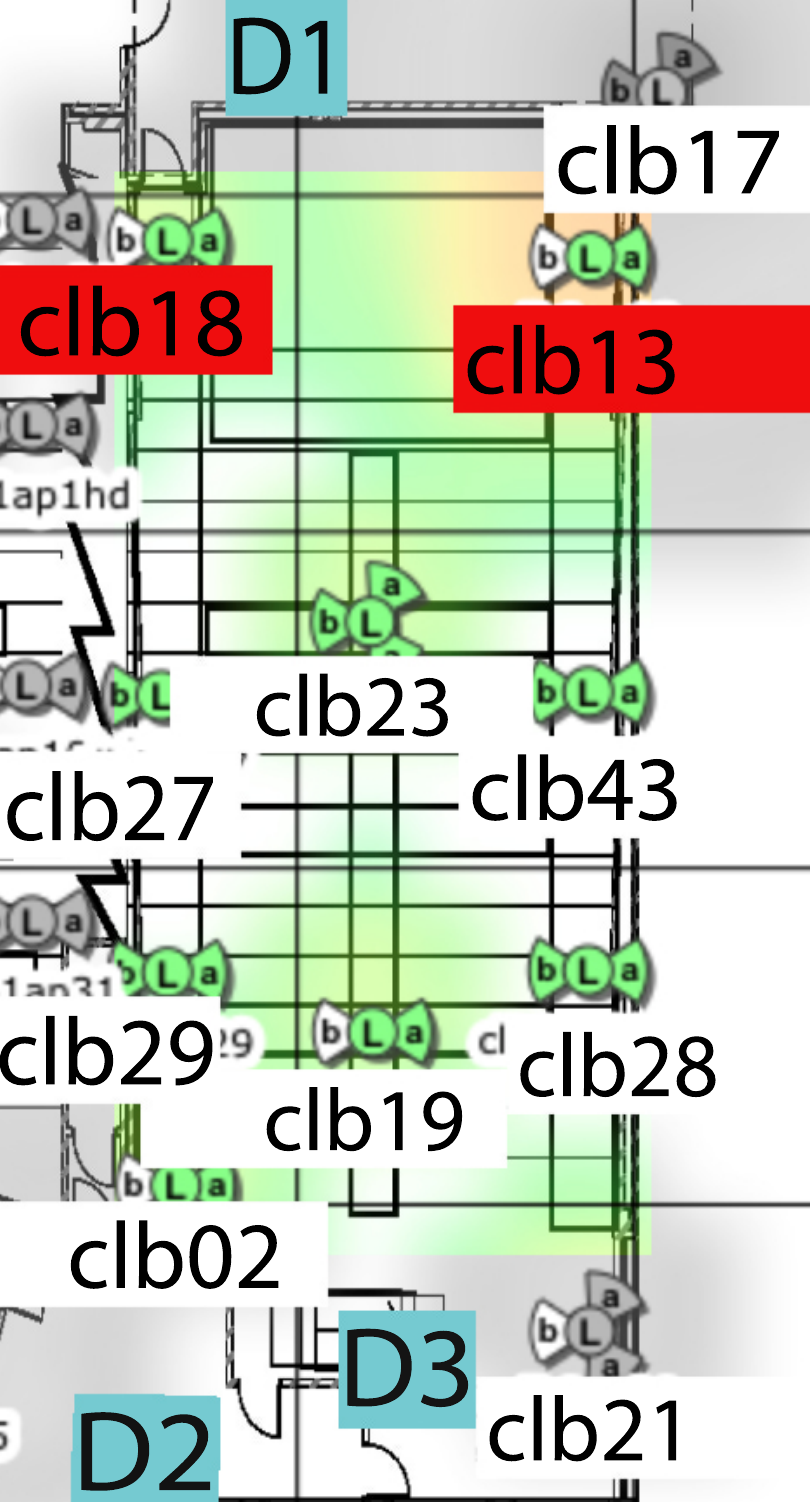}}\quad
							\label{CLB_L1}
						}
					}
					\hspace{.7cm}
					\mbox{
						\subfloat[{MAT228.}]{
							{\includegraphics[height=4.3cm]{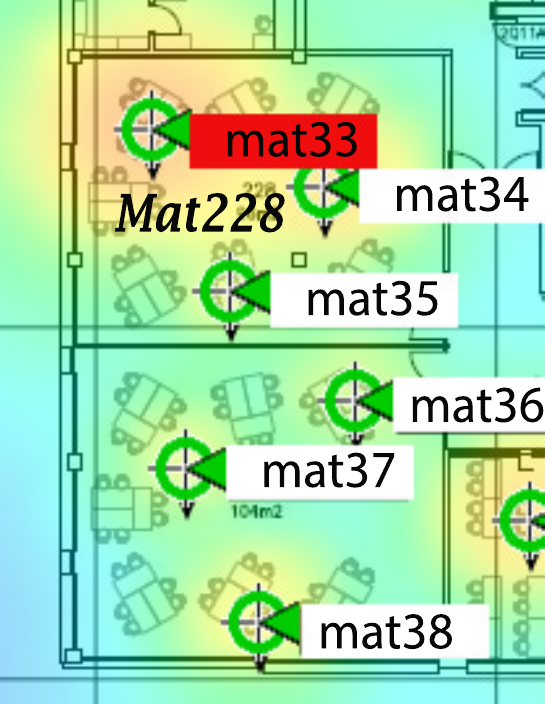}}\quad
							\label{map_228}
						}
					}
					\vspace{-3mm}
					\caption{Those APs that are located at corners (in red) of the room did not get mapped to their respective rooms.}
					\label{f_maps}
				\end{center}
				\vspace{-4mm}
			\end{figure*}

			The 5 classrooms  of our study include one very large lecture theater (\ie MatA), two large lecture theaters (\ie MatB and CLB8), one medium lecture room (\ie MatC), and one smaller classroom (\ie Mat228). We show in Fig.~\ref{f_confusionMatrices}, the confusion matrix of AP mapping for the 5 classrooms. It is seen that the accuracy of mapping APs outside rooms (\ie true negative) is very high close to $100\%$, meaning that APs faraway from rooms are well distinguished and thus not mapped to any rooms. 
			For APs located inside classrooms, the rate of correctly mapped instances is relatively lower. 
			For example, in the largest lecture theater MatA  with 17 APs inside, the rate of correctly mapped APs inside (\ie true positive) is $79\%$ as shown in Fig.~\ref{f_MatAConfMat}. For room CLB8 with 10 APs, this metric $80\%$ as shown in Fig.~\ref{f_CLB8ConfMat}. This is mainly because these rooms have APs which serve a small number of (or zero) enrolled students in the class -- these APs are located at the border/corner of rooms, and thus get misclassified. We highlight these APs by red color in Fig. \ref{MATA} and \ref{CLB_L1} for MatA and CLB8, respectively.
			We note that the rate of true positive gets slightly better (close to 90\%) for lecture rooms with fewer APs (\ie MatB with 4 APs and MatC with 3 APs).
			Surprisingly, the smallest room Mat228 with 3 APs displays the lowest true positive  rate -- we found that one of these 3 APs (\ie mat33 highlighted by red in Fig. 9(c)) was configured by the highest value of power level, (thus serving most of users in the classroom), while the power for other two APs  was set to default auto which is the recommended power setting. This inconsistent configuration results in small values of \(fracClass\) and \(classFrac\) features computed for the other two APs in the room, leading to an incorrect classification. 
	
			Lastly, we compute the impact of class duration on mapping APs to rooms. For short duration classes (\ie less than two hours) the accuracy of mapping APs is 81\% while it gets slightly better up to 86\% for long duration classes (\ie 2 hours or more). This is because our temporal features for longer classes become distinctly large, allowing our method to perform better in mapping APs to rooms.

\section{Modeling classroom occupancy}\label{sec:WiFiOccupancy}
	
	Given the list of student identifiers of enrolled students for a particular class or classes and the WiFi session data for the campus during that class, we identify the APs that the occupants of a particular room get connected. Then the WiFi session data from such APs that are selected for a particular room is used to estimate the number of people in that room. In this section we explain the feature extraction, method and results of estimating occupancy.

	\subsection {Feature Selection for WiFi Users}\label{S5_A}
		In general, bystanders would often differ from room occupants in the way they use WiFi. For instance, the duration of connection during a particular class is useful in determining the WiFi user's occupancy in that class. In our previous work~\cite{wifiLCN2018}, we identified the following set of features (extracted from WiFi sessions data) to distinguish room occupants from bystanders. For each user, we can compute the following features.
		\begin{enumerate}
			\item RSSI - Average of RSSI values across number of sessions associated with a user during the class of interest. {For instance, bystanders are expected to receive less signal strength compared to occupants.}
			\item Arrival delay - Time difference between the class start time and the WiFi user's first appearance in WiFi during the class of interest. {For instance, a student who attends a lecture is more likely to arrive to the classroom around the start of the class, and hence expected to have low arrival delays.}
			\item Number of sessions - Number of associations during the class of interest. {For example, there is a high chance for a lecture attendee to have multiple associations during the class due to inconsistent WiFi connectivity of mobile devices as highlighted in~\cite{Melfi2011}.}
			\item Number of devices - Number of devices used to connect to WiFi during the class of interest. {For instance, a bystander walking past a room is more likely to get connected only with their mobile phone while class attendees would probably have multiple devices (mobile phone, tablet, and laptop) connected to WiFi.}
		\end{enumerate}
		
		Also, we derive two other time-related features From WiFi data as follows. 
	
		\begin{enumerate}
			\setItemnumber{5}
			\item Percentage of `in time' (\(t_{in}\)) - Percentage of a user's WiFi access that occurred inside the class time during the class of interest. By considering the association and disassociation times of a session we removed the overlapping sessions by a single user to compute the non-overlapping connected time during a class. {Bystanders walking past the room have less connected time to WiFi.}
			\setItemnumber{6}
			\item Percentage of `out time' (\(t_{out}\)) - Percentage of user's WiFi access that occurred outside the class of interest. This is normalized by subtracting the class duration from the time in which the lectures are usually scheduled during the day (9am - 9pm) on our campus. {Bystanders connecting to APs inside a room are working in nearby offices or study spaces would typically have high \(t_{out}\) values.} 
		\end{enumerate}

\begin{figure*}[t!]
			\includegraphics[width=0.98\textwidth, height = 3.5cm]{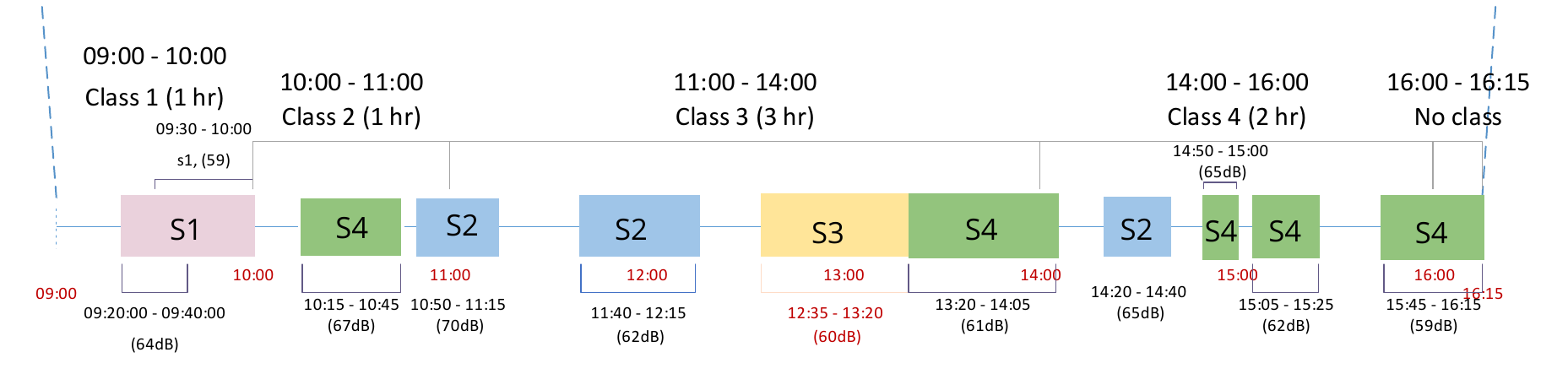}
			\vspace{-3mm}
			\caption{Daily trace of four users in session logs of APs in a classroom.}
			\label{f_exampleWiFiUsers}
			\vspace{-4mm}
		\end{figure*}

				To better understand these features, we illustrate in Fig. \ref{f_exampleWiFiUsers} a time trace of WiFi association with APs in a sample classroom from four selected users (S1...S4) -- each colored box represents the time interval over which a user connects to APs inside this room. The corresponding features are computed and summarized in Table \ref{t_ComputedFeatures}.  

		\begin{table}[t!]
		\renewcommand{\arraystretch}{1.3}
		\caption{Features computed for sample WiFi users.}
		\vspace{-3mm}
		\label{t_ComputedFeatures}
		\centering
		\adjustbox{max width = \columnwidth}{
			\begin{tabular}{|c|c|c|c|c|}
				\hline
				\textbf{User} & \textbf{Class duration} & \textbf{\(t_{in}\)}& \textbf{\(t_{out}\)} & \textbf{RSSI (dB)}\\
				\hline
				S1 & 1-hour & 40/60 = $66.7$\% & $0.0$\% & 61.5\\
				\hline
				S2 & 3-hour & 50/180 = $27.8$\% & 30/540 = $5.6$\%& 66.0\\
				\hline
				S3 & 3-hour & 45/180 = $25.0$\% & $0.0$\% & 60.0\\
				\hline
				S4 & 3-hour & 40/180 = $22.2$\% & 85/540 = $15.7$\%& 62.0\\
				\hline
		\end{tabular}}
	\end{table}

		User S1 connects to WiFi with multiple devices, having two overlapping sessions;
		S2 probably has two classes (\ie class3 and class4) scheduled in the same room on that day; S3 has one device only connected with WiFi during a class; user S4 is seen throughout the day, hence likely to be someone who is working/studying in proximity area, but may not be inside the room. During class1 which lasted one hour, user S1 has two connections; one from 9:20am to 9:40am, and another one from 9:30am to 10:00am. We compute the non-overlapping connected time during this class to be $40$ minutes. Rest of that day, S1 is not seen connected to any AP inside or nearby this room beyond the class1. 
		Similarly, during class3 which lasted for three hours, user S2 has two sessions having spent $50$ minutes in class and has an out of class time of $30$ minutes (10 minutes from 10:50am to 11:00am plus 20 minutes from 14:20pm - 14:40pm). 
		Another user, S3 has spent $45$ minutes in class3 and does not reappear beyond the class3 -- hence has a \(t_{out}\) of $0$ minutes. The WiFi user S4 is seen for $40$ minutes during class3, however this user has $85$ minutes connection out of the class3 during that day.
		
\begin{figure*}[t!]
			\begin{center}
				\mbox{
					\subfloat[{Percentage `in time' is higher for occupants than bystanders.}]{
						{\includegraphics[width=0.25\textwidth]{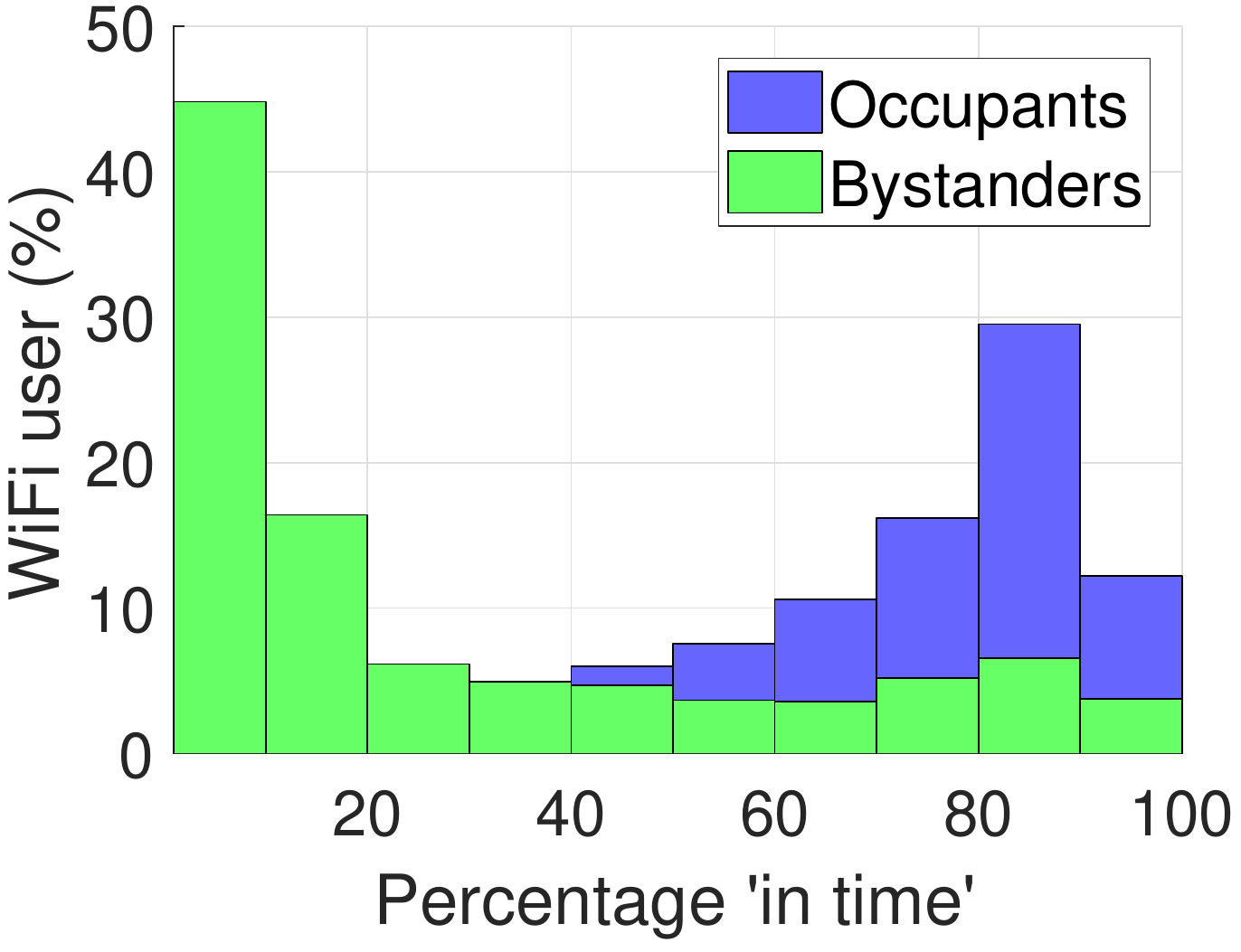}}\quad
						\label{f_featureInTime}
					}
				}
				\hspace{2mm}
				\mbox{
					\subfloat[{Occupants have lower percentage `out time'}]{
						{\includegraphics[width=0.25\textwidth]{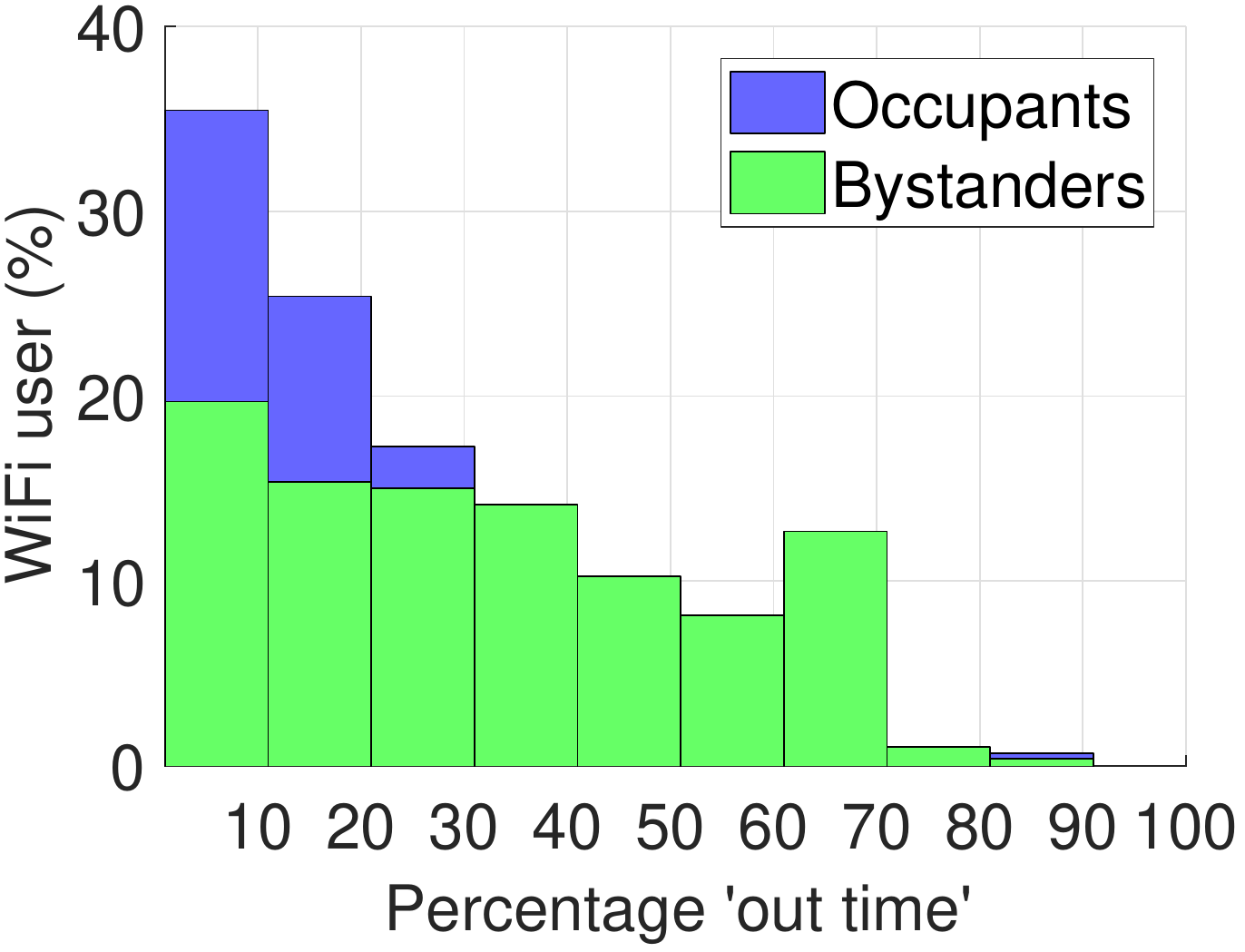}}\quad
						\label{f_featureOutTime}
					}
				}
				\hspace{2mm}
				\mbox{
					
					\subfloat[{Most of the occupants are first seen closer to class start time}]{
						{\includegraphics[width=0.25\textwidth]{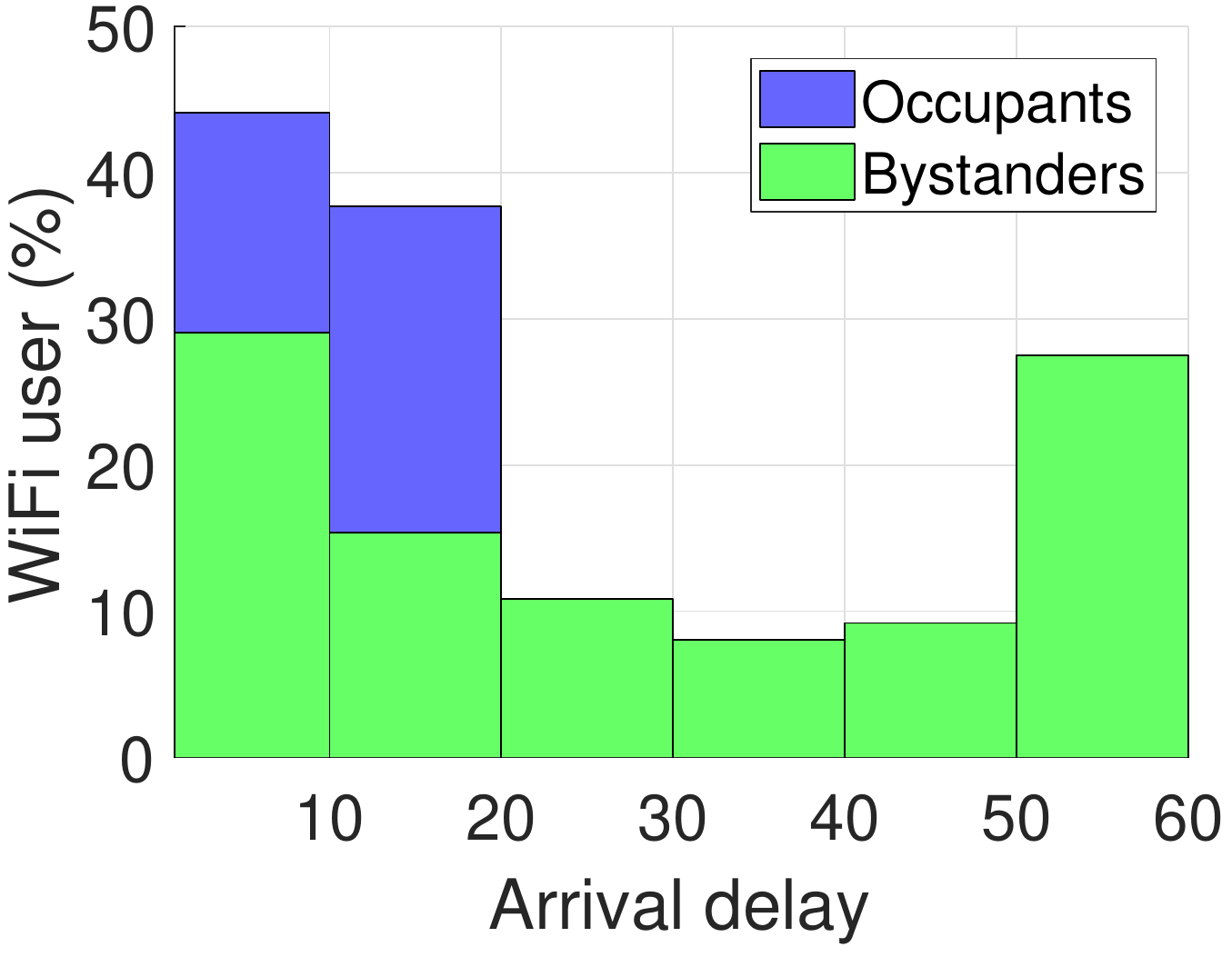}}\quad
						\label{f_featureArrivalDelay}
					}
				}
				\hspace{2mm}
				\mbox{
					\subfloat[{Nearly 95\% of bystanders connected with only a single device during the class}]{
						{\includegraphics[width=0.25\textwidth]{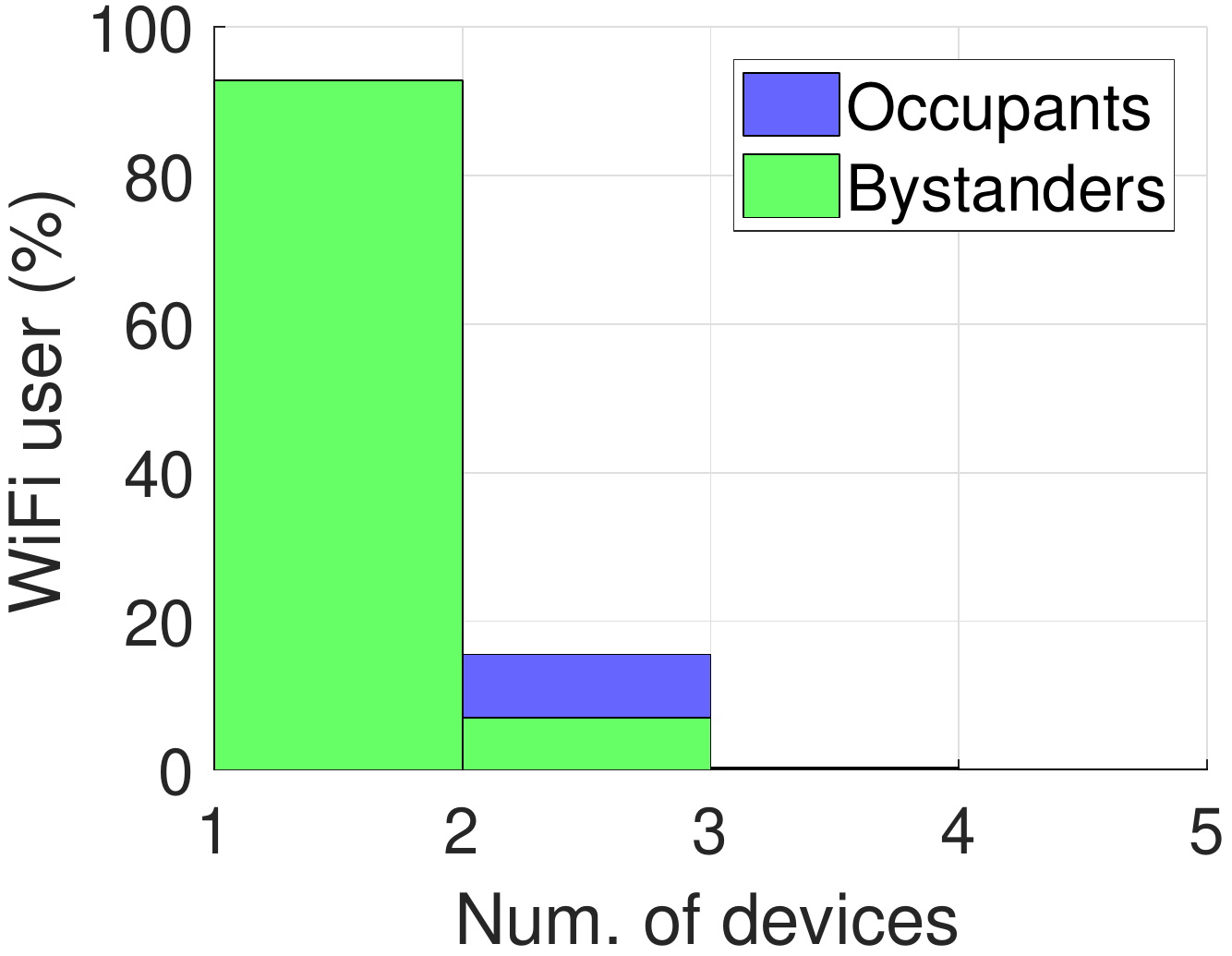}}\quad
						\label{f_featureNoDevices}
					}
				}
				\hspace{2mm}
				\mbox{
					\subfloat[{Majority of bystanders has only one session during class time}]{
						{\includegraphics[width=0.25\textwidth]{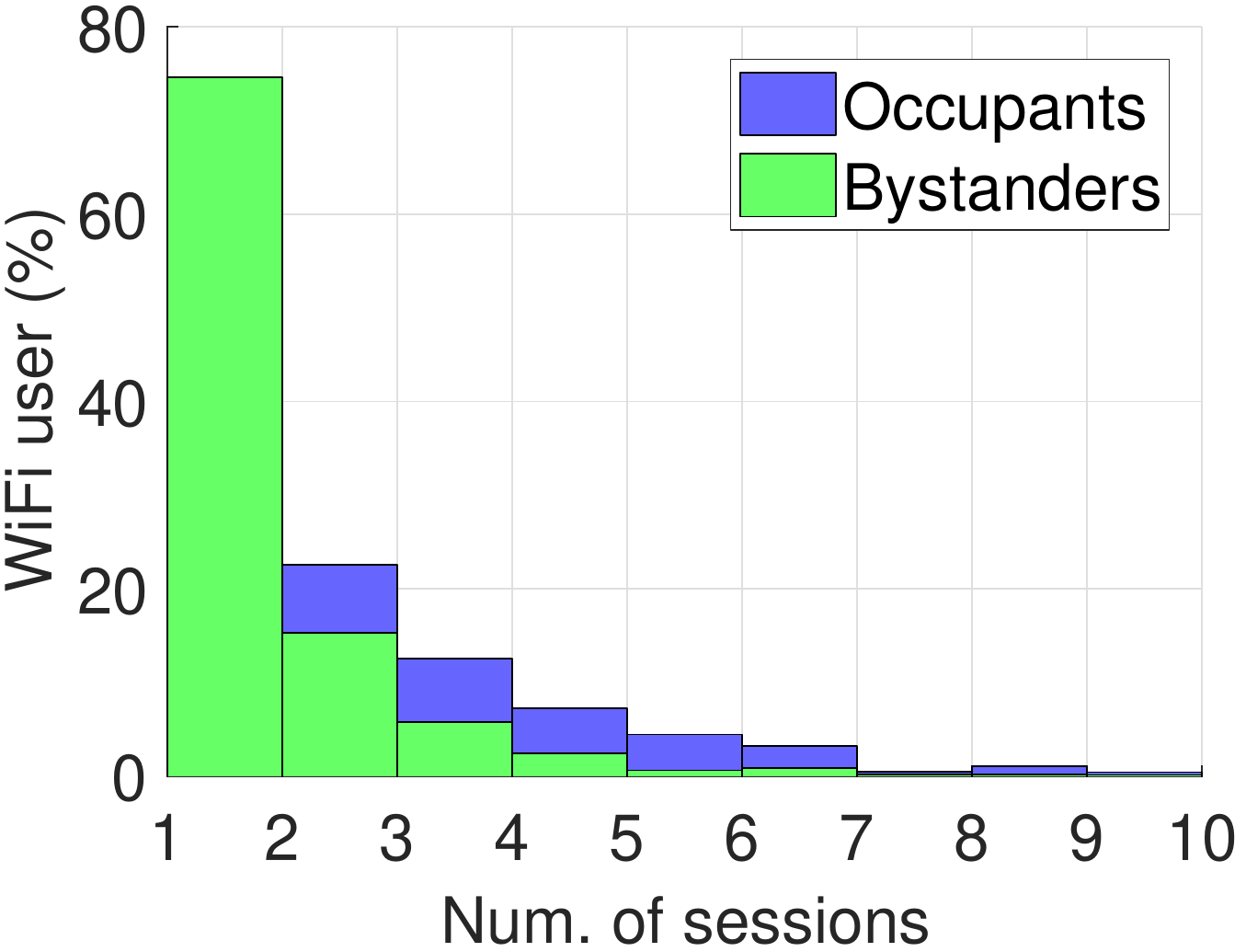}}\quad
						\label{f_featureNoSessions}
					}
				}
				\hspace{2mm}
				\mbox{
				\subfloat[{No visible difference in RSSI of occupants and bystanders.}]{
					{	\includegraphics[width=0.25\textwidth]{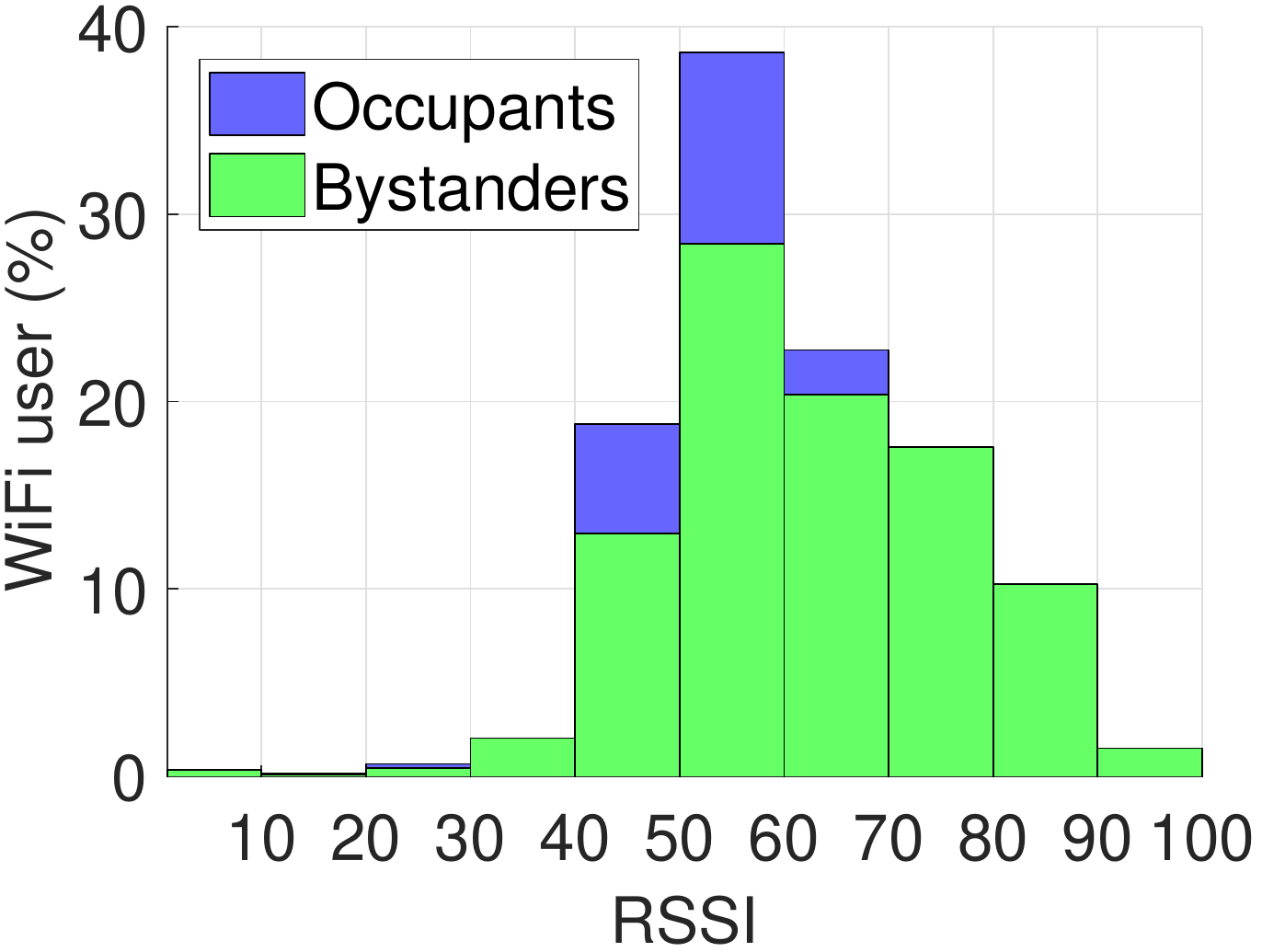}}\quad
					\label{f_featureRSSI}
					}
				}
				\caption{Histogram of features (occupants vs. bystanders)}
				\label{f_regFeatures}
			\end{center}
			
		\end{figure*}

		In Fig.~\ref{f_regFeatures}, we show the distribution of identified features for the two WiFi user groups (\ie room occupants in blue and bystanders in green) using a dataset of 20,000 WiFi users across 2700 classes. Looking at these plots, we can visually distinguish (to a great extent) the two groups by individual features (\ie \(t_{in}\), \(t_{out}\), arrival delay, number of devices, number of sessions, and average RSS) though there are some overlaps -- this shows that our features collectively capture the property of each user group.
		Considering Fig.~\ref{f_featureInTime}, occupants display a mean \(t_{in}\) of $67.9\%$ which is more than double the mean \(t_{in}\) (\ie $27.3\%$) for bystanders. Similarly, occupants of a room can be characterized by lower \(t_{out}\) (\ie $3.0\%$), and lower `arrival delay' (\ie $13.1$ minutes) compared to those of bystanders (\ie $25.1\%$ and $29.1$ minutes, respectively) as shown in Fig.~\ref{f_featureOutTime} and  Fig.~\ref{f_featureArrivalDelay}. 
		Furthermore, occupants on average display slightly more devices (\ie $1.47$) and more sessions (\ie $2.19$) compared to bystanders (\ie $1.08$ and $1.34$) as shown in Fig~\ref{f_featureNoDevices} and Fig.~\ref{f_featureNoSessions} respectively. 
		In terms of RSSI shown in Fig.\ref{f_featureRSSI}, we don't see a significant difference between occupants and bystanders (\ie mean value of $59.4$ vs. $66.4$). This is probably because that devices typically get connected to the AP with the strongest signal regardless of location. We also note that the received signal strength varies by a number of factors such as device type (\eg laptop, mobile phone) and device manufacturer. Additionally, the RSSI recorded in WiFi logs is an average value computed over the whole session.

	\subsection{Supervised Learning for Estimating Classroom Occupancy}\label{S5_AB}
	
\begin{figure}[h]
			\centering
			\includegraphics[width=0.9\textwidth, height=1.65cm]{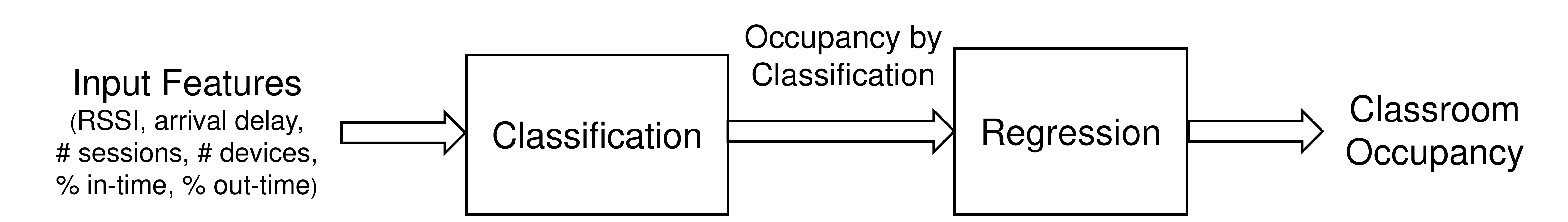}
			
			\caption{System architecture for classroom occupancy estimation.}
			\label{f_methodArchitecture}
		\end{figure}

		In this section we outline our two-step approach for estimating classroom occupancy. Firstly, we classify individual WiFi users as occupant or bystander using the six features described in \S\ref{S5_A}. To train our classifier model,  we extract the six features for each WiFi user, and obtain  users' label by checking the WiFi session logs against the class list.   
		Secondly, we employ a regression algorithm to predict the room occupancy using the count of occupants predicted by the classifier model. The ground-truth data for the regression was obtained by the actual count of the room occupants. The regression step compensates for the room occupants who are not captured by the WiFi logs. It is important to note that, nearly 18\% of the students on average (from the 40 classes in initial analysis), do not connect to wireless network during a class. Fig.~\ref{f_methodArchitecture} illustrates an overview of our proposed approach.
		
		\subsubsection{Classification of WiFi users:\\}
		
			We use the collected dataset of 20,000 WiFi users across 2700 classes and apply widely used binary data classification techniques, namely logistic regression, SVM (Support Vector Machine) and LDA ({Linear Discriminant Analysis), to distinguish room occupants from bystanders.

			For each of WiFi user IDs (unique identifier appears in WiFi data), we extracted the features: (1) Percentage of `in time' (\(t_{in}\)); (2) Percentage of `out time' (\(t_{out}\)); (3) Arrival delay; (4) Number of sessions; (5) Number of devices; (6) RSSI, as defined in \S\ref{S5_A}. We now rank the features using univariate feature selection method with F-test (a built-in function of Python sklearn library) for numerical variables. As shown in Fig.~\ref{f_featureRanking}, the feature \(t_{in}\) contains the highest information followed by (in order) the features \(t_{out}\), Arrival delay, Number of devices, Number of sessions, and RSSI.
			These features are fed as inputs to the model that classifies a WiFi user as an occupant or a bystander. The list of enrolled students for 12 classes are collected as the ground truth for classification. Based on the assumption that students who appear in both the class list and the WiFi session logs for the class are in fact inside the room, we labeled such WiFi users as occupants and others as bystanders.
			
			We showed in our previous work~\cite{wifiLCN2018}, that the LDA classification displays the best performance among classifiers we used. It correctly classified  room occupants and bystanders $85\%$ and $83$\% of the time respectively.
			
		\subsubsection {Regression Analysis:\\}
			
			The occupancy computed by the LDA classification only accounts for room occupants who connected to the WiFi network. However, we know that there are occupants (those with no device, or with only 3G/4G-enabled devices) whose traces are not found in WiFi session data of the classroom. As observed earlier (in Fig. \ref{f_LDAOccuVSgroundTruth}), there is a linear correlation between the room occupants count and the WiFi users count by LDA, we now develop a univariate linear regression model that takes WiFi users count by LDA as input and generates the classroom occupancy as output. The regression model corrects the occupancy estimated by LDA classification, yielding a value closer to the actual classroom occupancy.

			\begin{figure}
				\centering
				\begin{minipage}{.45\textwidth}
					\centering
					\includegraphics[width=\textwidth]{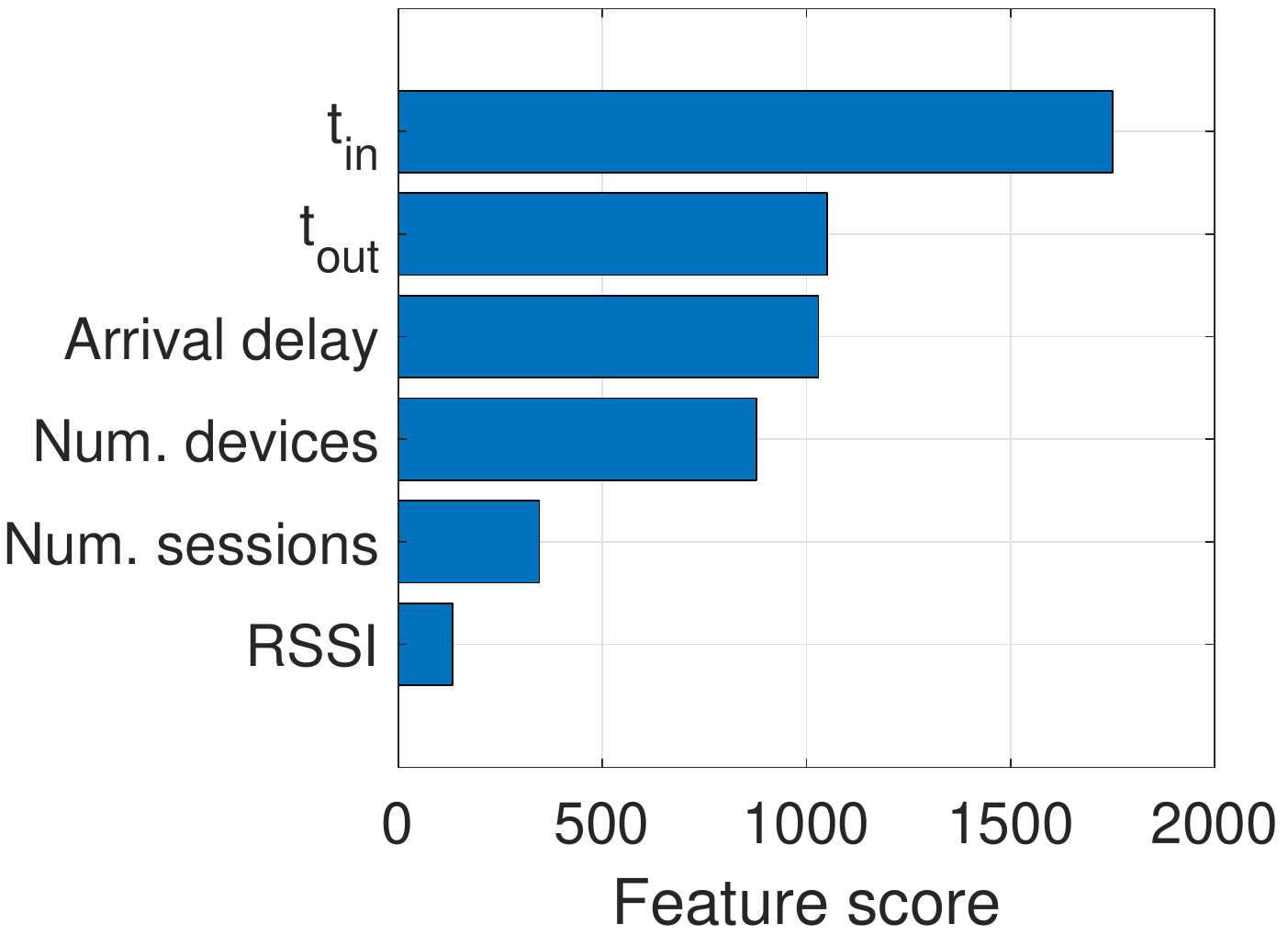}
					\caption{Feature ranking using univariate feature selection.}
					\label{f_featureRanking}
				\end{minipage}%
				\hspace{0.5cm}
				\begin{minipage}{0.45\textwidth}
					\centering
					\includegraphics[width=\textwidth]{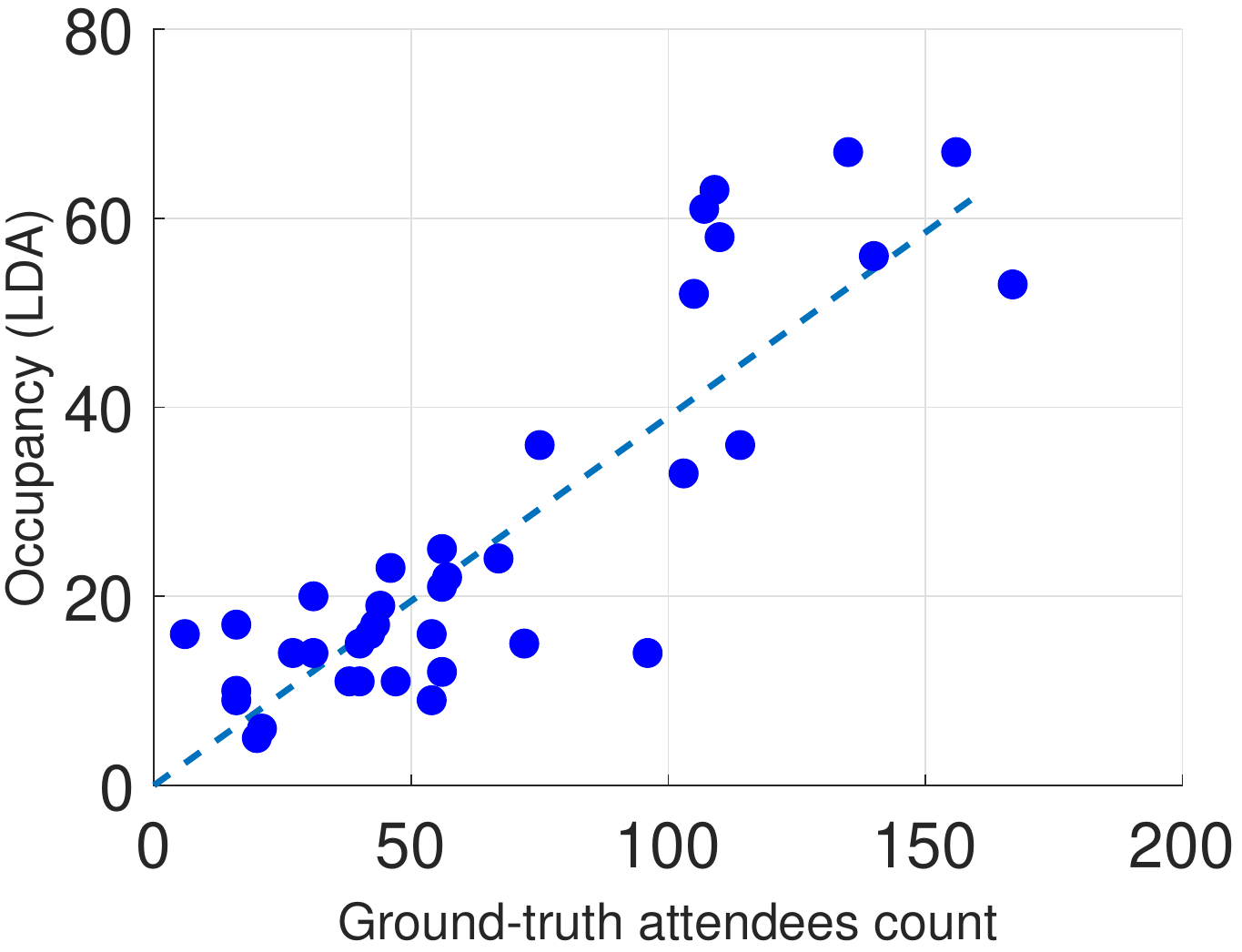}
					\caption{Room occupancy count and the WiFi user count are linearly correlated.}
					\label{f_LDAOccuVSgroundTruth}
				\end{minipage}
				
			\end{figure}

			We extended the data set from our previous study~\cite{wifiLCN2018} to collect 2700 classes during 2017-July-31 to 2017-October-27 (\ie{sem2-2017}) and 2018-February-26 to 2018-June-1 (\ie{sem1-2018}). The data were spanning across different courses and 7 classrooms on our campus. In the sample, $46\%$ of classes lasted one hour, $45\%$ lasted two hours, $5\%$ lasted three hours, $2\%$ lasted one and a half hours, and $1\%$ lasted four hours or two and a half hours. The rooms are scheduled for lectures most of the time while paper-based exams are also occasionally possible, therefore we expected anomalous periods with little WiFi use. However, we omitted the data from weeks when classes were not held (\eg mid-semester break). 
			For each class, we predicted individual WiFi user's presence in the room through classification and computed the number of occupants to be fed to regression analyzer as the input variable. {The regression training set was labeled using the actual count of room occupants.}
			We evaluated the performance of LR (Linear Regression) and SVR (Support Vector Regression) in estimating room occupancy in our previous work~\cite{wifiLCN2018}, and showed that both LR and SVR regression algorithms result in similar prediction performance.
		
		\subsection{Occupancy Estimation Results}
			In this section, we present the performance of our method to estimate classroom occupancy. We employ a two-step supervised learning approach.
			For all classrooms considered in our study, we identified those APs to which occupants get connected using our mapping method explained in previous Section.
			In what follows next, we show the results of LDA-based classification followed by LR-based prediction.

			\subsubsection{Performance Comparison:\\}\label{sec:perfComp}
				We now compare the performance of our method with special-purpose occupancy sensing (\ie beam counting) and prior work.
			
				\textbf{WiFi Sensing vs. Beam Counting:}
					In a parallel research to our work~\cite{Tara}, the same rooms considered in our study were instrumented with EvolvePlus wireless beam counters to estimate the room occupancy.  {We compare the error rate of occupancy estimation by directly applying linear regression to: (a) WiFi counts and (b) Enrolled WiFi counts, (c) standalone LDA classification, (d) LDA classification followed by linear regression (our method), and (e) Beam counters.} The WiFi Counts and Enrolled WiFi Counts are defined as the unique number of student identifiers and the unique number of enrolled student identifiers appeared in WiFi data during the class of interest respectively as termed by \(Occupancy_{WiFi}\) and \(Occupancy_{EnrolledWiFi}\) in \S\ref{sec:WiFiSensing}. We compute the occupancy output of LDA classification by summing up the number of WiFi users predicted as occupants while occupancy output by regression is computed by using the output of the LDA classification as the input to linear regression. The beam counter consists of a pair of sensors which are positioned across a doorway, each generates an IR beam. They are used to count the number of people passing through the beam in each direction.   We computed classroom occupancy from the data generated from beam counters by subtracting the total exits from the total entries across all doorways of a classroom during a particular class. Since it is intuitive to interpret results in percentage terms we used the symmetric mean absolute percentage error (sMAPE) (\ref{eq_smape}) in our comparison. 
			
					\vspace{-0.1cm}
					\begin{equation}
					\label{eq_smape}%
					sMAPE=\frac{100\%}{n}\sum_{i=1}^{n}\frac{\mid F_i - A_i \mid}{\mid F_i \mid + \mid A_i\mid}
					\end{equation}
					
					where \(A_i\) is the actual value, \(F_i\) is the forecast value for $i$\textsuperscript{th} regression input -- there are  $n$ inputs. We show in Table \ref{t_ErrorOccuEstimate} the value of sMAPE for various approaches.
					
					\begin{table}[t!]
						\renewcommand{\arraystretch}{1.3}
						\caption{{Error rate (sMAPE) for various methods of estimating occupancy across all rooms.}} 
						\label{t_ErrorOccuEstimate}
						\centering
						\adjustbox{max width=0.9\columnwidth}{
							\begin{tabular}{|c|c|c|c|c|c|}
								\hline
								& (a) WiFi Counts & (b) Enrolled WiFi Counts & (c) LDA & (d) Our method & (e) Beam Counters\\
								\hline
								sMAPE & $26.3$\% & $24.1$\% & $20.15$\% & $13.1$\% &$13.0$\% \\
								\hline
						\end{tabular}}
						\vspace{-0.4cm}
					\end{table}
			
					We see that the largest error is obtained when we directly model occupancy using WiFi Counts and the error reduces when filtered non-enrolled connections using the class lists -- using linear regression model with enrolled students as input. We achieve a lower error when a classification method is used. The objective of the classification is to remove the bystanders who corrupt the room occupancy estimation in a dense campus environment by connecting from outside the particular room. To compensate for room occupants who are not captured by WiFi we proposed employing a regression step. Regression after classification (\ie column `Our method') yielded better accuracy than standalone LDA classification displaying the importance of having a two-stage approach so as to remove bystanders and also to capture the actual room occupants who are not captured by WiFi. A closer look at the predictions of regression showed that it inflates the result of classification such that it gets closer to the actual occupancy. 
					
					In our previous work~\cite{wifiLCN2018}, the lowest percentage error was obtained with beam counters, however introduction of AP mapping to classrooms and extension of the dataset improved the performance of our method to become comparable to dedicated beam-counting sensors. Typically, beam-counting sensors can only be used for closed spaces (with doorways), and yield acceptable accuracy when doorways are narrow  -- beam-counters fail to count a group of people walking side-by-side in/out, specially for rooms with wider doorways \cite{Tara}. On the other hand, WiFi-based sensing seems more generic in terms of scope since the infrastructure is available in all spaces (open and closed) across the university campus. Also, room settings do not affect the accuracy of WiFi-based estimation.

				\textbf{Comparing our method with prior works:}
					We compare the performance of our method with that of state-of-the-art methods in Table \ref{t_MAERelatedWork}. In prior work, errors are computed in terms of mean absolute error (MAE). We, instead, normalize MAE ($N_{MAE}$) by dividing it by the corresponding sample size of each study. Therefore, numbers shown in Table \ref{t_MAERelatedWork} reflect their sample size.} In Table \ref{t_MAERelatedWork}, we also show a cost figure for deployment, maintenance and computational complexity of each method. As shown, lowest $N_{MAE}$ of 0.09 is obtained for camera and ambient sensing methods \cite{Paci2015}, however this method incurs a very high cost. Our method outperforms \cite{Raykov2016}, \cite{Sgouropoulos} and \cite{Yoshida2015} when error, cost, and number of occupants are collectively considered. Majority of methods in prior work were only evaluated for relatively smaller rooms (\ie capacity of up to 40) and none of them mentioned the scalability of their method (up to what occupancy level their method achieves a reasonable accuracy). The accuracy of our method, instead, slightly varies for different levels of occupancy (from $8.8$\% to $13.8$\%), as shown in Table \ref{t_generalization}. Given the performance, our method which comes at zero cost and is tested in rooms with occupants ranging from 0 to 500, seems more palatable.

					\begin{table}[!t]
						\renewcommand{\arraystretch}{1.3}
						\caption{Error comparison (prior work vs. our method).}
						\label{t_MAERelatedWork}
						\vspace{-0.5mm}
						\centering
						\adjustbox{max width=0.7\columnwidth}{
							\begin{tabular}{|c|c|c|c|c|}
								\hline
								Sensing Method &Occupants &Normalized MAE $(N_{MAE}$) & Cost \\
								\hline
								Camera + Ambient sensing(~\cite{Paci2015}) &0 - 150 &$0.09$ & High\\
								\hline
								Our method (WiFi)  &0 - 500 & $0.10$  & \textbf{Zero}\\
								\hline
								Raspberry Pi + WiFi APs(~\cite{Yoshida2015}) &0 - 8 &$0.12$ & Low\\
								\hline				
								PIR(~\cite{Raykov2016}) &0 - 14 &$0.14$ & Medium\\
								\hline				
								Camera(~\cite{Sgouropoulos}) &0 -  8 &$0.29$ & High \\
								\hline				
						\end{tabular}}
						\vspace{-2mm}
					\end{table}

			\subsubsection{Robustness of our Approach:\\}
			
				In this section, we analyze the performance of our method at various conditions of occupancy levels and room capacities. First, we compute the error in estimating classroom occupancy for short (less than 2-hours) and longer (2-hours or more) classes separately. The percentage error (sMAPE) is found to be 10.9\% and 11.9\% respectively for short and long classes. This shows that that class duration does not have a significant impact on occupancy estimation. We believe this is because our features for classifying WiFi users (\ie percentage in time, percentage out time, arrival delay, number of sessions, number of devices, and RSSI) are independent of class duration.
				
				Next, we quantify the error of our estimation with respect to occupancy level and room capacity as shown in Table 5. Considering class occupancy levels in Table \ref{occu_level}, the error of our method varies from 8.8\% to 13.8\% with a mean of 11.4\% and variance of 2\%. Similarly, for rooms with different capacities, shown in Table \ref{room_capacity}, the estimation errors fall between 8.6\% to 15.2\% with a mean of 11.4\% and variance of 2.5\%. In summary, the estimation error is fairly consistent (with slight variations) across classes of varying occupancy levels and room sizes.
				
				\begin{table}[htbp]
					\small
					\centering
					\caption{Average percentage error (sMAPE) of our method by occupancy-level}
					\label{t_generalization}
						\adjustbox{max width=0.45\columnwidth}{
							{\begin{tabular}{|c|c|}
									\hline
									Occupancy Level & Average sMAPE \\
									\hline
									0 - 100 & 13.8\%\\
									\hline
									101 - 200 & 8.8\%\\
									\hline
									201 - 300 & 13.1\%\\
									\hline				
									301 - 400 & 10.8\%\\
									\hline				
									401 - 500 & 10.5\%\\
									\hline
								\end{tabular}
								\label{occu_level}}}			
						\end{table}
					
				\begin{table}[htbp]
					\small
					\centering
					\caption{Average percentage error (sMAPE) of our method by room capacity.}
					\label{t_generalization}	
					\adjustbox{max width=0.45\columnwidth}{
					{\begin{tabular}{|l|c|c|}
						\hline
						& Capacity & Average sMAPE\\
						\hline
						Room 1 (Mat227) & 42 & $9.1$\%\\
						\hline
						Room 2 (Mat228) & 42 & $8.6$\%\\
						\hline
						Room 3 (MatC) & 110 &$9.7$\%\\
						\hline
						Room 4 (CLB8) & 231 &$13.6$\%\\
						\hline
						Room 5 (MatB) & 246 & $15.2$\%\\
						\hline
						Room 6 (MatA) & 472 &$11.4$\%\\
						\hline
						Room 7 (CLB7) & 497 & $12.3$\%\\
						\hline
					\end{tabular}
					\label{room_capacity}}}
				\end{table}

\section{Discussion}\label{sec:discussion}
	
	In this study, we have developed methods to first map APs to classrooms, and next estimate classrooms occupancy with the use of WiFi session data of their corresponding APs. The main advantage of our method is that we use data from existing WiFi infrastructure without needing new specialized hardware, thus saving costs of procurement, installation, and maintenance. We have shown that the performance of our method is comparable to beam counter sensors used in selected rooms of our university campus. Our results demonstrate the generality of our method which performs fairly consistently across classrooms of various size, duration, and attendance-level. On the other hand, one may argue that our method requires additional sources of information on class timetabling and enrollment list. This dependency would prevent our method to estimate the occupancy of social or open spaces. Another concern is related to the privacy of data from WiFi session logs, even though we obtained ethics approval from our university for this study. Lastly, there is a body of works on WiFi localization which promise to yield an accurate estimation of room occupancy \cite{Vasisht2016, Jiang2012}. However, these methods demand analysis of wireless channel information which is not available in our dataset. Considering the limitations of our work, one possible future work (given the same dataset) would be estimating occupancy for extended set of spaces in which activities do not adhere to a fixed timetable.

\section{Conclusion}\label{sec:concl}
	Quantitative measures for learning space utilization and student attendance  are of importance to university managers. 
	In this paper, we have proposed and evaluated machine learning-based methods to estimate classrooms occupancy using data collected from a dense wireless network in a large university campus. 
	We have analyzed real session logs of 70 APs from our campus WiFi network to understand coverage of APs and dynamics of users connections to APs. We, then, identified two features for each AP and thereby developed a clustering-based method to automatically map APs to their respective rooms. Lastly, we employed LDA followed by a regression to first classify WiFi users as room occupants and bystanders, and then estimate the occupancy count of a room. Our WiFi sensing method displayed a palatable accuracy compared to special-purpose beam counters.

\bibliographystyle{ACM-Reference-Format}
\bibliography{WiFiAPclassroom}

\end{document}